\documentclass[
  journal=pasa,
  manuscript=research-article,
  year=2025,
  volume=XX,
]{cup-journal}

\usepackage{amsmath}
\usepackage[nopatch]{microtype}
\usepackage{booktabs}
\usepackage{microtype,siunitx,booktabs}
\usepackage{makecell}
\usepackage{hyperref} 
\hypersetup{colorlinks,citecolor=blue,linkcolor=blue,urlcolor=blue}
\usepackage[nolist]{acronym}
\usepackage[version=3]{mhchem}
\usepackage{mathtools}
\usepackage{adjustbox}
\usepackage[normalem]{ulem}
\usepackage{xcolor}
\usepackage{comment}
\usepackage{amssymb}
\usepackage{upgreek}
\usepackage{amsfonts}
\usepackage{pifont} 
\usepackage{tikz} 
\usepackage{graphicx}
\usepackage{float}
\usepackage{footmisc}

\begin{acronym}[AWGN]
\acro{2MASS}{Two Micron All Sky Survey}
\acro{2MASX}{Two Micron All Sky Survey Extended Source Catalogue}
\acro{30Dor}[30 Dor]{30~Doradus}
\acro{AGN}{active galactic nucleus}
\acrodefplural{AGN}{active galactic nuclei}
\acro{AGN}{active galactic nuclei}
\acro{ATCA}{Australia Telescope Compact Array}
\acro{ATESP}{Australia Telescope ESO Slice Project}
\acro{ATNF}{Australia Telescope National Facility}
\acro{ATOA}{Australia Telescope Online Archive}
\acro{AT20G}{Australia Telescope 20 GHz Survey}
\acro{ASKAP}{Australian Square Kilometre Array Pathfinder}
\acro{ARC}{Australian Research Council}
\acro{BETA}{Boolardy Engineering Test Array}
\acro{BL}[BL Lac]{BL Lacertae Objects}
\acro{CASS}{CSIRO Astronomy and Space Science}
\acro{CABB}{Compact Array Broadband Back-end}
\acro{CD}{core degenerate}
\acro{CHII}[{\sc CHii}]{Compact \textsc{Hii}}
\acro{CC}{Core Collapse}
\acro{Cas A}{Cassiopeia A}
\acro{CASDA}{CSIRO ASKAP Science Data Archive}
\acro{CARTA}{Cube Analysis and Rendering Tool for Astronomy}
\acro{CMB}{Cosmic Microwave Background}
\acro{CSIRO}{Commonwealth Scientific and Industrial Research Organisation}
\acro{CSM}{circumstellar material}
\acro{CSS}{Compact Steep Spectrum}
\acro{DD}{double degenerate}
\acro{DR3}{Data Release 3}
\acro{DS9}[\textsc{DS9}]{\textsc{SAOImage DS9}}
\acro{DSA}{diffusive shock acceleration}
\acro{DSS}{Digital Sky Survey}
\acro{EBHIS}{Effelsberg--Bonn H{\sc i} Survey}
\acro{EM}{Electromagnetic}
\acro{EMU}{Evolutionary Map of the Universe}
\acro{eROSITA}{extended R\"Ontgen Survey with an Imaging Telescope Array}
\acro{ev}[eV]{electronvolt\acroextra{: 1 eV $\approx 1.6 \times 10^{-19}$ J}}
\acro{FITS}[\textsc{Fits}]{Flexible Image Transport System}
\acro{FRB}{Fast Radio Bursts}
\acro{FSRQ}{Flat Spectrum Radio Quasars}
\acro{FWHM}{Full Width at Half-Maximum}
\acro{GASS}{Galactic All-Sky Survey}
\acro{GB}{giant branch}
\acro{GLEAM}{GaLactic Extragalactic All-sky MWA}
\acro{GRB}{Gamma-ray burst}
\acro{GPS}{Gigahertz Peak Spectrum}
\acro{HEMT}{High Electron Mobility Transistor}
\acro{H.E.S.S.}{High Energy Stereoscopic System}
\acro{HFP}[HFP]{High-Frequency Peaker}
\acro{HPBW}{Half Power Beam Width} 
\acro{HzRGs}{High Redshift Radio Galaxies}
\acro{pHFP}[pHFP]{Potential High Frequency Peaker}
\acro{HI}[H{\sc i}]{Neutral Atomic Hydrogen} 
\acro{HST}{\textit{Hubble Space Telescope}}
\acro{ICM}{intracluster medium}
\acro{IAU}{International Astronomical Union}
\acro{IFRSs}{Infrared Faint Radio Sources}
\acro{ISM}{Interstellar Medium}
\acro{IFRS}{Infrared Faint Radio Source}
\acro{IR}{infrared}
\acro{IRAC}{Infrared Array Camera}
\acro{JY}[Jy]{Jansky\acroextra{, 1 Jy = $10^{-26} \times \mathrm{W~ m}^{-2}~\mathrm{Hz}^{-1}$}} 
\acro{LLS}{Largest Linear Size}
\acro{LFAA}{Low-Frequency Aperture Array}
\acro{LMC}{Large Magellanic Cloud}
\acro{LOFAR}{Low-Frequency Array}
\acro{LSO}[LSOs]{Large Scale Objects}
\acro{LBV}{Luminous Blue Variable}
\acro{MACHO}{Massive Astrophysical Compact Halo Objects}
\acro{MC}[MC]{Magellanic Cloud}    
\acro{MCELS}{Magellanic Cloud Emission Line Survey}
\acro{MW}{Milky Way}
\acro{MIRIAD}[\textsc{Miriad}]{Multichannel Image Reconstruction, Image Analysis and Display}
\acro{MIPS}{Multiband Imaging Photometer for Spitzer}
\acro{MIR}{Mid-Infrared}
\acro{MIT}{Massachusetts Institute of Technology}
\acro{MOST}{Molonglo Observatory Synthesis Telescope}
\acro{MGPS-2}{second epoch Molonglo Galactic Plane Survey}
\acro{MQS}{Magellanic Quasars Survey}
\acro{MRC}{Molonglo Reference Catalogue of Radio Sources}
\acro{MWA}{Murchison Widefield Array}
\acro{NRAO}{National Radio Astronomy Observatory}
\acro{NSR}{Nova Super Remnant}
\acro{NVSS}{NRAO VLA Sky Survey}
\acro{OPAL}{Online Proposal Applications \& Links}
\acrodefplural{ORC}{Odd Radio Circles}
\acro{ORC}{Odd Radio Circle}
\acro{OVV}{Optically Violent Variable Quasars}    
\acro{PACS}{Photodetector Array Camera and Spectrometer}
\acro{PAF}{Phased Array Feed}
\acro{pc}{parsec\acroextra{: 1 pc $\simeq 3.09 \times 10^{16}$ m}}
\acro{PMN}{Parkes-MIT-NRAO}
\acro{PN}{Planetary Nebula}
\acro{PNe}{Planetary Nebulae}
\acro{PWN}{pulsar-wind nebula}
\acrodefplural{PWN}[PWNe]{pulsar-wind nebulae}
\acro{QSO}{Quasi-Stellar Object}
\acro{RA}{Right Ascension}
\acro{RFI}{Radio-Frequency Interference}
\acro{RMS}[RMS]{Root Mean Squared}
\acro{RM}{rotation measure}
\acro{SARAO}{South African Radio Astronomy Observatory}
\acro{SDSS}{Sloan Digital Sky Survey}
\acro{SED}{spectral energy distribution}
\acro{SGPS}{Southern Galactic Plane Survey}
\acro{SHS}{SuperCOSMOS H$\alpha$ Survey}
\acro{SI}[$\alpha$]{Spectral Index\acroextra{, $S \propto \nu^\alpha$}}
\acro{SKA}{Square Kilometre Array}
\acro{SMB}{Super Massive Blackholes}
\acro{SMC}{Small Magellanic Cloud}
\acro{SN}{Supernova}
\acro{SUMSS}{Sydney University Molonglo Sky Survey}
\acro{SNR}{supernova remnant}
\acrodefplural{SNR}{Supernova Remnants}
\acro{SD}{single-degenerate}
\acro{SUMSS}{Sydney University Molonglo Sky Survey}
\acro{SMBH}{Super Massive Black Hole}
\acro{POSSUM}{Polarisation Sky Survey of the Universe's Magnetism}
\acro{SMGPS}{SARAO MeerKAT Galactic Plane Survey}
\acro{SMASH}{Survey of the Magellanic Stellar History}
\acro{TOPCAT}[\textsc{Topcat}]{Tool for OPerations on Catalogues And Tables}
\acro{USS}{Ultra Steep Spectrum}
\acro{YSOs}{young stellar objects}
\acro{WBAC}{Wide-Band Analogue Correlator}
\acro{WD}{white dwarf}
\acro{WIFES}[WiFeS]{Wide-Field Spectrograph}
\acro{WISE}{Wide-Field Infrared Survey Explorer}
\acro{WR}{Wolf-Rayet}
\acro{VLA}{Very Large Array} 
\acro{VLBI}{Very Long Baseline Interferometry} 
\acro{VLSR}[\textbf{$v_{lsr}$}]{Velocity in the Line of Sight}
\end{acronym}

\newcommand{\ergcm}[1]{erg\,cm$^{-2}$\,s$^{-1}$}
\def\HI{\hbox{H{\sc i}}}
\def\HII{\hbox{H{\sc ii}}}

\newcommand{\He}{He\,{\sc ii}}
\newcommand{\Halpha}{H${\alpha}$}

\newcommand{\D}{$^\circ$}

\newcommand{\kms}{km\,s$^{-1}$}
\newcommand{\g}{G305.4--2.2}
\newcommand{\farcm}{\mbox{\ensuremath{.\mkern-4mu^\prime}}}
\newcommand{\farcs}{\mbox{\ensuremath{.\!\!^{\prime\prime}}}}
\newcommand{\fdg}{\mbox{\ensuremath{.\!\!^\circ}}}

\def\HII{\hbox{H{\sc ii}}}
\def\HI{\hbox{H{\sc i}}}

\def\arcmin{\hbox{$^\prime$}}
\def\arcsec{\hbox{$^{\prime\prime}$}}
\def\kms{km\,s$^{-1}$}

\title{TELEIOS (\g) $-$ THE MYSTERY OF A PERFECTLY SHAPED NEW GALACTIC SUPERNOVA REMNANT}

\author{M. D. Filipovi\'c}
\affiliation{Western Sydney University, Locked Bag 1797, Penrith South DC, NSW 2751, Australia}
\email[M. D. Filipovi\'c]{m.filipovic@westernsydney.edu.au, Z. J. Smeaton, 19594271@student.westernsydney.edu.au}

\author{Z. J. Smeaton}
\affiliation{Western Sydney University, Locked Bag 1797, Penrith South DC, NSW 2751, Australia}

\author{R. Kothes}
\affiliation{Dominion Radio Astrophysical Observatory, Herzberg Astronomy \& Astrophysics, National Research Council Canada, P.O. Box 248, Penticton, BC V2A 6J9, Canada}

\author{S. Mantovanini}
\affiliation{International Centre for Radio Astronomy Research, Curtin University, Bentley, WA 6102, Australia}

\author{P. Kosti\'c}
\affiliation{Astronomical Observatory, Volgina 7, 11060 Belgrade, Serbia}

\author{D. Leahy}
\affiliation{Department of Physics and Astronomy, University of Calgary, Calgary, Alberta, T2N IN4, Canada}

\author{A. Ahmad}
\affiliation{Western Sydney University, Locked Bag 1797, Penrith South DC, NSW 2751, Australia}

\author{G. E. Anderson}
\affiliation{International Centre for Radio Astronomy Research, Curtin University, Bentley, WA 6102, Australia}

\author{M. Araya}
\affiliation{Escuela de F\'isica, Universidad de Costa Rica, San Jos\'e, 11501-2060, Costa Rica}

\author{B. Ball}
\affiliation{Department of Physics, University of Alberta, 4-181 CCIS, Edmonton, Alberta T6G 2EI, Canada}

\author{W. Becker}
\affiliation{Max-Planck-Institut f\"ur extraterrestrische Physik, Gie\ss enbachstra\ss e 1, 85748 Garching, Germany}
\alsoaffiliation{Max-Planck-Institut f\"ur Radioastronomie, Auf dem H\"ugel 69, 53121, Bonn, Germany}

\author{C. Bordiu}
\affiliation{INAF\,$-$\,Osservatorio Astrofisico di Catania, Via S. Sofia 78, I-95123, Catania, Italy}

\author{A. C. Bradley}
\affiliation{Western Sydney University, Locked Bag 1797, Penrith South DC, NSW 2751, Australia}

\author{R. Brose}
\affiliation{School of Physical Sciences and Centre for Astrophysics \& Relativity, Dublin City University, D09 W6Y4 Glasnevin, Ireland}
\alsoaffiliation{Astronomy \& Astrophysics Section, School of Cosmic Physics, Dublin Institute for Advanced Studies, DIAS Dunsink Observatory, Dublin D15 XR2R, Ireland}
\alsoaffiliation{Institute of Physics and Astronomy, University of Potsdam, 14476 Potsdam-Golm, Germany}

\author{C. Burger-Scheidlin}
\affiliation{Astronomy \& Astrophysics Section, School of Cosmic Physics, Dublin
Institute for Advanced Studies, DIAS Dunsink Observatory, Dublin D15
XR2R, Ireland}
\alsoaffiliation{School of Physics, University College Dublin, Belfield, Dublin, D04 V1W8, Ireland}

\author{S. Dai}
\affiliation{Australia Telescope National Facility, CSIRO, Space and Astronomy, PO Box 76, Epping, NSW 1710, Australia}
\alsoaffiliation{Western Sydney University, Locked Bag 1797, Penrith South DC, NSW 2751, Australia}

\author{S. Duchesne}
\affiliation{Australia Telescope National Facility, CSIRO Space and Astronomy, PO Box 1130, Bentley, WA 6151, Australia}

\author{T. J. Galvin}
\affiliation{Australia Telescope National Facility, CSIRO Space and Astronomy, PO Box 1130, Bentley, WA 6151, Australia}

\author{A. M. Hopkins}
\affiliation{School of Mathematical and Physical Sciences, 12 Wally's Walk, Macquarie University, NSW 2109, Australia}

\author{N. Hurley-Walker}
\affiliation{International Centre for Radio Astronomy Research, Curtin University, Bentley, WA 6102, Australia}

\author{B. S. Koribalski}
\affiliation{Australia Telescope National Facility, CSIRO, Space and Astronomy, PO Box 76, Epping, NSW 1710, Australia}
\alsoaffiliation{Western Sydney University, Locked Bag 1797, Penrith South DC, NSW 2751, Australia}

\author{S. Lazarevi\'c}
\affiliation{Western Sydney University, Locked Bag 1797, Penrith South DC, NSW 2751, Australia}
\alsoaffiliation{Australia Telescope National Facility, CSIRO, Space and Astronomy, PO Box 76, Epping, NSW 1710, Australia}
\alsoaffiliation{Astronomical Observatory, Volgina 7, 11060 Belgrade, Serbia}

\author{P. Lundqvist}
\affiliation{The Oscar Klein Centre, Department of Astronomy, Stockholm University, AlbaNova, SE-10691 Stockholm, Sweden}

\author{J. Mackey}
\affiliation{Astronomy \& Astrophysics Section, School of Cosmic Physics, Dublin Institute for Advanced Studies, DIAS Dunsink Observatory, Dublin D15 XR2R, Ireland}
\alsoaffiliation{School of Physics, University College Dublin, Belfield, Dublin, D04 V1W8, Ireland}

\author{P. Martin}
\affiliation{IRAP, Universit\'e de Toulouse, CNRS, CNES, F-31028 Toulouse, France}

\author{P. McGee}
\affiliation{School of Physics, Chemistry and Earth Sciences, The University of Adelaide, Adelaide 5005, Australia}

\author{A. Mitra\v{s}inovi\'c}
\affiliation{Astronomical Observatory, Volgina 7, 11060 Belgrade, Serbia}

\author{J. L. Payne}
\affiliation{Western Sydney University, Locked Bag 1797, Penrith South DC, NSW 2751, Australia}

\author{S. Riggi}
\affiliation{INAF\,$-$\,Osservatorio Astrofisico di Catania, Via S. Sofia 78, I-95123, Catania, Italy}

\author{K. Ross}
\affiliation{ICRAR, Australian SKA Regional Centre (AusSRC), Bentley 6102, Australia}

\author{G. Rowell}
\affiliation{School of Physics, Chemistry and Earth Sciences, The University of Adelaide, Adelaide 5005, Australia}

\author{L. Rudnick}
\affiliation{Minnesota Institute for Astrophysics, University of Minnesota, Minneapolis, MN, 55455, USA}

\author{H. Sano}
\affiliation{Faculty of Engineering, Gifu University, 1-1 Yanagido, Gifu 501-1193, Japan}
\alsoaffiliation{National Astronomical Observatory of Japan, Mitaka, Tokyo 181-8588, Japan}

\author{M. Sasaki}
\affiliation{Dr Karl Remeis Observatory, Erlangen Centre for Astroparticle Physics, Friedrich-Alexander-Universit\"{a}t Erlangen-N\"{u}rnberg, Sternwartstra{\ss}e 7, 96049 Bamberg, Germany}

\author{R. Soria}
\affiliation{INAF-Osservatorio Astrofisico di Torino, Strada Osservatorio 20, I-10025 Pino Torinese, Italy}
\alsoaffiliation{Sydney Institute for Astronomy, School of Physics A28, The University of Sydney, Sydney, NSW 2006, Australia}

\author{D. Uro\v{s}evi\'c}
\affiliation{Department of Astronomy, Faculty of Mathematics, University of Belgrade, Studentski trg 16, 11000 Belgrade, Serbia}

\author{B. Vukoti\'c}
\affiliation{Astronomical Observatory, Volgina 7, 11060 Belgrade, Serbia}

\author{J. West}
\affiliation{Department of Physics and Astronomy, University of Calgary, Calgary, Alberta, T2N IN4, Canada}


\begin{document}

\begin{abstract}

We present the serendipitous radio-continuum discovery of a likely Galactic \ac{SNR} \g. 
This object displays a remarkable circular symmetry in shape, making it one of the most circular Galactic \acp{SNR} known. 
Nicknamed Teleios due to its symmetry, it was detected in the new \ac{ASKAP} \ac{EMU} radio--continuum images with an angular size of 1320\arcsec$\times$1260\arcsec\ and PA\,=\,0\D. While there is a hint of possible H$\alpha$ and gamma-ray emission, Teleios is exclusively seen at radio--continuum frequencies. 
Interestingly, Teleios is not only almost perfectly symmetric, but it also has one of the lowest surface brightnesses discovered among Galactic \acp{SNR} and a steep spectral index of $\alpha$=--0.6$\pm$0.3. 
Our best estimates from \HI\ studies and the $\Sigma$--D relation place Teleios as a type~Ia \ac{SNR} at a distance of either $\sim$2.2\,kpc (near-side) or $\sim$7.7\,kpc (far-side). 
This indicates two possible scenarios, either a young (under 1000~yr) or a somewhat older \ac{SNR} (over 10000~yr). 
With a corresponding diameter of 14/48\,pc, our evolutionary studies place Teleios at the either early or late Sedov phase, depending on the distance/diameter estimate. 
However, our modelling also predicts X-ray emission, which we do not see in the present generation of eROSITA images. 
We also explored a type~Iax explosion scenario that would point to a much closer distance of $<$1\,kpc and Teleios size of only $\sim$3.3\,pc, which would be similar to the only known type~Iax remnant SN1181. Unfortunately, all examined scenarios have their challenges, and no definitive \ac{SN} origin type can be established at this stage.
Remarkably, Teleios has retained its symmetrical shape as it aged even to such a diameter, suggesting expansion into a rarefied and isotropic ambient medium. 
The low radio surface brightness and the lack of pronounced polarisation can be explained by a high level of ambient \ac{RM}, with the largest \ac{RM} being observed at Teleios's centre. 

\end{abstract}

\section{Introduction}
\label{sec:introduction}

There is no doubt that \acp{SNR} are essential objects in the evolution of every galaxy, as they enrich and impact the structure and physical properties of the surrounding \ac{ISM}~\citep{book2}. The census of the Galactic \ac{SNR} population is well-known to be incomplete~\citep{2013A&A...549A.107F,2021A&A...651A..86D,2023MNRAS.524.1396B}, as only some 300+ such objects are currently established~\citep{Green,Ferrand2012, Green2024_updatedcatalogue}. As many as up to $\sim$2000 additional Galactic \acp{SNR} are expected to remain undiscovered in the \ac{MW} \citep{2022ApJ...940...63R}. Recently, \citet{2024arXiv240916607A}, \citet{2023MNRAS.524.1396B} and \citet{2019PASA...36...45H,2019PASA...36...48H} demonstrated that a significant number of these missing Galactic \acp{SNR} may have a low surface brightness or be located in complex regions where clear distinctions from other source types (e.g. \HII~regions and \ac{PN}) can prove challenging. Moreover, bright, small-sized (compact), and presumably young \acp{SNR} are not likely to be found in abundance due to their rapid expansion and small size~\citep{2021Univ....7..338R}. However, we have recently found one such object -- the young Galactic \ac{SNR} Perun \citep[G329.9--0.5; ][]{2024MNRAS.534.2918S}. 

In several recent studies with the new generation of radio telescopes such as \ac{ASKAP} and MeerKAT, a number of new \acp{SNR} have been discovered, including the circumgalactic \ac{SNR} J0624--6948~\citep{2022MNRAS.512..265F}, the Galactic \acp{SNR} G288.8--6.3~\citep[Ancora;][]{2023AJ....166..149F,2024A&A...684A.150B}, G181.1--9.5~\citep{2017A&A...597A.116K}, G278.94+1.35 \citep[Diprotodon;][]{2024PASA...41..112F}, and G121.1--1.9 \citep{2023MNRAS.521.5536K}, as well as the Galactic \ac{SNR} candidates G308.73+1.38~\citep[Raspberry;][]{2024RNAAS...8..107L} and G312.65+2.87~\citep[Unicycle;][]{2024RNAAS...8..158S}. These examples demonstrate the ability of newer generation radio-telescopes to discover these Galactic \acp{SNR} in abundance. 
They are mainly located outside the Galactic Plane, where they can preserve their original circular shape for longer. Presumably, they are expanding in a low-density environment, thus resulting in a lower surface brightness than typical \acp{SNR}. 

We present the \ac{ASKAP} radio-continuum detection of a new Galactic \ac{SNR}, \g, nicknamed Teleios (Greek $\tau\epsilon\lambda\epsilon\iota o\varsigma$ -- perfect), due to its almost perfectly circular shape.

\section{DATA}
\label{sec:data}

\subsection{Radio-continuum observations}

    \subsubsection{ASKAP}
\begin{figure*}
    \centering
    \includegraphics[width=0.91\linewidth]{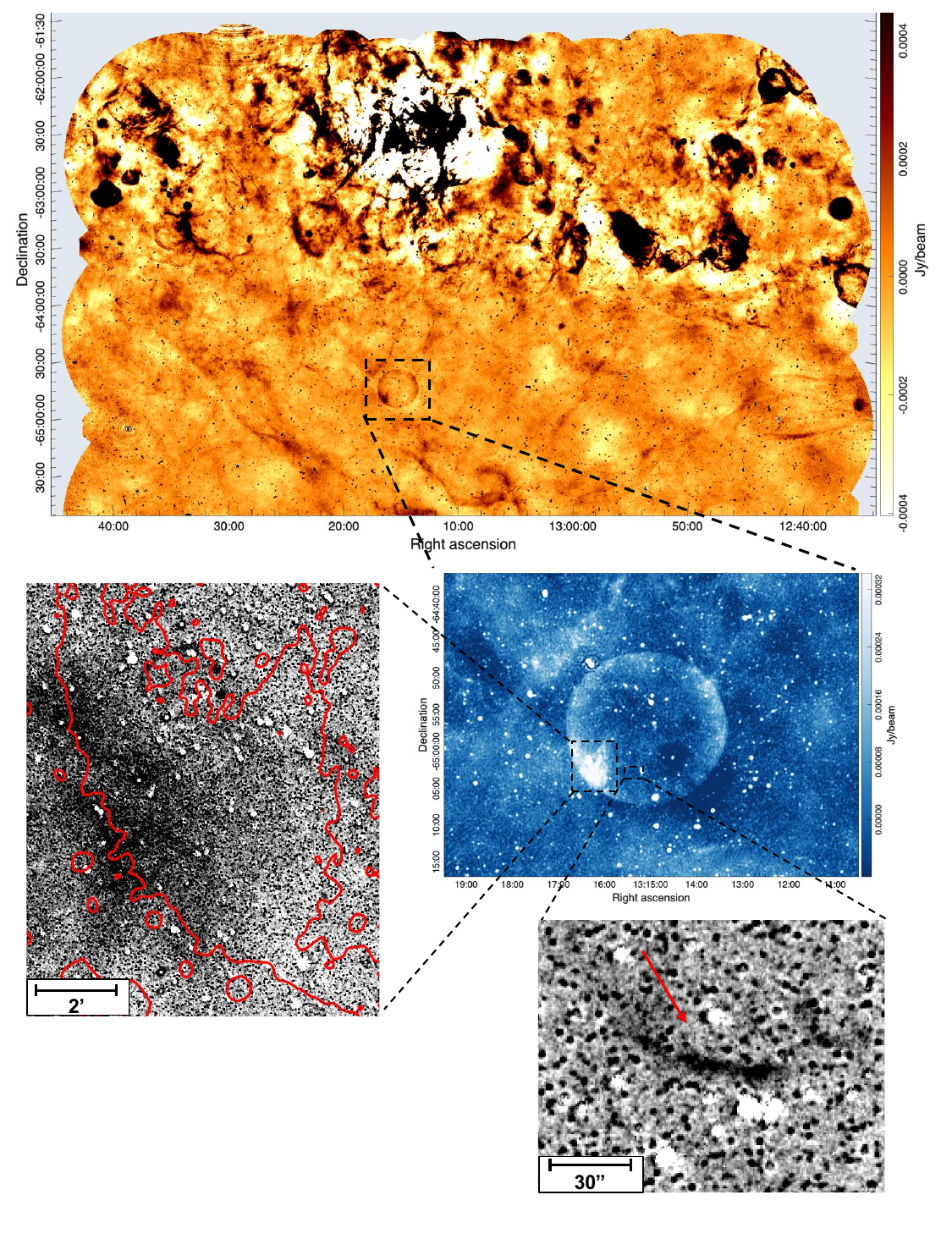}
    \caption{
    \ac{ASKAP} 943.5\,MHz radio-continuum image of Teleios and the surrounding environment showing the Galactic plane (top) with a zoomed-in inset of the same image (middle right). The \Halpha\ optical images are shown in the left and bottom insets. The bottom right inset shows a thin line of optical emission (marked with a red arrow) as a possible sign of Teleios's reverse shock. The left inset shows the \Halpha\ emission corresponding with the south-eastern patch of radio emission. The contour is from the ASKAP image at 100\,$\mu$Jy beam$^{-1}$. Both radio images have a convolved restoring beam of 15$\times$15~arcsec$^2$ and an rms noise level of $\sim$15$-$20\,$\mu$Jy beam$^{-1}$. All images have linearly scaled colour bars. The \Halpha\ images were created as described in Section~\ref{sec:data} and have scale bars in the bottom left corner.}
    \label{fig1}
\end{figure*}

The \ac{ASKAP} observation of Teleios was conducted as part of the \ac{EMU}~\citep[Hopkins et al. submitted]{Norris2011, Norris2021} and \ac{POSSUM}~\citep{2010AAS...21547013G} surveys, using 36 antennas at a frequency of 943.5\,MHz with a bandwidth of 288\,MHz. The observation was taken on 7$^{\rm th}$~May~2024, and the data is available through the \ac{CASDA}\footnote{\url{https://research.csiro.au/casda}}, in scheduling block SB62225 (EMU\_1309-64 tile). The data reduction used the standard ASKAPSoft pipeline as described in~\citet[]{2019ascl.soft12003G}, which produces both a multifrequency synthesis (MFS) band-averaged Stokes~I image for the EMU survey, and full Stokes~I, Q, U and V frequency cubes with 1\,MHz channels for POSSUM. Primary beam correction in all Stokes parameters is performed using beam models derived from standard observatory holography observations. This correction mitigates leakage from Stokes~I into Stokes~Q and U at around the 1~per cent level or less over most of the field. The resulting images (Figure~\ref{fig1}) have a restoring beam of 15$\times$15\,arcsec$^2$ and achieve a \ac{RMS} noise sensitivity of $\sigma$=$\sim$15\,$\mu$Jy\,beam$^{-1}$ for Stokes I and $\sim$15\,$\mu$Jy beam$^{-1}$ for polarised intensity. \ac{ASKAP} short baselines recover spatial scales up to $\sim$20\,arcmin at $\sim$1\,GHz (Hopkins et al. in prep.).

\begin{figure}
    \centering
    \includegraphics[width=\linewidth]{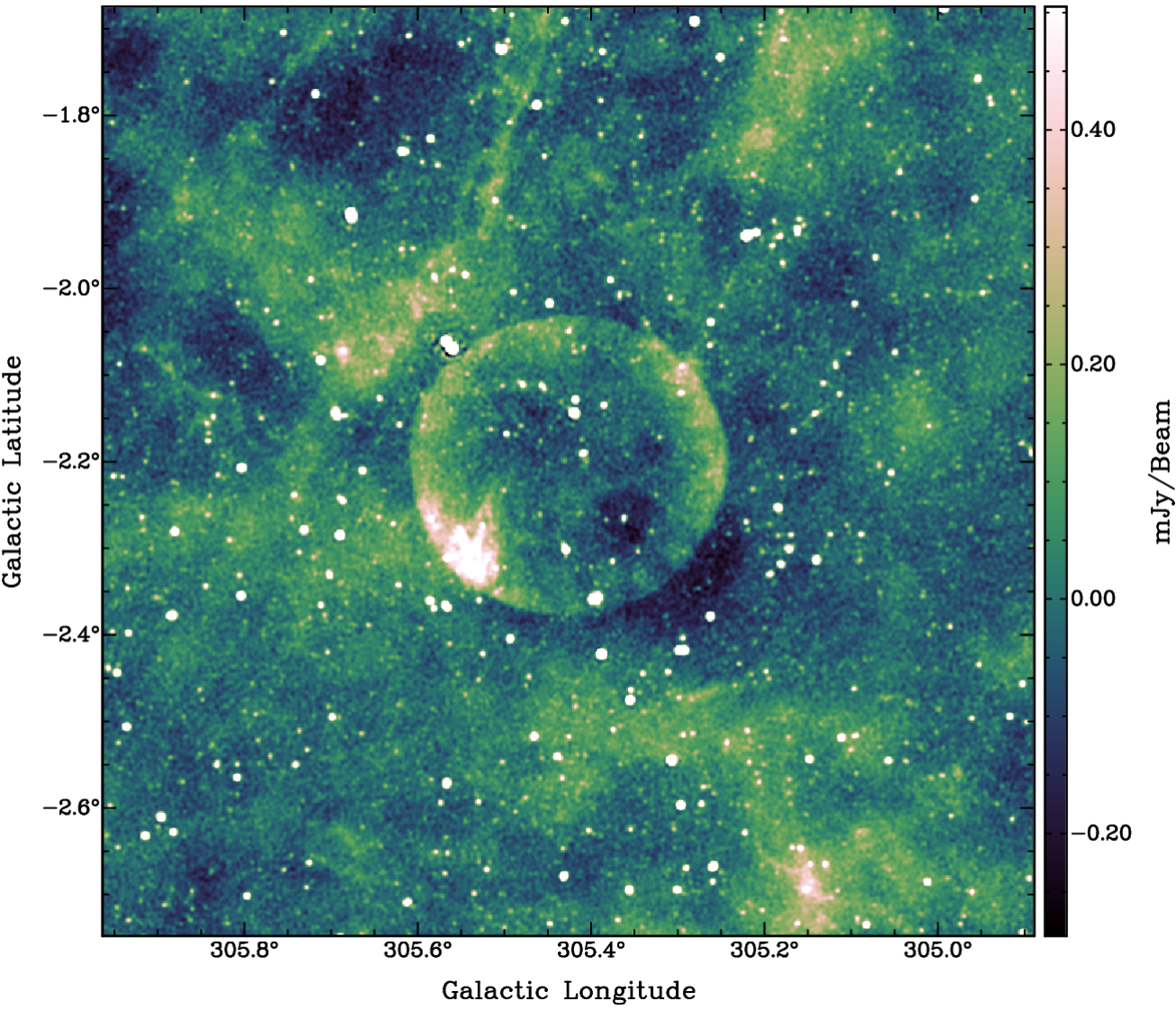}
    \includegraphics[width=\linewidth]{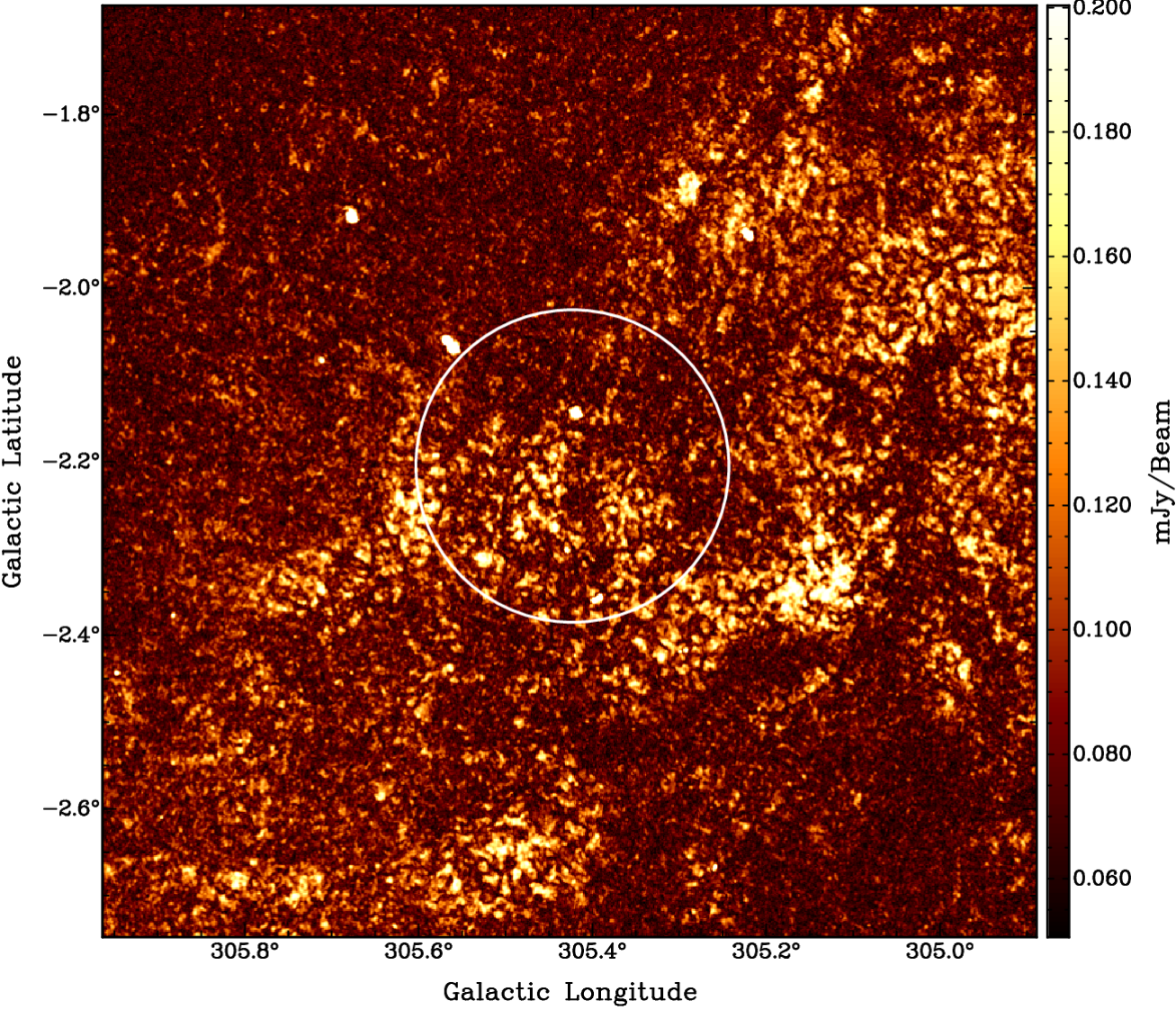}
    \includegraphics[width=\linewidth]{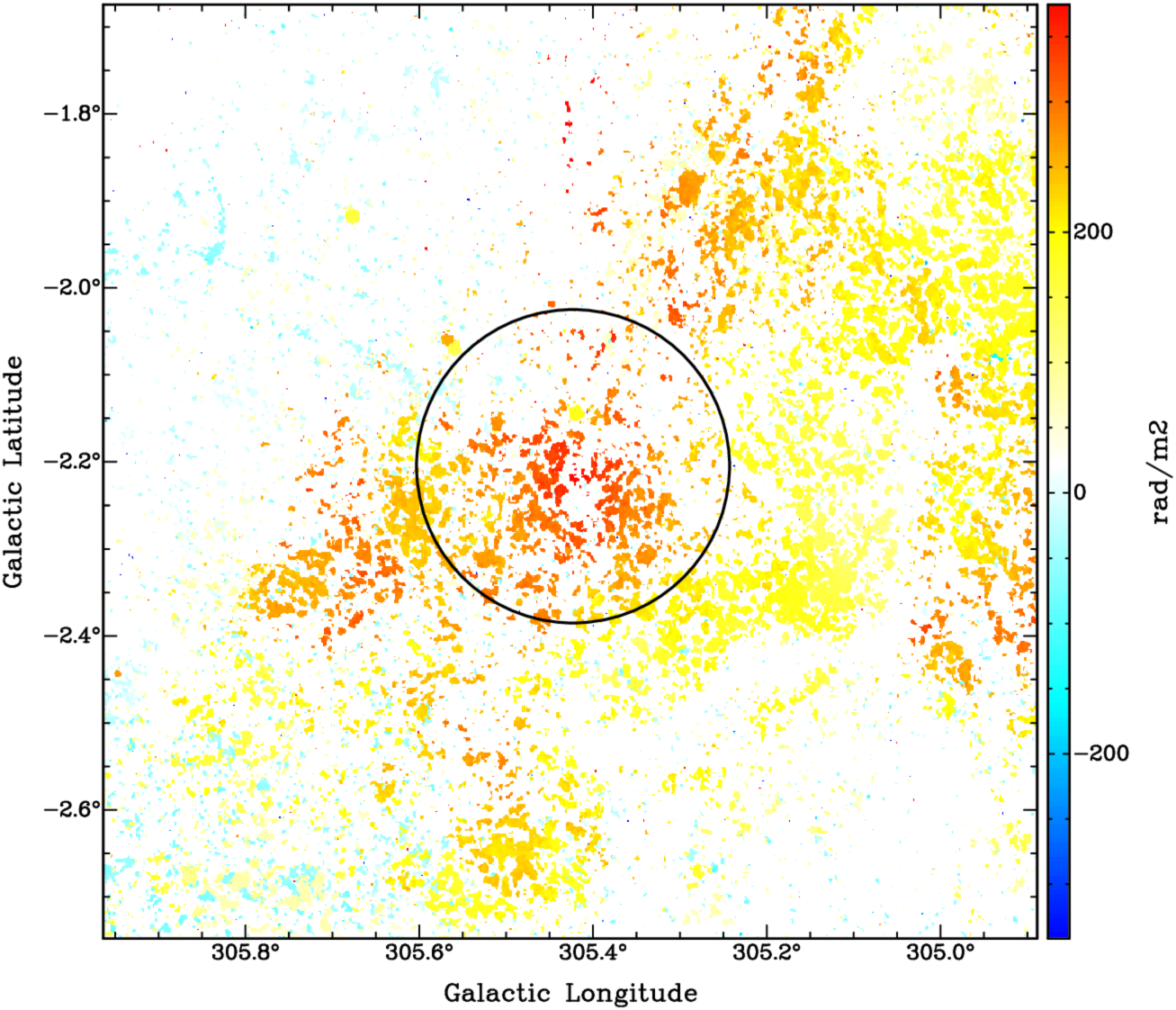}
    \caption{\ac{ASKAP} radio images of Teleios as Stokes~I (top), polarised intensity (PI) (middle) and \ac{RM} (bottom).}
    \label{fig2}
\end{figure}

We find no detectable emission in the corresponding Stokes~V image.
To obtain the polarised intensity image shown in Figure~\ref{fig2} we used the \ac{RM} synthesis technique. We did not use the Fourier transform method from the \ac{POSSUM} pipeline but the de-rotation technique as described in \citet{2023MNRAS.524.1396B}.

    \subsubsection{GLEAM-X}     
    
\begin{figure*}
    \centering
    \includegraphics[width=0.99\linewidth]{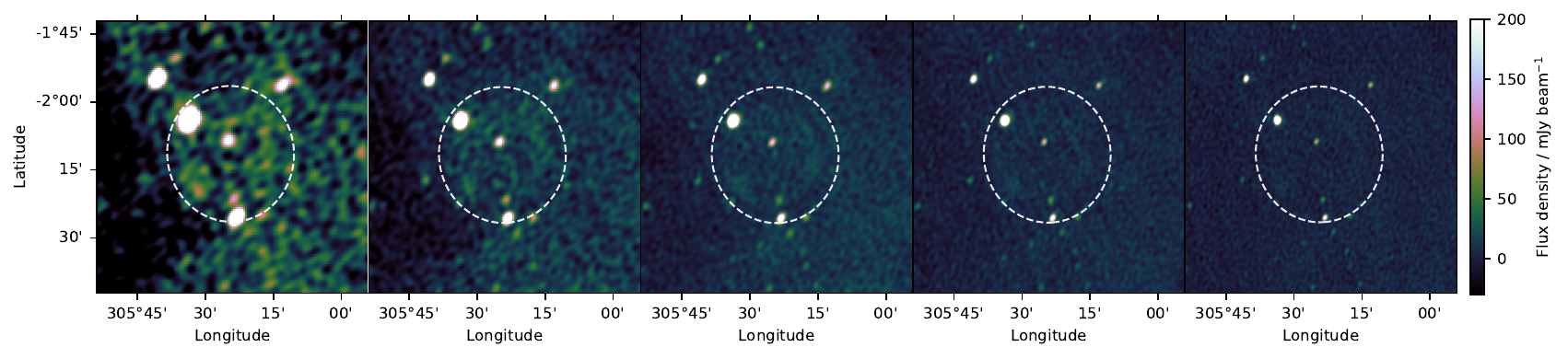}
    \caption{Region surrounding Teleios as observed by the \ac{MWA} respectively at 88, 118, 154, 185 and 216\,MHz. All the images are linearly scaled. }
    \label{fig:gleamx}
\end{figure*}

\begin{figure*}[ht]
    \centering
    \includegraphics[width=\linewidth]{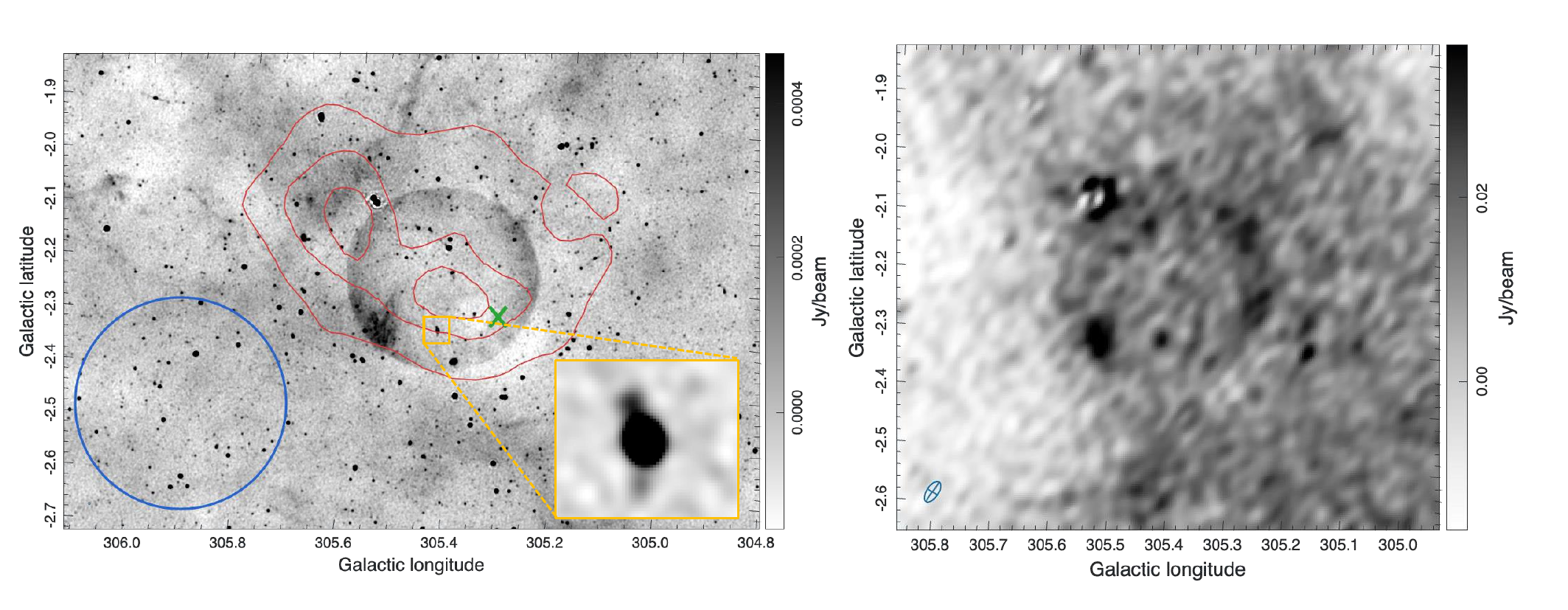}
\caption{{\bf Left: } \ac{ASKAP} radio image at 943.5\,MHz overlaid with H.E.S.S. $\gamma$--ray contours. The green cross marks the location of the Fermi point source candidate described in Section~\ref{subsubsec:fermi}. Orange-bordered inset shows the radio counterpart to the X-ray point source discussed in Section~\ref{subsubsec:Type Ia}. \ac{ASKAP} image has a convolved restoring beam of 15$\times$15~arcsec$^2$ and a local \ac{RMS} noise of 15\,$\mu$Jy\,beam$^{-1}$. H.E.S.S. has a mean point spread function (PSF) of 0\fdg2, shown by the blue circle in the bottom left. Contours are at significance levels of 2, 3 and 3.5 $\sigma$. {\bf Right: } \ac{MWA} broad-band radio image centred at 151.5\,MHz. \ac{MWA} image has a convolved restoring beam size of 144.9$\times$71.2~arcsec$^2$ with a P.A.\,=\,--35\fdg0, shown in the bottom left corner. The image has a local \ac{RMS} value of 14\,mJy\,beam$^{-1}$. Both images are linearly scaled, the right one has undergone point source subtraction as described in Section~\ref{subsec:radio continuum}.}
    \label{fig:hessgleamx}
\end{figure*}

The GaLactic and Extragalactic All-sky \ac{MWA} \citep[GLEAM; ][]{Wayth2015,Hurley2017} survey and GLEAM-eXtended \citep[GLEAM-X; ][]{Hurley2022,Ross2024} surveys have been conducted by the Murchison Widefield Array \citep[MWA, ][]{Tingay2013,Wayth2018} over a frequency range of 72--231\,MHz. A joint deconvolution of these surveys over the Southern Galactic Plane will be presented by \citet{2025PASA...42...21M}, which will be sensitive to angular scales from $45^{''} - 15^{\circ}$. A preliminary image from this effort shows a faint shell at the location of Teleios. The data achieved \ac{RMS} noise levels from 5--30\,mJy\,beam$^{-1}$ over the five 30-MHz bandwidth mosaics, which enabled flux density measurements (see Figures~\ref{fig:gleamx} and \ref{fig:hessgleamx} right). 
To enhance the S/N ratio, we used the Aegean software~\citep{Hancock2018} to find the point sources and removed them with the Aegean Residual (AeRes) package within Aegean. 
We then convolved the three images to achieve a uniform resolution before co-addition. The resulting image is centred at a frequency of 151.5\,MHz and features a beam size of $144.9 \times 71.2$~arcsec$^2$ with a P.A. of 145$^{\circ}$.

 \subsection{\HI\ observations}

We use \HI\ data from the HI4PI survey \citep{2016A&A...594A.116H}, which consists of data from the \ac{EBHIS} the Parkes \ac{GASS} surveys, as discussed in Section~\ref{subsec:HI}. The HI4PI survey resolution is 16\farcm2, and the average \ac{RMS} noise for the final data is 43\,mK.

\begin{figure*}[ht]
    \centering
    \includegraphics[width=\linewidth]{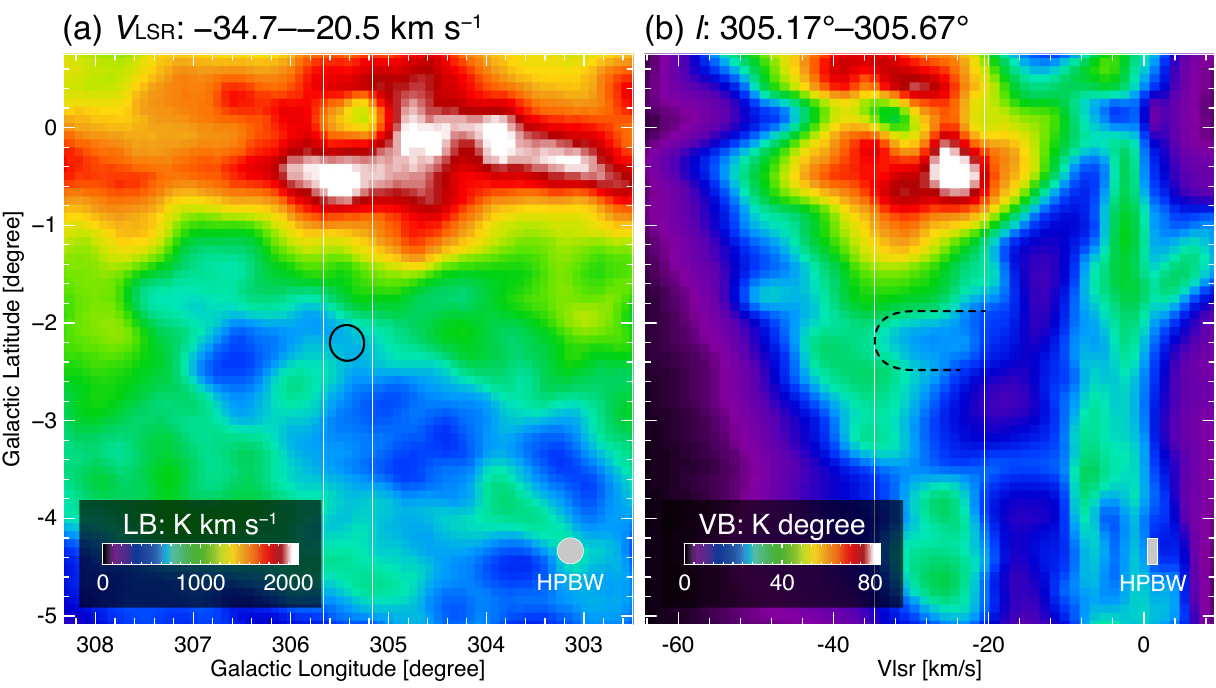}
    \caption{(a) Integrated intensity map of \HI\ obtained from HI4PI~\citep{2016A&A...594A.116H} at integrated velocity range --34.7\,\kms\ to --20.5\,\kms. The black circle represents Teleios's position and the beam size is shown in the bottom right. (b) Position--velocity ($p-v$) diagram of \HI\ integrated over the same velocity range and Galactic longitude range 305\fdg17 to 305\fdg67. The black dashed curve indicates a possible expanding \HI\ cavity centred at Teleios's Galactic latitude. The beam size is shown in the bottom right for both images.}
    \label{fig:HI}
\end{figure*}

 \subsection{Optical observations}

Some possible corresponding detection is that of H$\alpha$ emission from \ac{SHS}\footnote{\url{http://www-wfau.roe.ac.uk/sss/halpha/}} as examined in Section~\ref{sec:results}. This \Halpha\ image (Figure~\ref{fig1}, bottom right inset) was created by scaling by a factor of 0.65 to more closely match the typical star intensities in the corresponding short-R image. This short-R image was blurred slightly so that the PSFs more closely matched the \Halpha\ image, and the \Halpha\ image was then divided by this blurred short-R image.

\subsection{$\gamma$-ray observations}
 \label{subsec:gamma}

    \subsubsection{Fermi-LAT}
    \label{subsubsec:fermi}
We first analysed almost 16 years of Pass 8 \textit{Fermi}-LAT \citep{2009ApJ...697.1071A} SOURCE class data (from 2008 August to 2024 July) with the software fermitools (version 2.2.0) through the fermipy package (version 1.2.0) using the response functions P8R3\_SOURCE\_V3 and recommended cuts\footnote{See \url{https://fermi.gsfc.nasa.gov/ssc/data/analysis/documentation/Cicerone/Cicerone\_Data\_Exploration/Data\_preparation.html}}. We used front and back-converted events and, following \citet{2020ApJS..247...33A}, we used an energy-dependent maximum zenith angle cut such that a maximum of 90$^\circ$ was used in the energy range 0.4--500~GeV, 100$^\circ$ in the 1--500~GeV range and 105$^\circ$ for analyses above 5~GeV. We included all sources in the 4FGL-DR4 catalogue \citep{2022ApJS..260...53A} and used $20^\circ$-wide and $15^\circ$-wide regions of interest for analyses, for events below and above 1~GeV, respectively, around the location of Teleios in the centre of the region. We performed maximum likelihood fits to the spectral normalizations and calculated the test-statistic \citep[TS,][]{1996ApJ...461..396M} of the sources located within $5^\circ$ of the centre, including those of the Galactic diffuse emission (given by gll\_iem\_v07.fits) and the isotropic and residual cosmic-ray background (given by iso\_P8R3\_SOURCE\_V3\_v1.txt). The spectral index was also fit for sources catalogued within $3^\circ$ of the centre. We searched for new point sources with a TS~$>16$ to improve the model. There are no catalogued sources within $\sim1^\circ$ of the location of Teleios. In all these analyses, we saw a hint of point-like excess emission in the southwestern shell of Teleios.

We repeated the analyses using different event classes with different degrees of cosmic-ray background contamination using their corresponding response functions and isotropic diffuse models. We found that the (pre-trials) significance of the emission is maximized using events in the 10--500\,GeV energy range and the ULTRACLEAN class, which contains events having the highest probability of being photons (filtered with the parameter evclass $=512$). The source search algorithm in fermipy found a point source candidate at RA(J2000)\,=\,$13^{\text{h}}14^{\text{m}}2\farcs9$, Dec(J2000)\,=\, $-$65\D02\arcmin06\farcs0 (with a 95\%-confidence level positional uncertainty of $0.07$\D, position is shown by the green cross in Figure~\ref{fig:hessgleamx}, left). Using a power law spectrum with a fixed index of 2 the resulting TS of the source is 19.6, corresponding to a detection (pre-trials) significance of $3.7\sigma$ for three degrees of freedom. Assuming the source candidate is Galactic in origin, we derived a 95\%-confidence level upper limit on its 10--500~GeV luminosity of $2.6\times10^{32}\,d^2$~erg~s$^{-1}$. Here, $d$ is the distance in units of kpc.

\subsubsection{H.E.S.S.}
We inspected the publicly released data of very high-energy $\gamma$-ray emission from the H.E.S.S. Galactic plane survey (HGPS; \cite{HESS2018}).
Data presented in the HGPS consisted of nearly 2700-h of data taken between 2004 and 2013, although the exposure at the position around Teleios is very much unclear. Its proximity to the known H.E.S.S. source PSR\,B1259-63/LS\,2883 suggests that it could have been observed serendipitously, and additional data could be available due to more recent observations of the region since the release of the HGPS~\citep{2020A&A...633A.102H, 2024A&A...687A.219H}.
The HGPS contains data within an energy range of 0.2-100\,TeV and the size of the point-spread-function (PSF) for the observations is given to be around 5\,\arcmin.
Significance maps from the survey published with a 0\fdg2 oversampling radius show hints of excess spatially coincident with the \ac{SNR} shell. Contours indicating levels of 2, 3 and 3.5\,$\sigma$ significance are shown in Figure~\ref{fig:hessgleamx} (left). 

\subsection{Other data}
 
We searched for Teleios's signature in various multi-frequency surveys. These include optical (DSS2 and DECaPS DR2), IR (\ac{WISE} and Spitzer) and X-ray (eROSITA). There are no {\it Chandra} or {\it XMM-Newton} observations which cover this area. We also searched available Galactic \ac{SNR}, \ac{PN}, \HII\ region and \ac{LBV} catalogues. However, we found no sources or emissions matching Teleios.


\section{RESULTS}
 \label{sec:results}

\subsection{Teleios's morphology and classification}
 \label{subsec:morphology}
 
We clearly see Teleios's faint circular shell in our \ac{ASKAP} \ac{EMU} image (Figure~\ref{fig1}), and we see a possible hint of \Halpha\ and $\gamma$-ray emission, which draws immediate attention to the true nature of this object. 

Given Teleios's proximity to the Galactic Plane, circular morphology, and the fact that it is visible almost exclusively in radio-continuum (Section~\ref{sec:data}), we can exclude its classification as a \ac{PN}, \ac{LBV}, \ac{NSR}, \ac{WR} or (super)bubble. These source types should have a very prominent \ac{IR} or optical (narrow-band) appearance, which is lacking in the case of Teleios \citep{book2}. Also, apart from (super)bubbles, which would have a physical size in excess of 100\,pc \citep{2015A&A...573A..73K,2017ApJ...843...61S,2019A&A...621A.138K,2021ApJ...918...36Y}, all these Galactic source types would have angular sizes of less than a few arcminutes or would be located at distance $<$200\,pc. 

We also exclude classification as an \ac{ORC} \citep{galaxies9040083, Norris2021ORC, 10.1093/mnrasl/slab041, Gupta2022, Shabala2024, 2024arXiv240807727B}, since this object is very unlikely to be connected with anything extragalactic at larger distances of several hundred Mpc. We checked the \ac{ATNF} Pulsar Catalogue (\url{https://www.atnf.csiro.au/research/pulsar/psrcat/}), and the closest pulsar (J1309--6526) is $\sim$1\fdg8 away from Teleios's centre at a distance of 11.257\,kpc \citep{2017ApJ...835...29Y} and old age of 3.43$\times$10$^{8}$ years. We consider this pulsar as unrelated to Teleios.

Given the apparent circularity, we also considered that Teleios could be a Dyson Sphere \citep{2021map..book...11V,2020SerAJ.200....1W}. However, given that no \ac{IR} (Spitzer, WISE) emission can be detected anywhere within Teleios's boundaries, we also conclude this to be an unlikely scenario.

The circular appearance of Teleios and the prominence of its radio emission compared with other wavelengths are consistent with the \ac{SNR} hypothesis, and from now on, we will treat Teleios as a most likely Galactic \ac{SNR} (G305.4--2.2) especially given the evidence presented in Sections~\ref{subsec:radio continuum} and \ref{subsec:polarisation}. 

Tantalising hints of excess emission up to 3.75\,$\sigma$ in the H.E.S.S. $\gamma$-ray image (see Figure~\ref{fig:hessgleamx} left) might be in some smaller part related to nearby strong background sources. While we recognise that the H.E.S.S. $\gamma$-ray excess may be attributed to background fluctuations, together with hints of emission seen with \textit{Fermi}-LAT at HE $\gamma$-rays from the region, we believe that there may be a physical connection with Teleios that could be revealed by further high-sensitivity and high-resolution VHE $\gamma$-ray observations. 

Based on the publicly available flux maps, the $>$1 TeV integral photon flux for this region is estimated as $\sim$2.8$\times$10$^{-13}$\,ph\,cm$^{-2}$\,s$^{-1}$. Converting this to a luminosity gives values of 2.2$\times$10$^{32}$\,erg\,s$^{-1}$ for a distance of 2.2~kpc and 2.7$\times$10$^{33}$\,erg\,s$^{-1}$ for a distance of 7.7~kpc. This possible overlapping gamma-ray emission at GeV to TeV energies may come from protons and/or electrons which are accelerated and potentially still trapped within the \ac{SNR}. Similar to the work of \citep{2024PASA...41..112F}, who examined a similarly mature \ac{SNR} Diprotodon, this gamma-ray emission may be a combination of hadronic and leptonic emission mechanisms. If Teleios is located in a low-density \ac{ISM}, this may suggest that the leptonic component is dominant at this phase. If the emission is caused by a hadronic-dominant case, we can estimate the required proton energy $W$. Assuming an \ac{ISM} density $n\,=\,0.1$\,cm$^{-3}$, we calculate a proton energy of 3.7$\times$10$^{48}$\,erg (for 2.2~kpc distance) and 4.5$\times$10$^{49}$\,erg (for 7.7~kpc distance). 

We note some extended radio emission inside the southeastern edge of Teleios's shell, which is at odds with the almost perfect circular symmetry of the rest of the shell shape. We suggest that at least some parts of this region might be affected by an interaction of Teleios with local \ac{ISM} structures, as somewhat weak H$\alpha$ emission in this region is evident in Figure~\ref{fig1} (bottom left) even beyond the Teleios boundaries. This H$\alpha$ emission could be an unrelated \HII\ region or even a related \HII\ region, ionised by X-rays from the \ac{SN} shock where the radio emission is of thermal (free-free) nature. In the absence of a reliable spectral index for the exterior and interior emission in this location, we cannot 100\% rule out a priori that this enhanced patch of Teleios radio emission is free-free. 
However, if the weak and small size (compared to the entire \ac{SNR}) H$\alpha$ filament emission from Figure~\ref{fig1} (bottom right) is associated with Teleios, it might represent the reverse shock from the \ac{SNR}. \citet{Mckee1995} find a ratio between the blastwave and the reverse shock radii of 1.33 for an \ac{SNR} at the beginning of the Sedov phase. For Teleios, we find a ratio of 1.44, indicating a very young \ac{SNR}, in the Sedov phase. However, we caution this as this filament is tiny compared to the size of Teleios, and thus, if it is the reverse shock, then we would see an area at a slightly different evolutionary phase than the rest of Teleios, or an area with a slightly different environment.

On the other hand, if Teleios comes from the type~Ia \ac{SN} event, one should not expect to find hydrogen at the reverse shock.
Also, the reflected reverse shock could act as a secondary internal blast wave that will go through the whole ejecta and swept-up \ac{ISM}.
An alternative scenario could be as in the case of the \ac{LMC} \ac{SNR} DEM\,L71 where \citet{2003ApJ...590..833G} suggested that H$\alpha$ emission inside the rim of the Balmer-dominated collisionless shock is lumps of exterior neutral material ionised by \He\ $\lambda$304\,\AA\ \citep{2000ApJ...535..266G}. The fact that it is seen inside the radio shell could be a projection effect. 
A reflected reverse shock occurs for relatively evolved objects such as the \ac{LMC} \ac{SNR} DEM\,L71 with the age of 4360$\pm$90~years, which now suggests that if this H$\alpha$ emission is associated after all, then Teleios is not a youngish \ac{SNR}, which is in contradiction with the above speculation of a very young \ac{SNR}.

\subsection{Teleios's radio-continuum properties}
 \label{subsec:radio continuum}

In the same manner as described in \citet{2023AJ....166..149F}, we measured Teleios's radio--continuum properties, including its extension and flux density. As seen in Figure~\ref{fig1}, Teleios is a circular object centred at RA(J2000)\,=\,$13^{\text{h}}15^{\text{m}}1\farcs2$, Dec(J2000)\,=\, $-$64\D57\arcmin40\farcs7, some 2\fdg2 below the Galactic plane (Galactic G305.4--2.2). The ellipse axes are estimated to be 1320\arcsec\ by 1260\arcsec\ at PA=0\D\, which gives it a circularity $c$ = 95.4\%. 

We note that Teleios's exceptional circularity is unusual for an \ac{SNR} as we can find only a handful of similar \acp{SNR}. Some such examples are the newly discovered circumgalactic \ac{SNR} J0624--6948 \citep[][$e$ = 3.9\%]{2022MNRAS.512..265F} and several young (under 2000~yr old) Magellanic Cloud (MC) \acp{SNR} such as SN1987A \citep{2018ApJ...867...65C}, MC~SNR~J0509--6731 \citep{2014MNRAS.440.3220B,2018MNRAS.479.1800R}, N\,103B \citep{2019Ap&SS.364..204A}, and 1E0102 \citep{2024MNRAS.527.1444A}. Despite their young age, all the young (under 2000~years) Galactic \acp{SNR} such as, for example, the youngest G1.9+0.3 \citep{2020MNRAS.492.2606L,2023PASJ...75..970E} lack pronounced circular symmetry as shown by \citet{Ranasinghe:2019quc}.

One of the most perfectly circular ring-like sources seen in the sky is MAXI~J1348--630 \citep{2021A&A...647A...7L} -- a giant dust-scattering X-ray ring around the black hole transient.



\subsubsection{Radio spectral index}
 \label{sec:spectralindex}

The average spectrum for the bright ring of Teleios was estimated as follows using the \ac{ASKAP} 943.5\,MHz map in combination with the GLEAM-X 151.5\,MHz map. 
First we used the DiffuseFilter script\footnote{\url{https://gitlab.com/Sunmish/diffusefilter}}, based on the minimum/maximum filtering of~\citet{2002NewAR..46..101R} to remove the point sources in the field. We then perform background subtraction and then 
regridded the maps to make them the same, convolved the 943.5\,MHz image to match the 91\arcsec$\times$64\arcsec\ resolution at 151.5\,MHz, and blanked compact regions that were brighter than the ring at 151\,MHz. We measured the fluxes in three annuli, covering the radii 500\arcsec--650\arcsec\ (covering Teleios's bright ring), 650\arcsec--800\arcsec, and 800\arcsec--950\arcsec, (which we designate as A, B, and C, respectively). Because of large-scale background variations, we divided each annulus into eight sectors of 45\D. For each sector, the flux density from the bright Teleios ring, in practice, its excess over the background, was calculated by subtracting the flux in B from the flux in A, and then summing over all sectors. This yields an excess flux density for the ring of 0.17$\pm$0.03\,Jy at 943.5\,MHz and 0.50$\pm$0.17\,Jy at 151.5\,MHz, for a spectral index of $\alpha$=--0.6$\pm$0.3. The total flux densities of the ring would be higher if we did not remove the background.  However, this was necessary due to the missing short spacings from \ac{ASKAP} (it begins losing flux at 20\arcmin\ scale, and Teleios has an average diameter of 1290\arcsec\,=\,21.5\arcmin). It is not clear what the total spectral index would be if the entire structure were sampled at both frequencies.  

The errors were estimated using the scatter in the sector-by-sector differences between annulus B and annulus C. The errors are dominated by large-scale galactic background structures, and future observations and more detailed spectral modelling may allow these to be reduced. The most discrepant value was observed for the bright sector in the southeast, and dropping it led to an overall spectrum of --0.6. This does not affect the above overall spectral index, and confirmation of a possible flatter spectrum in the SE can be made with upcoming MeerKAT observations.

We also attempted to measure the total flux and spectral index of Teleios within $\approx$650\arcsec\ at both frequencies instead of just its bright ring. While an enhancement in brightness inside the ring is clearly visible at 151~MHz, it is completely absent at 943.5\,MHz, due to insufficient sampling at short baselines, and so this calculation was dropped. In addition, we attempted to measure the spectral index within the \ac{ASKAP} 288\,MHz band, but found extremely steep apparent spectra ($\alpha \sim -2.5)$, a problem for large angular size sources that is still under investigation within the \ac{EMU} collaboration.

Overall, after attempting several different methods to determine the spectral index across Teleios, the only reliable method we find is that for the excess in the bright ring, yielding $\alpha\,=\,-0.6\pm0.3$, as described above. This estimate has a high intrinsic uncertainty due to Teleios's low surface brightness. It was not possible to generate a detailed spectral index map due to the low signal-to-noise,  missing short spacings, and the presence of fluctuations on comparable brightness and size scales from the Galactic background. The smaller-scale radio-continuum variations in the spectra can not be determined at this time, without more sensitive observations such as the upcoming MeerKAT observations.


Teleios's spectral index of --0.6$\pm$0.3 is only slightly steeper compared to the observed value of --0.5$\pm$0.3 for shell-type \acp{SNR}, both within the \ac{MW} and in several nearby galaxies \citep{2012SSRv..166..231R, 2014SerAJ.189...15G, 2017ApJS..230....2B,2019A&A...631A.127M,book2,2023MNRAS.518.2574B,2023ApJS..265...53R,2024MNRAS.529.2443C}. 
Such a steep spectral index is expected for somewhat young ($<$2000~years) or very old \acp{SNR} \citep{2017ApJS..230....2B,2020A&A...634A..59B,2022MNRAS.515.4099K,2022A&A...661A.128D,2022ApJ...926..140S,2024A&A...689A...9D}.

This spectral index, combined with the measured angular size, gives us a surface brightness ($\Sigma_{1 \text{GHz}}$) of $\sim$5.1$\times$10$^{-23}$~W~m$^{-2}$~Hz$^{-1}$~sr$^{-1}$; making it one of the lowest surface-brightnesses of any known \acp{SNR}, similar to the Galactic \ac{SNR} Diprotodon \citep{2024PASA...41..112F}.

\subsection{Teleios's \HI\ appearance}
\label{subsec:HI}

We analyse archival \HI\ data from the HI4PI~\citep{2016A&A...594A.116H} to reveal any physical association with Teleios and the surrounding environment. HI4PI achieves an angular resolution of 16\arcmin, which is a modest size with respect to Teleios's shell size. This makes it difficult to resolve any small-scale \HI\ differences over Teleios's area, however, we find a possible \HI\ association in the form of an \HI\ cavity (Figure~\ref{fig:HI}).

This possible cavity is at a velocity range of $V_{LSR}$\,=\,--34.7 to --20.5\,\kms. Adopting the systemic velocity of $V_{LSR}$\,=\,--27$\pm$3\,\kms, this corresponds to kinematic distances of $\sim$2.2$\pm$0.3\,kpc (near side) and far-side at $\sim$7.7$\pm$0.3\,kpc\footnote{The kinematic distance was calculated using the values $R_{\odot}=8.5$\,kpc and $V=220$\,\kms, as recommended by the \ac{IAU}~\citep{1986MNRAS.221.1023K}.}.

The \HI\ integrated intensity map (Figure~\ref{fig:HI}a) shows a low-density \HI\ bubble ($<$0.01\,cm$^{-3}$ as per \citet{1977ApJ...218..377W}) at Teleios's location (indicated by the black circle), which corresponds with a possible expanding gas motion in the $p-v$ diagram (Figure~\ref{fig:HI}b). This cavity seen in the $p-v$ diagram correlates with Teleios's Galactic latitude and is indicated by the curved black dashed line. For an \ac{SNR} this can indicate an expanding \HI\ shell caused by initial \ac{SN} shockwaves or stellar winds from the progenitor carving out a bubble of rarefied space, as has been observed in other \acp{SNR} (e.g. Kes~75~\citep{Su_2009}, RCW 86~\citep{SANO20171}, N103B~\citep{Sano_2018}, CTB87~\citep{Liu_2018}).

It is, therefore, possible that this cavity represents such a \HI\ wind bubble. If Teleios's progenitor were able to carve out such a bubble prior to Teleios's expansion, then expansion into this rarefied but highly homogeneous environment could help explain Teleios's remarkable circularity.

\subsection{Teleios's polarisation and rotation measure}
 \label{subsec:polarisation}

In Figure~\ref{fig2}, we show Stokes~I (top), polarised intensity (PI) (middle), and \ac{RM} (bottom) images of Teleios and the area around it, created using the available \ac{ASKAP} \ac{POSSUM} data. The values for PI and \ac{RM} were taken from the peak of the Faraday Depth (FD) function in each pixel. Instead of a Fourier transform, we used the de-rotation method to produce the FD function for each pixel. To that end, we derotated the Stokes Q and U data in each frequency channel for each rotation measure. We probed an \ac{RM} range from $-2000$ to $+2000$~rad\,m$^{-2}$.

We did not find any polarised emission coincident with the SNR's total power emission, likely due to its low radio surface brightness and perhaps high internal Faraday rotation effects. Curiously, the overall \ac{RM} amplitude is highest at the SNR's centre and getting radially lower. There also seems to be a large foreground screen with very high \ac{RM} related to relatively high surface brightness polarisation signal almost all around the \ac{SNR}.

It seems like this Faraday screen is actually in the background of Teleios, and the \ac{SNR} is Faraday rotating its polarised emission. Since this additional \ac{RM} is highest in the centre of the remnant, we should find a magnetic field mainly along the line of sight there. This would be the case for either a young \ac{SNR} with a radial magnetic field or if the \ac{SNR} is expanding inside a Galactic magnetic environment with the field lines going mainly along the line of sight. However, the former should not add any Faraday rotation to the background polarisation if the \ac{SNR} is symmetric along the line of sight. And for the latter case, the \ac{RM} should again increase towards the edge of the source. Clearly, further studies are required.

\begin{figure}[ht]
    \centering
    \includegraphics[width=0.8\columnwidth]{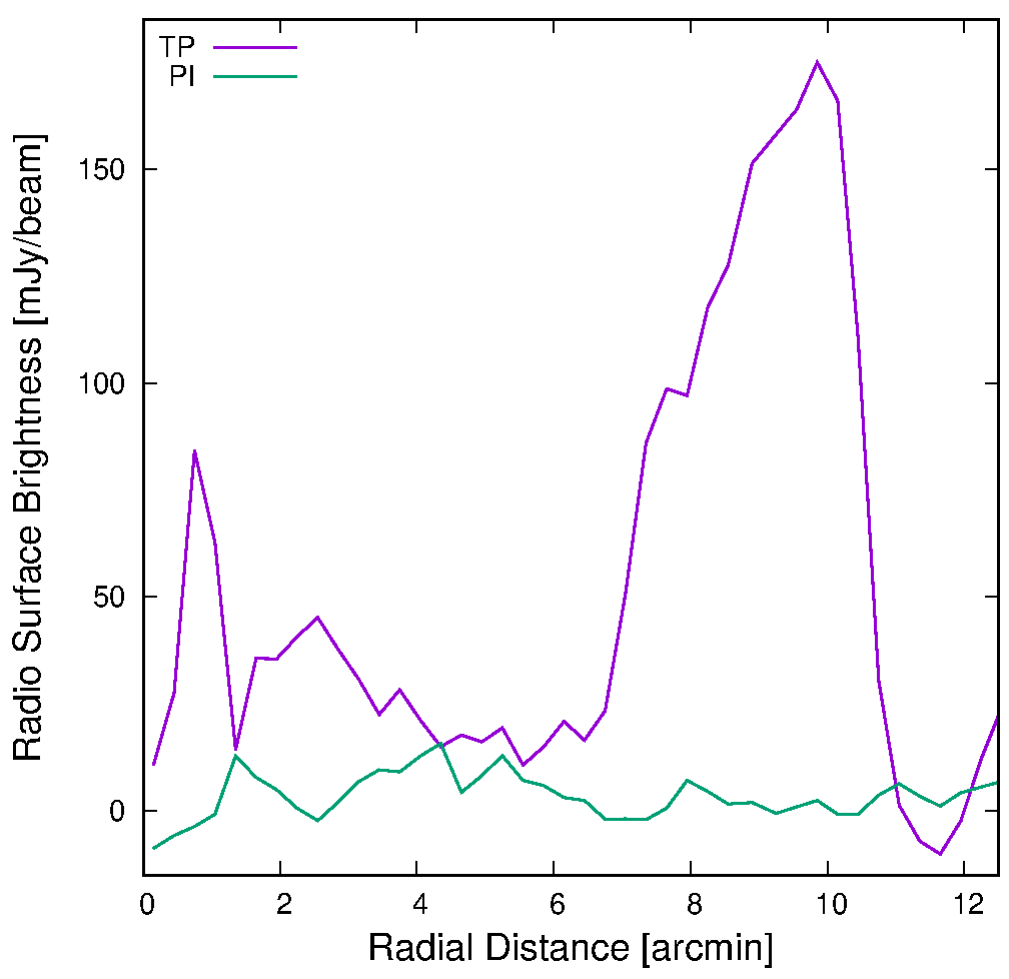}
    \caption{Radial profiles averaged over the western half of Teleios calculated for total power (TP) and polarized intensity (PI).}
    \label{fig:radprof}
\end{figure}

To get an estimate of the maximum fractional polarization of Teleios, we produced radial profiles for total power and the polarised emission, shown in Figure~\ref{fig:radprof}. We only calculated those profiles over the western half of the \ac{SNR} to avoid any contribution from the diffuse patch in the lower left corner of Teleios and the emission outside the top left area. Bright point-like sources were removed from the maps through Gaussian fitting.

The shell of the \ac{SNR} is clearly visible in the total power profile, but there is no obvious polarized emission. An estimate of the standard deviation of the PI values in the radio shell area results in about 3.5\,mJy\,beam$^{-1}$. If we now assume that the polarized emission must be lower than $3\sigma$=10.5\,mJy\,beam$^{-1}$, we derive a maximum fractional polarization of 6~\% for Teleios' radio shell. This is a very low value but certainly not unusual at this low frequency. It could also indicate that Teleios is in the transition phase between free and Sedov expansion like G11.2$-$0.3, which has only 2~\% integrated fractional polarization at a frequency of 32~GHz and clearly shows characteristics of both a free expanding and a Sedov type \ac{SNR} \citep{Kothes2001}. This would favour the near-side distance of 2.2\,kpc (see distance discussion in Section~\ref{subsec:dist}).

\section{DISCUSSION}
 \label{sec:discussion}

The most obvious characteristic of Teleios is its remarkable circular symmetry, coupled with a low surface brightness and a slightly steeper radio spectral index. Most \acp{SNR} exhibit some form of asymmetries, and there can be several different physical processes behind these. 

For a very young \ac{SNR} in the free-expansion phase, the ejecta freely expands into the surrounding medium, and this ejected mass is much greater than the swept-up \ac{ISM} mass. Thus, any initial asymmetries in the explosion and \ac{CSM} are present in this initial expansion. 

Once the ejecta has all been heated by the reverse shock, the remnant enters into the Sedov phase, where the shock begins being driven by this internal thermal pressure. The expansion is now subsonic with respect to the \ac{SNR} interior, and so the Sedov solution is effectively independent of the explosion geometry. This means that the pressure-driven expansion dominates over any inherent explosion asymmetries, and the expansion will relax into a roughly spherical shape. This expansion is expected to remain inherently symmetrical throughout the Sedov phase, and the remnant should only become asymmetric in the Sedov phase if there is very asymmetric \ac{ISM} or \ac{CSM} on length scales of $\sim$5--10\,pc. For example, in the case of the $\sim$10$^3$ M$_{\odot}$ bow shocks swept up by the stellar winds in~\citet[]{2015MNRAS.450.3080M}. 

The following phase after Sedov is the radiative phase, where the shell becomes thinner, and fragmentation of the shell can occur. The thinner shell also becomes more susceptible to instabilities, and so local density variations become more important. However, as the outer shock becomes radiative, we also expect strong optical line emission, which is not present in Teleios. 

Therefore, Teleios’s symmetrical shape and lack of pronounced optical emission indicate that it is likely in the Sedov evolutionary stage, and the shell results from the pressure-driven expansion. The fact that there are no large and obvious asymmetries in shape indicates that there is insufficient asymmetry in the swept-up material to distort the shape. This could be due to Teleios being a younger Sedov phase \ac{SNR}, as these possible asymmetries will become more apparent as Teleios expands further into the medium, or it could also be due to expansion into an isotropic but rarefied environment. Given the apparent size of Teleios ($\sim$21\farcm5) and position of 2\fdg2 below the Galactic Plane, we argue that the \ac{ISM} would have to be rarefied to a level of $<$0.1\,cm$^{-3}$ \citep{2020NatAs...4..910U} unless the \ac{SNR} is inside a wind-driven bubble where much lower densities of $\approx$0.02\,cm$^{-3}$ are possible. 

While the Sedov phase is expected to be the most symmetric, most Sedov phase \acp{SNR} still display some form of bilateral asymmetry due to the ambient magnetic field. As the remnant expands, it compresses the ambient magnetic field, resulting in some compression being parallel and some being perpendicular to this ambient field. The compression that occurs perpendicular causes brighter radio synchrotron emission, resulting in a typical bilateral radio morphology with opposite shells of brighter emission. Maintaining a perfectly symmetrical shape for a size greater than $\sim$10\,pc would require an unreasonably low magnetic field strength. The observed symmetry can, therefore, be explained as an orientation effect if we are viewing Teleios end-on, that is, viewing Teleios while the magnetic fields are in our line of sight. This would mean that the brighter emission where the expansion is perpendicular to the magnetic field would occur along the entire outer ring. This orientation would explain the apparent symmetry, as well as the lower surface brightness, as the radio synchrotron emission would be predominantly oriented perpendicular to our line of sight, thus making it appear fainter.

An example of the effect of a rarefied environment on \ac{SNR} morphology is the circumgalactic \ac{SNR} J0624--6948, which has a diameter of $\sim$72\,pc and a possible age of $\sim$2350~years. While it also has a prominent radio and X-ray emission \citep{2025A&A...693L..15S}, due to its position in the rarefied circumgalactic medium, this \ac{SNR} demonstrates how \acp{SNR} can preserve their circular shape in such rarefied environments of constant and uniform ambient density.

\subsection{Teleios's distance and position in \ac{MW}}
\label{subsec:dist}

\subsubsection{$\Sigma$--D method}
\begin{figure}[ht]
    \centering
    \includegraphics[width=\columnwidth]{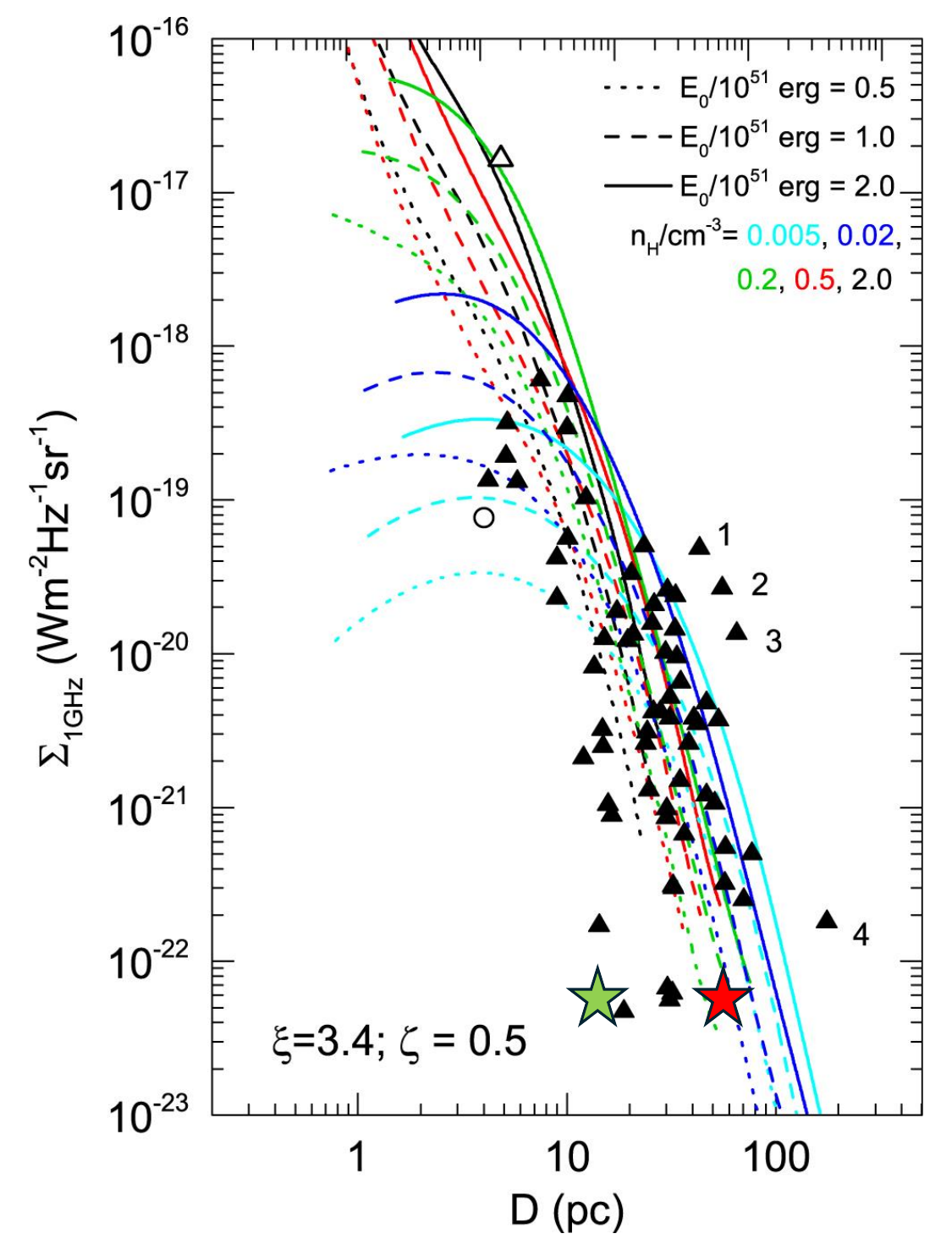}
    \caption{Radio surface brightness to diameter diagram for \acp{SNR} at frequency $\nu$\,=\,1\,GHz, obtained from \protect\citet[][Figure~3]{2018ApJ...852...84P}, shown as black triangles. Different line colours represent different ambient densities, while different line types represent different explosion energies. The open circle is young Galactic \ac{SNR} G1.9+0.3 \citep{2020MNRAS.492.2606L}, and the open triangle represents Cassiopeia~A. The numbers represent \acp{SNR} (1): CTB~37A, (2): Kes~97, (3): CTB~37B, and (4): G65.1+0.6. The stars represents Teleios at estimated surface brightness of 5.89$\times$10$^{-23}$W m$^{-1}$ Hz$^{-2}$ sr$^{-1}$. The red star corresponds to Teleios diameter of 48\,pc, and the green star with a diameter of 14\,pc. The image shows evolutionary tracks for representative cases with injection parameter $\xi$\,=\,3.4 and nonlinear magnetic field damping parameter $\zeta$\,=\,0.5.}
    \label{fig:SigmaD}
\end{figure}

Teleios's surface brightness estimate is compared to other Galactic and Magellanic Cloud (MC) shell-type \acp{SNR} (Figure~\ref{fig:SigmaD}). To estimate the most likely intrinsic diameter, we use the $\Sigma$--D method as described in \citet[][their Figure~3]{2018ApJ...852...84P}. We obtain a wide diameter range ($D$=30--150\,pc). This diameter range would also correspond to a wide distance range of approximately $d$\,=\,4.8--24.0~kpc.

We compare Teleios’s estimated surface brightness value with empirical calibration from \citet{2019SerAJ.199...23S}. For a fixed surface brightness of $5.1\times10^{-23}\,\mathrm{{Wm^{-2}Hz^{-1}sr^{-1}}}$ this empirical calibration gives a diameter probability distribution with the median value of $33$\,pc and a $95\%$ confidence interval range of $11-134$\,pc. This translates to a distance probability with a median value at $5$\,kpc and a $2-21$\,kpc as a $95\%$ confidence interval.

The orthogonal $\Sigma-D$ fit on the same calibration gives higher values, 48\,pc diameter and a corresponding distance of 8\,kpc. 

Given Teleios's somewhat unusual morphology as larger, rounder, and fainter than any other known \ac{SNR}, it means that the expectations to fit into the $\Sigma-D$ method would be challenging. Especially given that the $\Sigma-D$ method was not designed for or built off of extremely low surface brightness \acp{SNR}. Still, the above $\Sigma-D$ method results, despite the wide range, show a definite tendency when combined with other (see below) methods.

\subsubsection{Churchwell \ac{MW} model map}
To further constrain possible distances, and thus size, one could compare the Churchwell \ac{MW} model map \citep{Churchwell2009, 2013SerAJ.187...43F, Efremov2011, Hou2014, Vallee2017} with Teleios's Galactic longitude to estimate the most likely distances. 
As \ac{CC} \ac{SN} only occur for massive, short-lived stars, one can argue that they would typically occur in areas of massive star formation~\citep{Bartunov_1994, Anderson2017, Verberne2021} such as the Galactic spiral arms. Namely, Teleios's 
Galactic longitude of $l\sim$\,305\D\ passes tangentially through the inner Sagittarius arm at $\sim$2\,kpc and Scutum-Centaurus arm at approximately 4--8\,kpc, and then through the outer Sagittarius arm at about 11--14\,kpc. 
Using the same reasoning as in \citet{2024MNRAS.534.2918S}, this gives distances $\sim$2\,kpc, $\sim$6\,kpc or $\sim$12\,kpc, corresponding to $\sim$13\,pc, $\sim$38\,pc and $\sim$75\,pc diameters, respectively. It is also worth noting that if we assume that the \ac{MW} extends out to about 20\,kpc in that direction~\citep{Churchwell2009}, this would give a somewhat unrealistic maximum diameter of 125\,pc if it is a Galactic object. Therefore, this can be taken as the upper limit of our distance estimations.

\subsubsection{\HI\ method}
Finally, from our \HI\ study (Section~\ref{sec:data}), we infer a possible cavity in which Teleios could be expanding at the systemic velocity of V$_{\rm LSR}$ $\sim$--27$\pm$3\,\kms, which suggests the kinematic distance of $\sim$2.2$\pm$0.3\,kpc (near-side) or $\sim$7.7$\pm$0.3\,kpc (far-side). For these distances, we calculate physical sizes of 14\,pc (for 2.2\,kpc distance) and 48\,pc (for 7.7\,kpc distance). We deem that both of these sizes are realistic for a wind-blown cavity, and thus, for the remainder of the paper, we use both distances (2.2/7.7\,kpc) and corresponding diameters of 14/48\,pc as the most likely values. 
We note that the \HI\ diameter estimate of 48\,pc is in excellent agreement with the orthogonal fit distance estimate from empirical $\Sigma-D$ calibration  \citep{2019SerAJ.199...23S}. The derived distance probability from the same calibration gives $\approx30\%$ chance that Teleios's distance is $>8$\,kpc and a $\approx70\%$ for $<8$\,kpc values. 

Using the calculated radii of the wind-blown cavities, we can estimate a progenitor mass using the method of \citet{2013ApJ...769L..16C}. Comparing with their Table~1, we find corresponding progenitor masses of 28$M_\odot$ (for 14\,pc size) and 54$-$72$M_\odot$. Assuming the distance of $\sim$2.2\,kpc with a corresponding diameter of $\sim$14\,pc and an optimistic upper limit of a constant expansion speed of $\sim$7,000\,\kms\, we estimate Teleios's minimum age of $\sim$980~years ($\sim$1045~A.D.). This is a similar age to a well-known historical \ac{SNR} SN1054 \citep{10542,10541,10543}. We note that Teleios is at \mbox{Dec(J2000)$\sim$--65\D}, placing it in the Southern hemisphere where a very limited amount of firm astronomical events have been recorded in the past.

For a far distance of 7.7\,kpc, our estimated upper limit on the source $\gamma$-ray luminosity above 10\,GeV is $\sim 2\times10^{33}$~erg~s$^{-1}$, which is compatible with the luminosities of some GeV-emitting \acp{SNR}. For a closer distance of 2.2\,kpc our $\gamma$-ray luminosity upper limit is $0.6\times10^{33}$~erg~s$^{-1}$, which is relatively low but still comparable to those of GeV \acp{SNR} likely evolving in low-density environments \citep[e.g.,][]{2016ApJS..224....8A}.

\subsection{Teleios's \ac{SN} explosion type}

At the distances of the inner Sagittarius arms, Scutum--Centaurus and outer Sagittarius arms, Teleios's Galactic latitude (--2\fdg2) would place the object at $\sim$70--540\,pc (assuming distances of 2, 6 and 12\,kpc) below the Galactic plane. The most likely distances of $\sim$2.2\,kpc and $\sim$7.7\,kpc (from Section~\ref{subsec:HI}) places Teleios at $\sim$70\,pc or $\sim$300\,pc below the plane. Therefore, assuming that the \ac{MW}'s thin disk has a scale height of $\sim$220--450\,pc \citep{2016ARA&A..54..529B}, this would indicate that Teleios is located out of the densest Galactic plane regions if at a distance of 7.7~kpc.

Since the Galactic spiral arms represent the areas of the greatest star formation within the \ac{MW}, it is still possible that Teleios's progenitor, whether \ac{CC} or type~Ia \ac{SN}, originated from these areas \citep{2000A&A...358L..13D}. 
\citet[][Figure~3]{2018MNRAS.481L..21P} argue that the spatial distribution of upper main sequence stars shows enhancement corresponding to Galactic spiral arms. However, the distribution of red giants is smooth and can be described as an exponential disk (radial decrease in density) plus a decrease of stellar numbers as a function of heliocentric distance (due to it being a magnitude-limited sample). Therefore, a potential spatial correlation with the Galactic spiral arms would depend on Teleios's progenitor star type.

\subsubsection{Teleios as a \ac{CC} \ac{SN}}

At a distance of $\sim$70--540\,pc below the Galactic plane, one still can expect to find a significant number of massive stars above 8$M_\odot$. It is natural to expect that some of these massive stars may explode as \ac{CC} \ac{SN} and form \acp{SNR} that will expand in this more rarefied and uniform environment outside of the Galactic plane.


However, Teleios could also come from an isolated runaway massive star \ac{CC} \ac{SN} \citep{Bla93, OhKroPfl15, 2024A&A...687L...7W}. This would explain its location far from star-forming regions and 2\fdg2 below the Galactic Plane. \citet{2015MNRAS.447..598S} were able to explain the isolation of some \ac{LBV} stars using a similar scenario. 

In the case of Teleios, a massive star in a binary system could have been kicked from its birth cluster by the explosion of a (more massive) companion. The star then ``escapes'' in a random direction (in this case, perpendicular to the Galactic Plane), ending up in a rather rarefied medium, where it eventually explodes. Still, close to a perfectly symmetric explosion is challenging to achieve for a runaway scenario~\citep{2018ApJ...867...61Z}, not only because the ejecta will be faster on one side but also because a runaway massive star will create a bow shock, which will then disrupt the symmetry of the expanding ejecta \citep{2015MNRAS.450.3080M, 2024MNRAS.534.2918S}.

From Figure~\ref{fig1}, we can see that Teleios is actually located just ($\sim$1\D) below a massive \HII\ region complex. A similar case might be seen in Galactic \ac{SNR} DA530 \citep{2022ApJ...941...17B} located some 500\,pc above a major \HII\ region complex. 
DA530 is an \ac{SNR} that came from a \ac{CC} explosion and has a central neutron star with an X-ray \ac{PWN}. Teleios differs in morphology from DA530, as DA530 is an example of a symmetrically bilateral \ac{SNR}, potentially indicating an environmental difference.

\subsubsection{Teleios as a type~Ia \ac{SN}}
\label{subsubsec:Type Ia}
 
Teleios, at 2\fdg2 below the Galactic Plane, is away from any obvious and nearby star formation activity. Given its circular shape \citep{Ranasinghe:2019quc,2011ApJ...732..114L}, which is similar to circumgalactic \ac{SNR} J0624--6948 \citep{2022MNRAS.512..265F}, and its location outside the Galactic Plane \citep{2017MNRAS.471.1390H}, Teleios could be a type~Ia \ac{SN} explosion from a star that was formed (and lived) below the Galactic Plane. While it is possible for a more massive star to travel outside of the Galactic Plane and explode as a \ac{CC} \acp{SNR}, this scenario is more likely for a smaller Type~Ia progenitor. It is also suggested that Type~Ia \acp{SNR} are more symmetrical than their \ac{CC} counterparts~\citep{2011ApJ...732..114L}, however there is some debate about this relationship, particularly concerning radio morphology~\citet[][Leahy et al. in prep;] {2019AJ....158..149L}. However, if Teleios is located inside the cavity as suggested in Section~\ref{subsec:dist}, then the type~Ia scenario is no longer viable. While the type~Ia scenario does not preclude Teleios's location within a cavity, it is difficult to explain how such a large cavity would have formed. A white dwarf progenitor is not likely to be able to form an $\sim$50\,pc sized cavity, and this scenario would favour a more massive progenitor.

There have been previous attempts to use an \ac{SNR}'s circularity as an indicator of its \ac{SN} type, however we note that this can be a difficult correlation to draw, and circularity is typically a poor indicator of \ac{SN} type. For example, as shown in \citet{Ranasinghe:2019quc}, we see many \acp{SNR} with identical morphological features but coming from various explosion types.
\citet{2019NewAR..8701535S,2024OJAp....7E..31S} argues that all \ac{CC} \acp{SNR} are non-spherical because of the effect of jets, which will imprint ``ears'' onto the spherical bubble. As we do not see any sign of ``ears'' in Teleios, one could conclude that the thermonuclear (type~Iax) \ac{SN} is a more likely scenario. However, we note that the present generation of 3D \ac{SN} simulations doesn't easily produce (show) the \ac{SN} jets. Another caveat is that the jet-driven \ac{SN} explosion is more likely to be relevant for the morphology of younger \acp{SNR}. Further on, the most prominent \ac{SNR} with ``ears'' can be seen in Kepler's \ac{SNR} (G4.5+6.8), and that was definitely a type~Ia.
Also, \citet{2019NewAR..8701535S,2024OJAp....7E..31S} suggests that some (but not all) of the many possible explosion mechanisms of type~Iax do produce perfectly spherically symmetric remnants \citep[Table~1 and Section~4.2,][]{2024OJAp....7E..31S}. As the best example of a circular remnant, \citet{2024OJAp....7E..31S} takes \ac{SNR} Pa\,30, the likely remnant of SN\,1181. Certainly, Teleios is even more circular (the circularity of $c$ = 95.4\%) than Pa\,30 ($c$ = 90.8\%). 

We also discuss the possibilities of Teleios being a type~Ia as \ac{SD}, \ac{DD} or Iax.
Type~Iax has only recently been distinguished from type~Ia as they have lower explosion energy (0.01--0.1$\times10^{51}$\,erg), lower optical luminosity (--14 to --19~mag) and lower or absent X-ray emission. If Teleios indeed comes from a type~Iax \ac{SN}, this lower energy could be the reason why there is no X-ray detection \citep{2013ApJ...767...57F}. We note that there is one prominent eROSITA point source within Teleios's area (1eRASS J131507.1$-$650312). This source is also seen in the radio and appears to resemble a typical background \ac{AGN} with jet structure (see Figure~\ref{fig:hessgleamx}, left panel inset). Therefore, it is unlikely to be associated with Teleios, and we observe no corresponding diffuse X-ray emission in the eROSITA data.
As pointed out in \citet{2022MNRAS.511.2708S}, there should be dozens of so-far unidentified type~Iax remnants in the \ac{MW}. No definitive type~Iax identification has been proposed for \ac{MW} remnants, except probably for SN\,1181 (as discussed above). 

We also note that Teleios's circular shape is consistent with the theoretical ``lonely \ac{WD}'' scenarios presented in~\citet[]{2024RAA....24a5012S, 2024OJAp....7E..31S}. These scenarios include the \ac{CD} and the \ac{DD} with a long Merger to Explosion Delay (MED) time. The \ac{CD} scenario is when a \ac{WD} merges with a more massive companion, forming a massive \ac{WD}, which then explodes after the MED time. The \ac{DD} with a long MED time involves the merging of two \acp{WD}, and then the merger remnant explodes. If the MED time is long enough, then the remnant has relaxed, and \citet[]{2024OJAp....7E..31S} predicts a spherical, near Chandrasekhar mass explosion. \citet[]{2024OJAp....7E..31S} also predicts that these mergers can form \ac{PN} shells that can clear the surrounding \ac{ISM}. For massive and energetic \ac{PN} ejecta, as could be formed by a merger of a \ac{WD} with a relatively massive companion star (about 4-5\,M$_{\odot}$ in the CD scenario), if it were expanding at $\sim$50\,\kms, the expansion time to clean 24\,pc (radius) would be $\sim$500\,000~years (MED time). For this to happen, one needs to account for a very rarefied \ac{CSM}/\ac{ISM}. These scenarios could account for a spherical explosion and a rarefied \ac{CSM}/\ac{ISM}. These properties would impact Teleios's circularity, however if Teleios is in the Sedov phase, then it is more likely that the uniformity of the surrounding medium and Teleios's age contributes more to the observed symmetry rather than the initial explosion geometry.

\subsection{Teleios's possible \ac{SN} progenitor}
  \label{subsec:progenitor}

To better constrain Teleios's \ac{SN} explosion type (as type~Ia or Iax), we search the available GAIA DR3~\citep{GAIA2016, GAIA2023} data for a possible progenitor near Teleios's geometric centre. We investigate stars off of the main sequence; that is, a \ac{WD} (a.k.a. zombie) star could indicate a type~Iax scenario, and a red giant could indicate a \ac{SD} type~Ia scenario. 

We applied a set of parameters based on the GAIA magnitude and colour data to identify any potential post-SN explosion \ac{WD} or \ac{GB} star candidates (remnant star). Our criteria are based on ~\citet[their Figure~2]{2019A&A...631A.119L}. We used the restrictions G$_{\text{abs}} > 7$ and (G$_{\text{BP}}$--G$_{\text{RP}}$) $ < 0.4$ to search for potential \acp{WD} and G$_{\text{abs}} < 4$ and (G$_{\text{BP}}$--G$_{\text{RP}}$) $ > 0.4$ to search for potential \ac{GB} stars. 

Restricting this to Teleios's geometric centre and Galactic distances (assuming a maximum distance of 20\,kpc), we found 1 \ac{WD} candidate within 1\arcmin\ of Teleios's centre and 9 \ac{GB} star candidates within 30\arcsec. 

Analysing the GAIA proper motion data, we find that the \ac{WD} candidate (Gaia DR3~5858854017669128192) does not originate from the direction of Teleios's centre, and thus, we deem it somewhat unlikely to be a progenitor. Also, the measured parallax of 1.8981$\pm$1.7116~mas gives a distance 527$\pm$475\,pc implying $D$=3.3\,pc, which is at odds with the distance estimates from Section~\ref{subsec:HI}. However, the well-studied type~Iax remnant SN1181 has a similar diameter of about 1.82\,pc at the age of 843~years and expansion velocity of $\sim$1100\,\kms\ \citep{Fesen_2023}. Assuming the same expansion velocity of $\sim$1100\,\kms\ of Teleios's ring, we arrive at the \ac{SNR} age of $\sim$1467~year. 
This young age would argue that such events could be recorded in historical records. 
However, as mentioned above, Teleios can be seen only from the Southern Hemisphere, where a limited number of historical astronomical events have been recorded.

Conducting a similar proper motion analysis on the nine potential \ac{GB} star candidates, we find no stars that could have originated from Teleios's centre. 
We conclude that we could not find any suitable white dwarf or red giant candidates as a remnant star from the Teleios original explosion as \ac{SN}.

\subsection{Teleios's evolutionary state}
 \label{subsec:evolutionarystate}

To evaluate Teleios's possible evolutionary phase as type~Ia or type~Iax, we investigated two \ac{SNR} evolutionary models.

\subsubsection{Radio $\Sigma$--$D$ modeling}

\begin{figure*}[ht]
	\centering
	\includegraphics[width=0.32\textwidth]{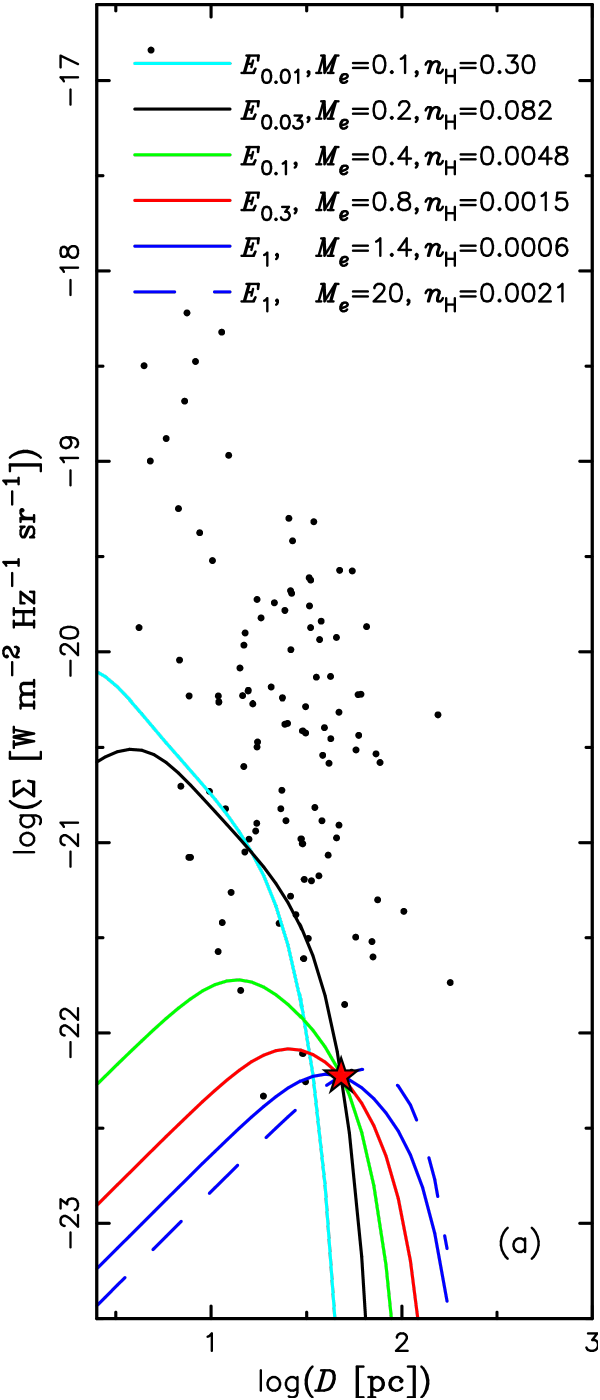}~~~~
	\includegraphics[width=0.32\textwidth]{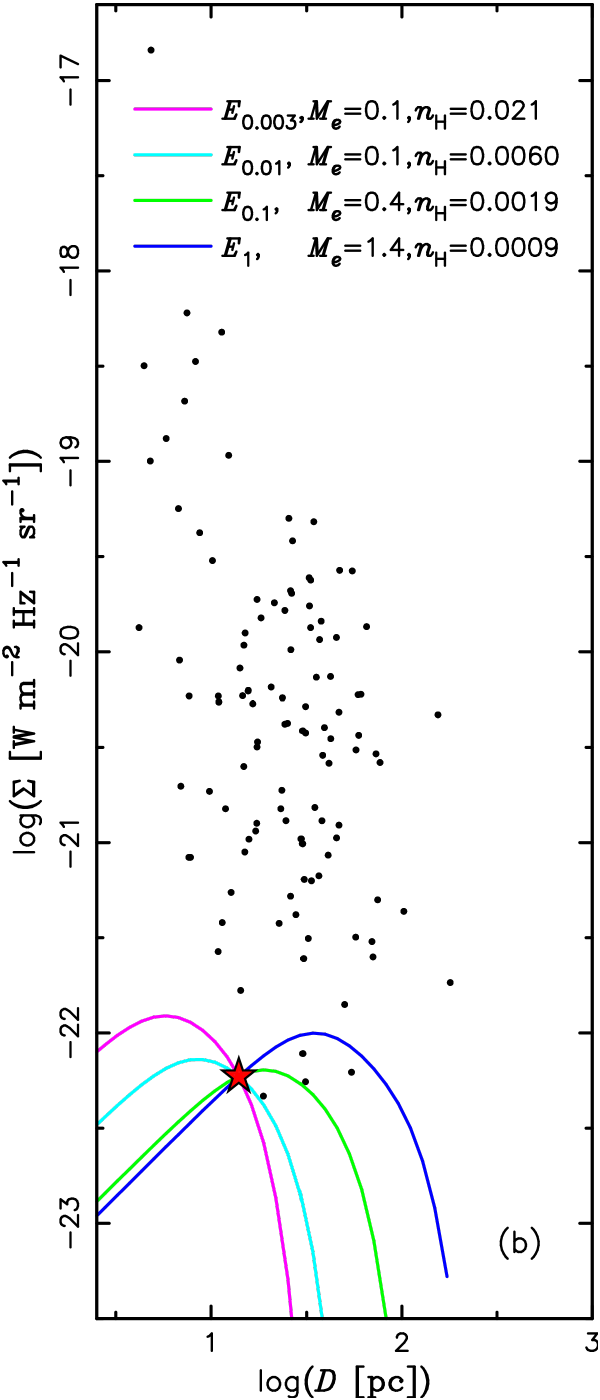}~~~~
	\includegraphics[width=0.32\textwidth]{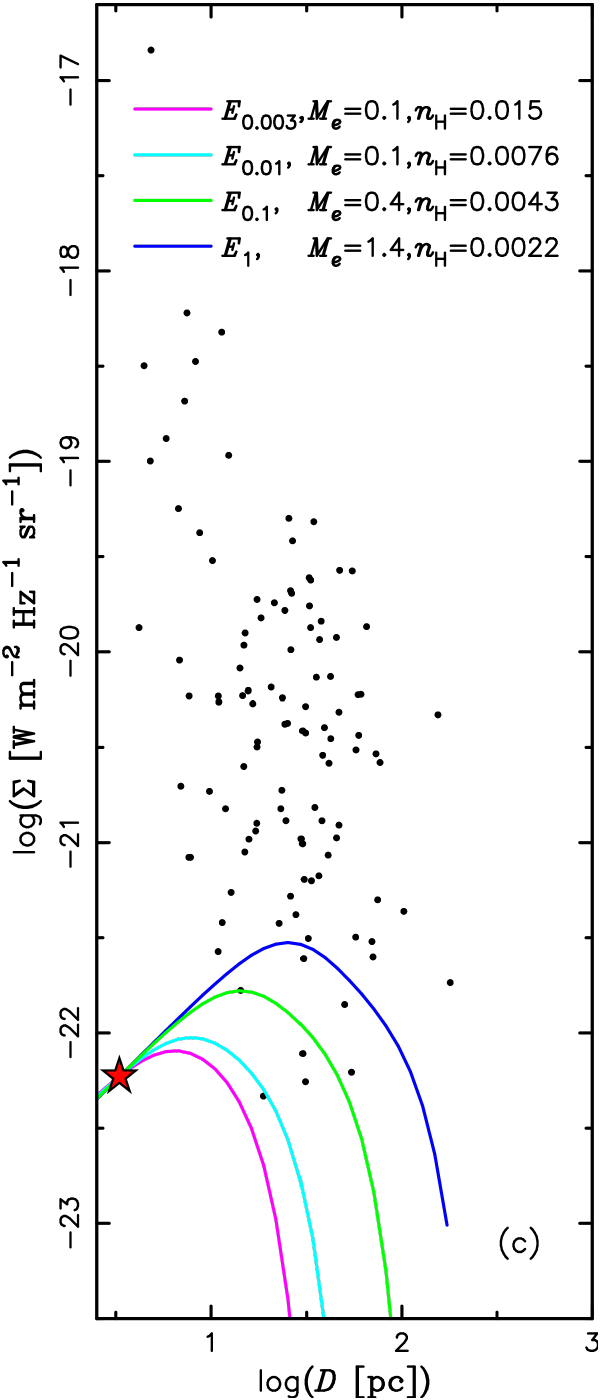}
	\caption{The evolutionary paths for Teleios, obtained using the emission model from \citet{Kosticetal2024}. The panels (a), (b), and (c) stand for the diameters $D=48$~pc, $D=7$~pc, and $D=3.3$~pc, respectively. The explosion energy (in ergs), ejecta mass (in solar masses, M$_{\odot}$) and ambient density (in cm$^{-3}$) for different models are displayed in the legend. The black points represent the Galactic \ac{SNR} sample from \citet{2019SerAJ.199...23S}. The red star marks the Teleios's position on $\Sigma$--$D$ plot.  Note: Density values below 0.001\,H\,cm$^{-3}$ are included in the modelling; however, are likely too low to exist within the Galaxy.}
	\label{fig:SDpaths}
\end{figure*}

Since Teleios's surface brightness is one of the lowest measured, it is expected to evolve in a very rarefied medium. This is certainly in contradiction with the relative proximity of Teleios to the Galactic plane, where densities lower than 0.01\,cm$^{-3}$ can hardly be expected. To estimate its possible ambient densities, we calculate several radio $\Sigma$--$D$ evolutionary paths for Teleios. Three cases for the \ac{SNR} diameter are considered: a) $D=48$\,pc, b) $D=14$\,pc (these two correspond to kinematic distances obtained from HI data), and c) $D=3.3$\,pc (the case of WD progenitor from Section~\ref{subsec:progenitor}). We combined different explosion energies and ejecta masses (all parameters are listed in Table~\ref{tab:evol_models}), in order to model different \ac{SN} scenarios: type~Ia ($E_0=10^{51}$\,erg, $M_e=1.4$\,M$_{\odot}$), type Iax ($E_0=3\times10^{48}$\,erg, $M_e=0.1$\,M$_{\odot}$), and four cases between these two, as well as \ac{CC} \ac{SN} of a massive star ($E_0=10^{51}$\,erg, $M_e=20$\,M$_{\odot}$). Not all of these models are applied to every diameter. The synchrotron emission is modelled with test-particle approximation (non-modified shock) of \ac{DSA} mechanism \citep{Axfordetal1977,Bell1978a,BlandfordOstriker1978}, from \citet{Kosticetal2024}, for a circularly symmetric shell-type \ac{SNR}. The lower limit for the shock thickness of 5\% of the radius (estimated from the radio image on Figure~\ref{fig1}) is used in the model. The evolution of \ac{SNR} shock velocity is calculated using the simple equation from \citet{FinkeDermer2012}:
\begin{equation}
    \label{eq:vs}
    \frac{v_s^2}{2} = \frac{k E_0}{k M_e + 4\pi R^3\rho_0/3},    
\end{equation}
which approximately covers the free expansion and Sedov phase (the constant $k$ is determined so the equation tends to Sedov solution for $R\rightarrow\infty$). Based on the results of the models, we propose the evolutionary phase of Teleios, as shown in Table~\ref{tab:evol_models}. The evolutionary paths of these models are shown on panels (a), (b), and (c) of Figure~\ref{fig:SDpaths}.

In the case a) ($D=48$\,pc), an \ac{SNR} with the lowest explosion energy ($E_{0.01}$) cannot reach the given diameter with the measured surface brightness for any density (the closest approach is at $n_{\textrm{H}}=0.3$\,cm$^{-3}$), so it is ruled out. The highest density (0.082\,cm$^{-3}$) is obtained in $E_{0.03}$ case, with \ac{SNR} being in a pressure-driven shell (PDS) phase. All other models give densities below 0.005\,cm$^{-3}$, with the lowest value of 0.0006\,cm$^{-3}$ for the SN type Ia model. In the case b) ($D=14$\,pc), the highest denisty (0.021\,cm$^{-3}$) is obtained for the lowest energy model ($E_{0.003}$), with the remnant still in Sedov phase. The SN Type Ia model results in the lowest ambient density (0.0009\,cm$^{-3}$), being in the ejecta-dominated phase. In case c) ($D=3.3$\,pc), we did not model the \ac{CC} \ac{SN} type as all model results require an asymmetric remnant with significant X-ray emission. As this is at odds with what is observed, this scenario is deemed very unrealistic. The highest density (0.015\,cm$^{-3}$) is obtained with the lowest explosion energy ($E_{0.003}$). Here, all models are in an ejecta-dominated phase, where the surface brightness of the \ac{SNR} increases with diameter. However, a very young \ac{SNR} would be a bright X-ray source (see the next chapter), making it an unlikely scenario.

As we can see, all the obtained densities are much lower than the value expected close to the Galactic plane. Since the test-particle \ac{DSA} mechanism gives lower emission than modified \ac{DSA}, and we use a lower limit for the shock thickness, as well as the equation (Equation~\ref{eq:vs}), which gives lower initial shock velocities, we conclude that these models give strict upper limits for the ambient densities. 

As an alternative possibility, the young \acp{SNR} with fast oblique shocks, where the ambient field inclination of the amplified magnetic field structures creates a superluminal configuration for magnetized electrons \citep[see][]{Zekovicetal2024}, may have a steeper electron momentum spectra than in the \ac{DSA} mechanism. This mechanism, known as quasi-periodic shock acceleration (QSA), can result in the spectral index in the range of $\alpha = -0.55$ to $-1.35$ and up to GeV. Such a steep spectrum could, in the case of a young \ac{SNR} (e.g. at $D=3.3$\,pc for Teleios), result in much lower synchrotron emission than through \ac{DSA} mechanism, which would possibly give rise to higher ambient densities than in our models.

\begin{table*}
	\centering
	\caption{The results of the radio $\Sigma$--$D$ evolutionary models. Note: Density values below 0.001 H cm$^{-3}$ are included in the modelling; however, are likely too low to exist within the Galaxy.}
	\begin{tabular}{llllllll}
		\multicolumn{8}{c}{a) $D=48$\,pc} \\
		\hline
		Energy & Ejecta & Density & Swept mass & Shock velocity & Age & Spectral & Phase \\
		(10$^{51}$\,erg) & (M$_{\odot}$) & (H\,cm$^{-3}$) & (M$_{\odot}$) & (km\,s$^{-1}$) & (yr) & Index & \\
		\hline
		\hline
		0.03 & 0.2 & 0.08 & 470 & 130 & 72000 & 0.55 & PDS \\
		0.1 & 0.4 & 0.005 & 30 & 970 & 11200 & 0.52 & Sedov \\
		0.3 & 0.8 & 0.001 & 8 & 2800 & 5200 & 0.51 & Sedov \\
		1 & 1.4 & 0.0006 & 3.5 & 6100 & 3100 & 0.50 & Ejecta/Sedov \\
		1 & 20 & 0.002 & 12 & 2000 & 10800 & 0.51 & Ejecta \\
		\hline
        \\
		\multicolumn{8}{c}{b) $D=14$\,pc} \\
		\hline
		Energy & Ejecta & Density & Swept mass & Shock velocity & Age & Spectral & Phase \\
		(10$^{51}$\,erg) & (M$_{\odot}$) & (H\,cm$^{-3}$) & (M$_{\odot}$) & (km\,s$^{-1}$) & (yr) & Index & \\
		\hline
		\hline
		0.003 & 0.1 & 0.02 & 3 & 490 & 7100 & 0.51 &  Sedov \\
		0.01 & 0.1 & 0.006 & 0.8 & 1550 & 2800 & 0.5 &  Sedov \\
		0.1 & 0.4 & 0.002 & 0.3 & 4550 & 1400 & 0.5 &  Ejecta \\
		1 & 1.4 & 0.0009 & 0.1 & 8300 & 800 & 0.5 &  Ejecta \\	
		\hline
        \\
		\multicolumn{8}{c}{b) $D=3.3$\,pc} \\
		\hline
		Energy & Ejecta & Density & Swept mass & Shock velocity & Age & Spectral & Phase \\
		(10$^{51}$\,erg) & (M$_{\odot}$) & (H\,cm$^{-3}$) & (M$_{\odot}$) & (km\,s$^{-1}$) & (yr) & Index & \\
		\hline
		\hline
		0.003 & 0.1 & 0.01 & 0.03 & 1650 & 940 & 0.5 &  Ejecta \\
		0.01 & 0.1 & 0.008 & 0.01 & 3100 & 510 & 0.5 &  Ejecta \\
		0.1 & 0.4 & 0.004 & 0.008 & 5000 & 320 & 0.5 &  Ejecta \\
		1 & 1.4 & 0.002 & 0.004 & 8400 & 190 & 0.5 &  Ejecta \\	
		\hline
	\end{tabular}
	\label{tab:evol_models}
\end{table*}

\begin{table*}
\scriptsize
\centering
\caption{The results of the \ac{SNR} evolutionary models.}
\begin{tabular}{llllllllllll}
\multicolumn{12}{c}{a) distance=7.7\,kpc radius=24\,pc}\\
\hline
Density & Energy & Ejecta & age & EDtoST & STtoPDS & EM$_{FS}$ & kT$_{FS}$ & flux$_0$ & flux$_{2\times10^{21}}$ & flux$_{4\times10^{21}}$ & flux$_{7\times10^{21}}$ \\
(cm$^{-3}$) & ($10^{51}$erg) & (M$_{\odot}$)  & (yr) & (yr) & (yr) & ($10^{58}$cm$^{-3}$) & (K) & (erg/s/cm$^2$) &  (erg/s/cm$^2$) &  (erg/s/cm$^2$) &  (erg/s/cm$^2$) \\
\hline
\hline
0.01 & 0.003 & 1.4   & 115000 & 35000 & 53000 & 1.60E+03 & 1.50E+05 & 7.11E-14 & 2.36E-18 & 8.77E-21 & 2.81E-23 \\
0.01 & 0.01 & 1.4   & 56000 & 19000 & 69000 & 1.30E+03 & 9.90E+05 & 1.51E-08 & 4.14E-10 & 6.65E-11 & 7.96E-12 \\
0.01 & 0.1 & 1.4   & 17700 & 6100 & 113000 & 1.30E+03 & 5.80E+06 & 5.51E-08 & 2.24E-08 & 1.15E-08 & 5.27E-09 \\
0.01 & 1 & 1.4   & 5600 & 1900 & 185000 & 1.30E+03 & 1.40E+07 & 4.48E-08 & 2.49E-08 & 1.71E-08 & 1.14E-08 \\
\hline
0.01 & 0.1 & 0.2   & 16200 & 1200 & 113000 & 1.30E+03 & 5.50E+06 & 5.46E-08 & 2.18E-08 & 1.11E-08 & 4.97E-09 \\
0.01 & 0.1 & 20   & 29600 & 56000 & 113000 & 1.90E+03 & 4.90E+06 & 7.81E-08 & 3.01E-08 & 1.49E-08 & 6.46E-09 \\ 
\hline
0.001 & 0.1 & 1.4   & 8600 & 13000 & 420000 & 1.90E+03 & 7.10E+06 & 8.43E-08 & 3.70E-08 & 2.00E-08 & 9.94E-09 \\
0.1 & 0.1 & 1.4   & 53500 & 2800 & 30000 & 1.30E+03 & 6.70E+05 & 4.21E-09 & 4.22E-11 & 5.28E-12 & 4.76E-13 \\
0.3 & 0.1 & 1.4   & 122000 & 2000 & 16000 & 2.10E+03 & 1.00E+05 & 2.72E-16 & 8.85E-22 & 5.31E-25 & 2.24E-28 \\
\hline
\\
\multicolumn{12}{c}{b) distance=3.85\,kpc radius=12\,pc} \\
\hline
Density & Energy & Ejecta & age & EDtoST & STtoPDS & EM$_{FS}$ & kT$_{FS}$ & flux$_0$ & flux$_{2\times10^{21}}$ & flux$_{4\times10^{21}}$ & flux$_{7\times10^{21}}$ \\
(cm$^{-3}$) & ($10^{51}$erg) & (M$_{\odot}$)  & (yr) & (yr) & (yr) & ($10^{58}$cm$^{-3}$) & (K) & (erg/s/cm$^2$) &  (erg/s/cm$^2$) &  (erg/s/cm$^2$) &  (erg/s/cm$^2$) \\
\hline
\hline
0.01 & 0.01 & 1.4 &  14400 & 19000 & 69000 & 2.50E+02 & 4.50E+06 & 4.03E-08 & 1.47E-08 & 7.10E-09 & 2.96E-09 \\
0.01 & 0.1 & 1.4 & 4500 & 6100 & 113000 & 2.50E+02 & 1.10E+07 & 4.25E-08 & 2.20E-08 & 1.37E-08 & 8.28E-09 \\
0.01 & 1 & 1.4   & 1430 & 1900 & 185000 & 2.50E+02 & 3.45E+07 & 8.51E-09 & 7.34E-09 & 9.94E-09 & 1.29E-08 \\
\hline
\\
\multicolumn{12}{c}{b) distance=15.4\,kpc radius=48\,pc} \\
\hline
Density & Energy & Ejecta & age & EDtoST & STtoPDS & EM$_{FS}$ & kT$_{FS}$ & flux$_0$ & flux$_{2\times10^{21}}$ & flux$_{4\times10^{21}}$ & flux$_{7\times10^{21}}$ \\
(cm$^{-3}$) & ($10^{51}$erg) & (M$_{\odot}$)  & (yr) & (yr) & (yr) & ($10^{58}$cm$^{-3}$) & (K) & (erg/s/cm$^2$) &  (erg/s/cm$^2$) &  (erg/s/cm$^2$) &  (erg/s/cm$^2$) \\
\hline
\hline
0.01 & 0.01 & 1.4   & 380000 & 19000 & 69000 & 1.10E+04 & 4.40E+04 & 8.50E-19 & 2.63E-25 & 2.25E-29 & 1.01E-33 \\
0.01 & 0.1 & 1.4 & 91500 & 6100 & 113000 & 1.10E+04 & 1.25E+06 & 5.40E-08 & 2.74E-09 & 5.23E-10 & 7.50E-11 \\
0.01 & 1 & 1.4   & 29000 & 1900 & 185000 & 1.10E+04 & 7.30E+06 & 1.23E-07 & 5.46E-08 & 2.97E-08 & 1.49E-08 \\
\hline
\\
\multicolumn{12}{c}{d) distance=2.2\,kpc radius=6.85\,pc} \\
\hline
Density & Energy & Ejecta & age & EDtoST & STtoPDS & EM$_{FS}$ & kT$_{FS}$ & flux$_0$ & flux$_{2\times10^{21}}$ & flux$_{4\times10^{21}}$ & flux$_{7\times10^{21}}$ \\
(cm$^{-3}$) & ($10^{51}$erg) & (M$_{\odot}$)  & (yr) & (yr) & (yr) & ($10^{58}$cm$^{-3}$) & (K) & (erg/s/cm$^2$) &  (erg/s/cm$^2$) &  (erg/s/cm$^2$) &  (erg/s/cm$^2$) \\
\hline
\hline
0.01 & 0.01 & 1.4   & 9580 & 8900 & 18500 & 4.02E+01 & 4.28E+06 & 1.92E-08 & 6.98E-09 & 3.39E-09 & 1.41E-09 \\
0.01 & 0.1 & 1.4 & 3030 & 2800 & 30300 & 4.05E+01 & 1.70E+07 & 1.32E-08 & 8.09E-09 & 6.20E-09 & 4.65E-09 \\
0.01 & 1 & 1.4   & 958 & 890 & 49600 & 4.10E+01 & 4.21E+07 & 1.38E-08 & 9.81E-09 & 8.63E-09 & 7.66E-09 \\
\hline
\\
\multicolumn{12}{c}{e) distance=7.7\,kpc radius=24\,pc n=10} \\
\hline
Density & Energy & Ejecta & age & EDtoST & STtoPDS & EM$_{FS}$ & kT$_{FS}$ & flux$_0$ & flux$_{2\times10^{21}}$ & flux$_{4\times10^{21}}$ & flux$_{7\times10^{21}}$ \\
(cm$^{-3}$) & ($10^{51}$erg) & (M$_{\odot}$)  & (yr) & (yr) & (yr) & ($10^{58}$cm$^{-3}$) & (K) & (erg/s/cm$^2$) &  (erg/s/cm$^2$) &  (erg/s/cm$^2$) &  (erg/s/cm$^2$) \\
\hline
\hline
0.01 & 0.01 & 1.4   & 55500 & 12600 & 68900 & 1.31E+03 & 9.98E+05 & 1.94E-08 & 5.94E-10 & 9.96E-11 & 1.12E-11 \\
0.01 & 0.1 & 1.4 & 17600 & 3990 & 11300 & 1.40E+03 & 5.81E+06 & 6.41E-08 & 2.70E-08 & 1.42E-08 & 6.60E-09 \\
0.01 & 1 & 1.4   & 5550 & 1260 & 185000 & 1.49E+03 & 1.39E+07 & 4.87E-08 & 2.86E-08 & 2.05E-08 & 1.40E-08 \\
\hline
\end{tabular}
\label{tab:SNR_models}
\end{table*}

\subsubsection{Evolutionary model for \ac{SNR} including X-ray emission} 
We can apply the \cite{2019AJ....158..149L} evolutionary models for circularly symmetric \ac{SNR} to estimate its evolutionary state.  Important input parameters are \ac{SNR} radius, explosion energy, ejected mass and \ac{ISM} density.
The reference distance is 7.7\,kpc (radius 24\,pc) for Teleios, but we test distances half and twice as large as well as the nearest distance of 2.2\,kpc. For energy, we test values of $3\times10^{48}$, $10^{49}$, $10^{50}$ and $10^{51}$\,erg, thus allowing the possibility of a low-energy \ac{SN}. For ejected mass, we use 1.4\,M$_{\odot}$ for a type~Ia but also test lower (0.2\,M$_{\odot}$) and higher (20\,M$_{\odot}$) values. The environment is likely low density, so we test \ac{ISM} densities of 0.001, 0.01, 0.1 and 0.3\,cm$^{-3}$. For the ejecta power law index, we use n\,=\,7 for most cases, and n\,=\,10 for a subset.

The \ac{SNR} models yield several properties, including shock temperatures and emission measures for forward and reverse shocks and transition times that yield the evolutionary phase. The transition time from ejecta-dominated to Sedov-Taylor phase is labelled EDtoST, and from Sedov-Taylor to pressure-driven shell as STtoPDS. 
For all n\,=\,7 cases, the forward shock produces much more flux than the reverse shock, so we give the forward shock properties (emission measure, EM$_{FS}$, and temperature, kT$_{FS}$). 
We calculate the 0.2-10 keV flux from the shocked gas from EM$_{FS}$ and kT$_{FS}$ using the WebSpec tool at HEASARC\footnote{\url{heasarc.gsfc.nasa.gov}} for an APEC hot plasma spectrum with interstellar absorption. The total column of interstellar absorption in the direction of Teleios is $7.5\times10^{21}$cm$^{-2}$, and likely column densities for distances of 3.85, 7.7 and 15.4\,kpc are $2\times10^{21}$cm$^{-2}$, $4\times10^{21}$cm$^{-2}$ and $7\times10^{21}$cm$^{-2}$ so we calculate unabsorbed flux and absorbed (observed) fluxes (labeled flux$_{0}$, flux$_{2\times10^{21}}$,  flux$_{4\times10^{21}}$ and flux$_{7\times10^{21}}$).

The results of the models are given in Table~\ref{tab:SNR_models}, with case a) for the reference distance, b) for half that distance, c) for twice that distance and d) for the near distance case. We also give the results for n=10 for the reference distance, e) in Table~\ref{tab:SNR_models}.
The properties of the above set of models can be summarised as follows.
For the fiducial case ($10^{50}$\,erg, 1.4\,M$_{\odot}$, 0.01\,cm$^{-3}$), the age to reach a radius of 24\,pc is 17700\,yr and it is in the Sedov phase. It has forward shock temperature of $5.8\times10^6$\,K and emission measure of $1.3\times10^{61}$\,cm$^{-3}$. However, it would be a bright X-ray source in the 0.2--10\,keV band.

Changing the explosion energy to $10^{49}$ or $10^{51}$\,erg changes the age (to 56000 or 5600\,yr, resp.) and shock temperature (to $9.9\times10^5$\,K or $1.4\times10^7$\,K) but does not change the emission measure. Changing the ejecta mass to 0.2 or 20\,M$_{\odot}$ changes the age (to 16000 or 30000\,yr, resp.) and changes emission measure by a factor $<1.5$, shock temperature by a factor $<1.2$ and the X-ray flux by a factor $<2$. Decreasing the energy to $3\times10^{48}$\,erg, however, decreases the shock temperature enough that the \ac{SNR} is no longer detectable in X-rays.

Changing the \ac{ISM} density to 0.001\,cm$^{-3}$ similarly changes shock temperature and emission measure by small factors and the flux by a factor $<2$. However, increasing the density to 0.1\,cm$^{-3}$ puts the \ac{SNR} age at 53000~yr well in the PDS phase which starts at 30000~yr and reduces the shock temperature to $7\times10^5$\,K, which yields a much smaller absorbed X-ray flux, but still detectable by eROSITA. Increasing the density to 0.3\,cm$^{-3}$ however, decreases the shock temperature enough to make the \ac{SNR} undetectable in X-rays. The n=10 case with energy $10^{49}$\,erg, is the only case where the reverse shock flux is higher than the forward shock flux, but it is only brighter by a factor of $\sim$1 to 20, and does not change the conclusion that the \ac{SNR} would be a bright X-ray source. 

Decreasing the distance to either 3.85 or 2.2\,kpc results in a bright X-ray source for all cases. Increasing the distance decreases the shock temperature and makes the \ac{SNR} considerably fainter, mainly because of decreased shock temperature and partly because of increased \ac{ISM} absorption. Increasing the ejecta power-law index to n=10 (case e) yields slightly brighter X-ray emission in all cases in comparison to n=7 models, so it cannot give an \ac{SNR} with low enough X-ray flux.

In summary, to have a low X-ray flux (below $3\times10^{-13}$ erg cm$^{-2}$ s$^{-1}$), the shock temperature needs to be below $\sim3\times10^5$ K. The models can satisfy this if the energy is low ($\lesssim 10^{49}$\,erg) and either the distance is large ($\gtrsim 15$\,kpc) or the \ac{ISM} density is high ($\gtrsim 0.2$\,cm$^{-3}$), so that the \ac{SNR} is in an evolved state. In this case, the \ac{SNR} has transitioned from the Sedov-Taylor phase into the pressure-driven shell phase. This is certainly in contradiction with our initial suggestion that Teleios should be a youngish \ac{SNR}.

\subsection{What is Teleios?}

Teleios's unusually close-to-perfect circular shape, steep radio spectral index ($\alpha=-0.6\pm$0.3), weak polarisation signature and low surface brightness are rather challenging to reconcile with typical \ac{SNR} characteristics.
However, despite only confirmed radio-continuum emission, there is little doubt that Teleios is a Galactic \ac{SNR} as no other known source type could better fit its radio properties. 

We observe that Teleios is located in an environment with a high level of ambient \ac{RM}, with the largest \ac{RM} being observed at Teleios's centre (Section~\ref{subsec:polarisation}). This unusual \ac{RM} environment, combined with Teleios's remarkably low surface brightness, raises the possibility that this may be caused by Teleios's orientation. That is, if Teleios is being viewed end on, and if the ambient magnetic field is oriented along our line of sight, either towards or away from us, then the compression would be perpendicular to our line of sight. Thus, the typical synchrotron emission observed from most \acp{SNR} would be predominantly oriented perpendicular to our line of sight, resulting in Teleios's observed low surface brightness. This scenario would also explain the lack of polarisation from the shells, as this magnetic field orientation would result in increased Faraday rotation rather than synchrotron emission in our direction. It would also explain the observed circular symmetry, as the compression occurs perpendicular to our line of sight. 

This scenario is also supported by Teleios's possible Galactic location when compared with the magnetic field models of~\cite{West2016}. Namely, the~\cite{West2016} model of \ac{SNR}~G302.3+00.7 is the best comparison, as it is the example located closest to Teleios and so is the best representation of the local Galactic magnetic field in this area. Comparing Teleios with~\citet[their Figure~D.1]{West2016}, we see that the models for $\sim$6\,kpc show a round morphology similar to Teleios. This model demonstrates that these circular morphologies typically occur in regions where the Galactic magnetic field lines are orientated along our line of sight, i.e. we are viewing the Galactic magnetic field end on. 

We investigated Teleios's possible evolutionary state as \ac{SN} \ac{CC}, type~Ia or Iax explosions by applying two different theoretical models.
We have a challenge reconciling the observational evidence, such as Teleios's low surface--brightness and circular shape, steep spectral index (--0.6) and a possible size of D\,=\,14/48\,pc with any of the presented scenarios.
As in the case of only known type~Iax \ac{SNR} from SN1181 with $D=$1.8\,pc, Teleios's size of $D$=48\,pc at the distance of 7.7\,kpc is very much out of any acceptable range. If Teleios comes from type~Iax \ac{SN} explosion, it should be located at much closer distances than even our lower distance estimate of 2.2\,kpc and corresponding $D$=14\,pc. This immediately places WD Gaia DR3~5858854017669128192 in the spotlight as a possible remnant star at only 527\,pc distance, thus giving Teleios a diameter $D$=3.3\,pc (see Section~\ref{subsec:progenitor}). However, none of the other independent measurements place Teleios at this distance.
At the same time, neither of the two evolutionary methods (see Section~\ref{subsec:evolutionarystate}) could explain Teleios as a type~Ia \ac{SNR} without detectable X-ray emission.

Certainly, future high-resolution multi-frequency observations could determine a possible expansion velocity, which could more accurately indicate Teleios's properties.

\section{CONCLUSION}
\label{sec:conclusion}

We suggest that the low surface-brightness circular object \g, Teleios, detected in our new \ac{ASKAP} \ac{EMU} images, is most likely a new Galactic \ac{SNR} with spectral index of $\alpha=-0.6\pm$0.3, a diameter of either 14\,pc or 48\,pc and at a distance of $\sim$2.2 or 7.7\,kpc. 

We consider several different scenarios to explain Teleios's unusual properties, all of which have their challenges. We consider a \ac{CC} progenitor scenario, but this is deemed unlikely due to Teleios's distance from the Galactic Plane (70--540\,pc) and spherical symmetry. We also explored a type~Iax explosion scenario that would argue for a much closer distance ($<$1\,kpc) and sizes of only $\sim$3.3\,pc, which would be comparable to the only known type~Iax remnant SN1181. We also consider a Type~Ia scenario, which we argue to be the most likely. Although we note that the lack of detectable X-ray emission is puzzling, and the hint of H$\alpha$ emission projected within the remnant would be surprising for an SN~Ia.

We have made an exhaustive exploration of the possible evolutionary state of the SN based on its surface brightness, apparent size and possible distances. All possible scenarios have their challenges, especially considering the lack of X-ray emission that is expected to be detectable given our evolutionary modelling. While we deem the Type~Ia scenario the most likely, we note that no direct evidence is available to definitively confirm any scenario and new sensitive and high-resolution observations of this object are needed.


\begin{acknowledgement}

This scientific work uses data obtained from Inyarrimanha Ilgari Bundara, the \ac{CSIRO} Murchison Radio-astronomy Observatory. We acknowledge the Wajarri Yamaji People as the Traditional Owners and native title holders of the Observatory site. CSIRO’s \ac{ASKAP} radio telescope is part of the Australia Telescope National Facility (\url{https://ror.org/05qajvd42}). Operation of \ac{ASKAP} is funded by the Australian Government with support from the National Collaborative Research Infrastructure Strategy. \ac{ASKAP} uses the resources of the Pawsey Supercomputing Research Centre. Establishment of \ac{ASKAP}, Inyarrimanha Ilgari Bundara, the \ac{CSIRO} Murchison Radio-astronomy Observatory and the Pawsey Supercomputing Research Centre are initiatives of the Australian Government, with support from the Government of Western Australia and the Science and Industry Endowment Fund. We thank I. Sushch for scientific discussions which improved the paper.
We thank an anonymous referee for comments and suggestions that greatly improved our paper.

\end{acknowledgement}

\paragraph{Funding Statement}

MDF, GR and SL acknowledge \ac{ARC} funding through grant DP200100784.  
N.H.-W. is the recipient of an \ac{ARC} Future Fellowship (project number FT190100231).
HS acknowledges funding from JSPS KAKENHI Grant Number 21H01136.
DU and BA acknowledge the financial support provided by the Ministry of Science, Technological Development and Innovation of the Republic of Serbia through the contract 451-03-66/2024-03/200104 and for support through the joint project of the Serbian Academy of Sciences and Arts and Bulgarian Academy of Sciences  ``Optical search for Galactic and extragalactic supernova remnants''. BA additionally acknowledges the funding provided by the Science Fund of the Republic of Serbia through project \#7337 ``Modeling Binary Systems That End in Stellar Mergers and Give Rise to Gravitational Waves'' (MOBY).
RB acknowledges funding from the Irish Research Council under the Government of Ireland Postdoctoral Fellowship program. 
JM acknowledges support from a Royal Society-Science Foundation Ireland University Research Fellowship (20/RS-URF-R/3712).
CBS acknowledges support from a Royal Society Research Fellows Enhancement Award 2021 (22/RS-EA/3810).
JM and CBS acknowledge that this publication results from research conducted with the financial support of Taighde \'Eireann - Research Ireland under Grant numbers 20/RS-URF-R/3712, 22/RS-EA/3810.
SL, PK, AM and BV were supported by the Ministry of Science, Technological Development and Innovation of the Republic of Serbia (MSTDIRS) through contract no. 451-03-66/2024-03/200002 made with Astronomical Observatory (Belgrade).
RS is supported by INAF grant 1.05.23.04.04.

\paragraph{Data Availability Statement}

The data that support the plots/images within this paper and other findings of this study are available from the corresponding author upon reasonable request. The \ac{ASKAP} data used in this article are available through the \ac{CASDA}\footnote{\url{https://research.csiro.au/casda}}.

\printendnotes

\bibliography{Teleios}

\begin{thebibliography}{}
\expandafter\ifx\csname natexlab\endcsname\relax\def\natexlab#1{#1}\fi

\bibitem[{{Abdollahi} {et~al.}(2020){Abdollahi}, {Acero}, {Ackermann}, {Ajello}, {Atwood}, {Axelsson}, {Baldini}, {Ballet}, {Barbiellini}, {Bastieri}, {Becerra Gonzalez}, {Bellazzini}, {Berretta}, {Bissaldi}, {Blandford}, {Bloom}, {Bonino}, {Bottacini}, {Brandt}, {Bregeon}, {Bruel}, {Buehler}, {Burnett}, {Buson}, {Cameron}, {Caputo}, {Caraveo}, {Casandjian}, {Castro}, {Cavazzuti}, {Charles}, {Chaty}, {Chen}, {Cheung}, {Chiaro}, {Ciprini}, {Cohen-Tanugi}, {Cominsky}, {Coronado-Bl{\'a}zquez}, {Costantin}, {Cuoco}, {Cutini}, {D'Ammando}, {DeKlotz}, {de la Torre Luque}, {de Palma}, {Desai}, {Digel}, {Di Lalla}, {Di Mauro}, {Di Venere}, {Dom{\'\i}nguez}, {Dumora}, {Fana Dirirsa}, {Fegan}, {Ferrara}, {Franckowiak}, {Fukazawa}, {Funk}, {Fusco}, {Gargano}, {Gasparrini}, {Giglietto}, {Giommi}, {Giordano}, {Giroletti}, {Glanzman}, {Green}, {Grenier}, {Griffin}, {Grondin}, {Grove}, {Guiriec}, {Harding}, {Hayashi}, {Hays}, {Hewitt}, {Horan}, {J{\'o}hannesson}, {Johnson}, {Kamae}, {Kerr}, {Kocevski}, {Kovac'evic'},
  {Kuss}, {Landriu}, {Larsson}, {Latronico}, {Lemoine-Goumard}, {Li}, {Liodakis}, {Longo}, {Loparco}, {Lott}, {Lovellette}, {Lubrano}, {Madejski}, {Maldera}, {Malyshev}, {Manfreda}, {Marchesini}, {Marcotulli}, {Mart{\'\i}-Devesa}, {Martin}, {Massaro}, {Mazziotta}, {McEnery}, {Mereu}, {Meyer}, {Michelson}, {Mirabal}, {Mizuno}, {Monzani}, {Morselli}, {Moskalenko}, {Negro}, {Nuss}, {Ojha}, {Omodei}, {Orienti}, {Orlando}, {Ormes}, {Palatiello}, {Paliya}, {Paneque}, {Pei}, {Pe{\~n}a-Herazo}, {Perkins}, {Persic}, {Pesce-Rollins}, {Petrosian}, {Petrov}, {Piron}, {Poon}, {Porter}, {Principe}, {Rain{\`o}}, {Rando}, {Razzano}, {Razzaque}, {Reimer}, {Reimer}, {Remy}, {Reposeur}, {Romani}, {Saz Parkinson}, {Schinzel}, {Serini}, {Sgr{\`o}}, {Siskind}, {Smith}, {Spandre}, {Spinelli}, {Strong}, {Suson}, {Tajima}, {Takahashi}, {Tak}, {Thayer}, {Thompson}, {Tibaldo}, {Torres}, {Torresi}, {Valverde}, {Van Klaveren}, {van Zyl}, {Wood}, {Yassine}, \& {Zaharijas}}]{2020ApJS..247...33A}
{Abdollahi}, S., {Acero}, F., {Ackermann}, M., {et~al.} 2020, ApJs, 247, 33

\bibitem[{{Abdollahi} {et~al.}(2022){Abdollahi}, {Acero}, {Baldini}, {Ballet}, {Bastieri}, {Bellazzini}, {Berenji}, {Berretta}, {Bissaldi}, {Blandford}, {Bloom}, {Bonino}, {Brill}, {Britto}, {Bruel}, {Burnett}, {Buson}, {Cameron}, {Caputo}, {Caraveo}, {Castro}, {Chaty}, {Cheung}, {Chiaro}, {Cibrario}, {Ciprini}, {Coronado-Bl{\'a}zquez}, {Crnogorcevic}, {Cutini}, {D'Ammando}, {De Gaetano}, {Digel}, {Di Lalla}, {Dirirsa}, {Di Venere}, {Dom{\'\i}nguez}, {Fallah Ramazani}, {Fegan}, {Ferrara}, {Fiori}, {Fleischhack}, {Franckowiak}, {Fukazawa}, {Funk}, {Fusco}, {Galanti}, {Gammaldi}, {Gargano}, {Garrappa}, {Gasparrini}, {Giacchino}, {Giglietto}, {Giordano}, {Giroletti}, {Glanzman}, {Green}, {Grenier}, {Grondin}, {Guillemot}, {Guiriec}, {Gustafsson}, {Harding}, {Hays}, {Hewitt}, {Horan}, {Hou}, {J{\'o}hannesson}, {Karwin}, {Kayanoki}, {Kerr}, {Kuss}, {Landriu}, {Larsson}, {Latronico}, {Lemoine-Goumard}, {Li}, {Liodakis}, {Longo}, {Loparco}, {Lott}, {Lubrano}, {Maldera}, {Malyshev}, {Manfreda}, {Mart{\'\i}-Devesa},
  {Mazziotta}, {Mereu}, {Meyer}, {Michelson}, {Mirabal}, {Mitthumsiri}, {Mizuno}, {Moiseev}, {Monzani}, {Morselli}, {Moskalenko}, {Negro}, {Nuss}, {Omodei}, {Orienti}, {Orlando}, {Paneque}, {Pei}, {Perkins}, {Persic}, {Pesce-Rollins}, {Petrosian}, {Pillera}, {Poon}, {Porter}, {Principe}, {Rain{\`o}}, {Rando}, {Rani}, {Razzano}, {Razzaque}, {Reimer}, {Reimer}, {Reposeur}, {S{\'a}nchez-Conde}, {Saz Parkinson}, {Scotton}, {Serini}, {Sgr{\`o}}, {Siskind}, {Smith}, {Spandre}, {Spinelli}, {Sueoka}, {Suson}, {Tajima}, {Tak}, {Thayer}, {Thompson}, {Torres}, {Troja}, {Valverde}, {Wood}, \& {Zaharijas}}]{2022ApJS..260...53A}
{Abdollahi}, S., {Acero}, F., {Baldini}, L., {et~al.} 2022, ApJs, 260, 53

\bibitem[{{Acero} {et~al.}(2016){Acero}, {Ackermann}, {Ajello}, {Baldini}, {Ballet}, {Barbiellini}, {Bastieri}, {Bellazzini}, {Bissaldi}, {Blandford}, {Bloom}, {Bonino}, {Bottacini}, {Brandt}, {Bregeon}, {Bruel}, {Buehler}, {Buson}, {Caliandro}, {Cameron}, {Caputo}, {Caragiulo}, {Caraveo}, {Casandjian}, {Cavazzuti}, {Cecchi}, {Chekhtman}, {Chiang}, {Chiaro}, {Ciprini}, {Claus}, {Cohen}, {Cohen-Tanugi}, {Cominsky}, {Condon}, {Conrad}, {Cutini}, {D'Ammando}, {de Angelis}, {de Palma}, {Desiante}, {Digel}, {Di Venere}, {Drell}, {Drlica-Wagner}, {Favuzzi}, {Ferrara}, {Franckowiak}, {Fukazawa}, {Funk}, {Fusco}, {Gargano}, {Gasparrini}, {Giglietto}, {Giommi}, {Giordano}, {Giroletti}, {Glanzman}, {Godfrey}, {Gomez-Vargas}, {Grenier}, {Grondin}, {Guillemot}, {Guiriec}, {Gustafsson}, {Hadasch}, {Harding}, {Hayashida}, {Hays}, {Hewitt}, {Hill}, {Horan}, {Hou}, {Iafrate}, {Jogler}, {J{\'o}hannesson}, {Johnson}, {Kamae}, {Katagiri}, {Kataoka}, {Katsuta}, {Kerr}, {Kn{\"o}dlseder}, {Kocevski}, {Kuss}, {Laffon}, {Lande},
  {Larsson}, {Latronico}, {Lemoine-Goumard}, {Li}, {Li}, {Longo}, {Loparco}, {Lovellette}, {Lubrano}, {Magill}, {Maldera}, {Marelli}, {Mayer}, {Mazziotta}, {Michelson}, {Mitthumsiri}, {Mizuno}, {Moiseev}, {Monzani}, {Moretti}, {Morselli}, {Moskalenko}, {Murgia}, {Nemmen}, {Nuss}, {Ohsugi}, {Omodei}, {Orienti}, {Orlando}, {Ormes}, {Paneque}, {Perkins}, {Pesce-Rollins}, {Petrosian}, {Piron}, {Pivato}, {Porter}, {Rain{\`o}}, {Rando}, {Razzano}, {Razzaque}, {Reimer}, {Reimer}, {Renaud}, {Reposeur}, {Rousseau}, {Saz Parkinson}, {Schmid}, {Schulz}, {Sgr{\`o}}, {Siskind}, {Spada}, {Spandre}, {Spinelli}, {Strong}, {Suson}, {Tajima}, {Takahashi}, {Tanaka}, {Thayer}, {Thompson}, {Tibaldo}, {Tibolla}, {Torres}, {Tosti}, {Troja}, {Uchiyama}, {Vianello}, {Wells}, {Wood}, {Wood}, {Yassine}, {den Hartog}, \& {Zimmer}}]{2016ApJS..224....8A}
{Acero}, F., {Ackermann}, M., {Ajello}, M., {et~al.} 2016, ApJs, 224, 8

\bibitem[{{Alsaberi} {et~al.}(2019){Alsaberi}, {Barnes}, {Filipovi{\'c}}, {Maxted}, {Sano}, {Rowell}, {Bozzetto}, {Gurovich}, {Uro{\v{s}}evi{\'c}}, {Oni{\'c}}, {For}, {Manojlovi{\'c}}, {Wong}, {Galvin}, {Kavanagh}, {Ralph}, {Crawford}, {Sasaki}, {Haberl}, {Maggi}, {Tothill}, \& {Fukui}}]{2019Ap&SS.364..204A}
{Alsaberi}, R.~Z.~E., {Barnes}, L.~A., {Filipovi{\'c}}, M.~D., {et~al.} 2019, Ap\&SS, 364, 204

\bibitem[{{Alsaberi} {et~al.}(2024){Alsaberi}, {Filipovi{\'c}}, {Dai}, {Sano}, {Kothes}, {Payne}, {Bozzetto}, {Brose}, {Collischon}, {Crawford}, {Haberl}, {Hill}, {Kavanagh}, {Knies}, {Leahy}, {Macgregor}, {Maggi}, {Maitra}, {Manojlovi{\'c}}, {Mart{\'\i}n}, {Matthew}, {Ralph}, {Rowell}, {Ruiter}, {Sasaki}, {Seitenzahl}, {Tokuda}, {Tothill}, {Uro{\v{s}}evi{\'c}}, {van Loon}, {Velovi{\'c}}, \& {Vogt}}]{2024MNRAS.527.1444A}
{Alsaberi}, R. Z.~E., {Filipovi{\'c}}, M.~D., {Dai}, S., {et~al.} 2024, \mnras, 527, 1444

\bibitem[{Anderson {et~al.}(2017)Anderson, Wang, Bihr, Rugel, Beuther, Bigiel, Churchwell, Glover, Goodman, Henning, Heyer, Klessen, Linz, Longmore, Menten, Ott, Roy, Soler, Stil, \& Urquhart}]{Anderson2017}
Anderson, L.~D., Wang, Y., Bihr, S., {et~al.} 2017, A\&A, 605, A58

\bibitem[{{Anderson} {et~al.}(2024){Anderson}, {Camilo}, {Faerber}, {Bietenholz}, {Bordiu}, {Bufano}, {Chibueze}, {Cotton}, {Ingallinera}, {Loru}, {Rigby}, {Riggi}, {Thompson}, {Trigilio}, {Umana}, \& {Williams}}]{2024arXiv240916607A}
{Anderson}, L.~D., {Camilo}, F., {Faerber}, T., {et~al.} 2024, arXiv e-prints, arXiv:2409.16607

\bibitem[{{Atwood} {et~al.}(2009){Atwood}, {Abdo}, {Ackermann}, {Althouse}, {Anderson}, {Axelsson}, {Baldini}, {Ballet}, {Band}, {Barbiellini}, {Bartelt}, {Bastieri}, {Baughman}, {Bechtol}, {B{\'e}d{\'e}r{\`e}de}, {Bellardi}, {Bellazzini}, {Berenji}, {Bignami}, {Bisello}, {Bissaldi}, {Blandford}, {Bloom}, {Bogart}, {Bonamente}, {Bonnell}, {Borgland}, {Bouvier}, {Bregeon}, {Brez}, {Brigida}, {Bruel}, {Burnett}, {Busetto}, {Caliandro}, {Cameron}, {Caraveo}, {Carius}, {Carlson}, {Casandjian}, {Cavazzuti}, {Ceccanti}, {Cecchi}, {Charles}, {Chekhtman}, {Cheung}, {Chiang}, {Chipaux}, {Cillis}, {Ciprini}, {Claus}, {Cohen-Tanugi}, {Condamoor}, {Conrad}, {Corbet}, {Corucci}, {Costamante}, {Cutini}, {Davis}, {Decotigny}, {DeKlotz}, {Dermer}, {de Angelis}, {Digel}, {do Couto e Silva}, {Drell}, {Dubois}, {Dumora}, {Edmonds}, {Fabiani}, {Farnier}, {Favuzzi}, {Flath}, {Fleury}, {Focke}, {Funk}, {Fusco}, {Gargano}, {Gasparrini}, {Gehrels}, {Gentit}, {Germani}, {Giebels}, {Giglietto}, {Giommi}, {Giordano}, {Glanzman},
  {Godfrey}, {Grenier}, {Grondin}, {Grove}, {Guillemot}, {Guiriec}, {Haller}, {Harding}, {Hart}, {Hays}, {Healey}, {Hirayama}, {Hjalmarsdotter}, {Horn}, {Hughes}, {J{\'o}hannesson}, {Johansson}, {Johnson}, {Johnson}, {Johnson}, {Johnson}, {Kamae}, {Katagiri}, {Kataoka}, {Kavelaars}, {Kawai}, {Kelly}, {Kerr}, {Klamra}, {Kn{\"o}dlseder}, {Kocian}, {Komin}, {Kuehn}, {Kuss}, {Landriu}, {Latronico}, {Lee}, {Lee}, {Lemoine-Goumard}, {Lionetto}, {Longo}, {Loparco}, {Lott}, {Lovellette}, {Lubrano}, {Madejski}, {Makeev}, {Marangelli}, {Massai}, {Mazziotta}, {McEnery}, {Menon}, {Meurer}, {Michelson}, {Minuti}, {Mirizzi}, {Mitthumsiri}, {Mizuno}, {Moiseev}, {Monte}, {Monzani}, {Moretti}, {Morselli}, {Moskalenko}, {Murgia}, {Nakamori}, {Nishino}, {Nolan}, {Norris}, {Nuss}, {Ohno}, {Ohsugi}, {Omodei}, {Orlando}, {Ormes}, {Paccagnella}, {Paneque}, {Panetta}, {Parent}, {Pearce}, {Pepe}, {Perazzo}, {Pesce-Rollins}, {Picozza}, {Pieri}, {Pinchera}, {Piron}, {Porter}, {Poupard}, {Rain{\`o}}, {Rando}, {Rapposelli}, {Razzano},
  {Reimer}, {Reimer}, {Reposeur}, {Reyes}, {Ritz}, {Rochester}, {Rodriguez}, {Romani}, {Roth}, {Russell}, {Ryde}, {Sabatini}, {Sadrozinski}, {Sanchez}, {Sander}, {Sapozhnikov}, {Parkinson}, {Scargle}, {Schalk}, {Scolieri}, {Sgr{\`o}}, {Share}, {Shaw}, {Shimokawabe}, {Shrader}, {Sierpowska-Bartosik}, {Siskind}, {Smith}, {Smith}, {Spandre}, {Spinelli}, {Starck}, {Stephens}, {Strickman}, {Strong}, {Suson}, {Tajima}, {Takahashi}, {Takahashi}, {Tanaka}, {Tenze}, {Tether}, {Thayer}, {Thayer}, {Thompson}, {Tibaldo}, {Tibolla}, {Torres}, {Tosti}, {Tramacere}, {Turri}, {Usher}, {Vilchez}, {Vitale}, {Wang}, {Watters}, {Winer}, {Wood}, {Ylinen}, \& {Ziegler}}]{2009ApJ...697.1071A}
{Atwood}, W.~B., {Abdo}, A.~A., {Ackermann}, M., {et~al.} 2009, ApJ, 697, 1071

\bibitem[{{Axford} {et~al.}(1977){Axford}, {Leer}, \& {Skadron}}]{Axfordetal1977}
{Axford}, W.~I., {Leer}, E., \& {Skadron}, G. 1977, in International Cosmic Ray Conference, Vol.~11, International Cosmic Ray Conference, 132

\bibitem[{{Ball} {et~al.}(2023){Ball}, {Kothes}, {Rosolowsky}, {West}, {Becker}, {Filipovi{\'c}}, {Gaensler}, {Hopkins}, {Koribalski}, {Landecker}, {Leahy}, {Marvil}, {Sun}, {Bufano}, {Carretti}, {Ingallinera}, {Van Eck}, \& {Willis}}]{2023MNRAS.524.1396B}
{Ball}, B.~D., {Kothes}, R., {Rosolowsky}, E., {et~al.} 2023, MNRAS, 524, 1396

\bibitem[{Bartunov {et~al.}(1994)Bartunov, Tsvetkov, \& Filimonova}]{Bartunov_1994}
Bartunov, O.~S., Tsvetkov, D.~Y., \& Filimonova, I.~V. 1994, Publications of the Astronomical Society of the Pacific, 106, 1276

\bibitem[{{Bell}(1978)}]{Bell1978a}
{Bell}, A.~R. 1978, MNRAS, 182, 147

\bibitem[{{Blaauw}(1993)}]{Bla93}
{Blaauw}, A. 1993, in Astronomical Society of the Pacific Conference Series, Vol.~35, Massive Stars: Their Lives in the Interstellar Medium, ed. J.~P. {Cassinelli} \& E.~B. {Churchwell}, 207

\bibitem[{{Bland-Hawthorn} \& {Gerhard}(2016)}]{2016ARA&A..54..529B}
{Bland-Hawthorn}, J., \& {Gerhard}, O. 2016, ARA\&A, 54, 529

\bibitem[{{Blandford} \& {Ostriker}(1978)}]{BlandfordOstriker1978}
{Blandford}, R.~D., \& {Ostriker}, J.~P. 1978, ApJl, 221, L29

\bibitem[{{Booth} {et~al.}(2022){Booth}, {Kothes}, {Landecker}, {Brown}, {Gray}, {Foster}, \& {Greisen}}]{2022ApJ...941...17B}
{Booth}, R.~A., {Kothes}, R., {Landecker}, T., {et~al.} 2022, ApJ, 941, 17

\bibitem[{{Bordiu} {et~al.}(2024){Bordiu}, {Filipovic}, {Umana}, {Cotton}, {Buemi}, {Bufano}, {Camilo}, {Cavallaro}, {Cerrigone}, {Dai}, {Hopkins}, {Ingallinera}, {Jarrett}, {Koribalski}, {Lazarevic}, {Leto}, {Loru}, {Lundqvist}, {Mackey}, {Norris}, {Payne}, {Rowell}, {Riggi}, {Rizzo}, {Ruggeri}, {Shabala}, {Smeaton}, {Trigilio}, \& {Velovic}}]{2024arXiv240807727B}
{Bordiu}, C., {Filipovic}, M.~D., {Umana}, G., {et~al.} 2024, arXiv e-prints, arXiv:2408.07727

\bibitem[{{Bozzetto} {et~al.}(2014){Bozzetto}, {Filipovi{\'c}}, {Uro{\v{s}}evi{\'c}}, {Kothes}, \& {Crawford}}]{2014MNRAS.440.3220B}
{Bozzetto}, L.~M., {Filipovi{\'c}}, M.~D., {Uro{\v{s}}evi{\'c}}, D., {Kothes}, R., \& {Crawford}, E.~J. 2014, MNRAS, 440, 3220

\bibitem[{{Bozzetto} {et~al.}(2017){Bozzetto}, {Filipovi{\'c}}, {Vukoti{\'c}}, {Pavlovi{\'c}}, {Uro{\v{s}}evi{\'c}}, {Kavanagh}, {Arbutina}, {Maggi}, {Sasaki}, {Haberl}, {Crawford}, {Roper}, {Grieve}, \& {Points}}]{2017ApJS..230....2B}
{Bozzetto}, L.~M., {Filipovi{\'c}}, M.~D., {Vukoti{\'c}}, B., {et~al.} 2017, Astrophys. J. Suppl., 230, 2

\bibitem[{{Bozzetto} {et~al.}(2023){Bozzetto}, {Filipovi{\'c}}, {Sano}, {Alsaberi}, {Barnes}, {Boji{\v{c}}i{\'c}}, {Brose}, {Chomiuk}, {Crawford}, {Dai}, {Ghavam}, {Haberl}, {Hill}, {Hopkins}, {Ingallinera}, {Jarrett}, {Kavanagh}, {Koribalski}, {Kothes}, {Leahy}, {Lenc}, {Leonidaki}, {Maggi}, {Maitra}, {Matthew}, {Payne}, {Pennock}, {Points}, {Reid}, {Riggi}, {Rowell}, {Sasaki}, {Safi-Harb}, {van Loon}, {Tothill}, {Uro{\v{s}}evi{\'c}}, \& {Zangrandi}}]{2023MNRAS.518.2574B}
{Bozzetto}, L.~M., {Filipovi{\'c}}, M.~D., {Sano}, H., {et~al.} 2023, MNRAS, 518, 2574

\bibitem[{{Brose} {et~al.}(2020){Brose}, {Pohl}, {Sushch}, {Petruk}, \& {Kuzyo}}]{2020A&A...634A..59B}
{Brose}, R., {Pohl}, M., {Sushch}, I., {Petruk}, O., \& {Kuzyo}, T. 2020, \aap, 634, A59

\bibitem[{{Burger-Scheidlin} {et~al.}(2024){Burger-Scheidlin}, {Brose}, {Mackey}, {Filipovi{\'c}}, {Goswami}, {Guillen}, {de O{\~n}a Wilhelmi}, \& {Sushch}}]{2024A&A...684A.150B}
{Burger-Scheidlin}, C., {Brose}, R., {Mackey}, J., {et~al.} 2024, A\&A, 684, A150

\bibitem[{{Cendes} {et~al.}(2018){Cendes}, {Gaensler}, {Ng}, {Zanardo}, {Staveley-Smith}, \& {Tzioumis}}]{2018ApJ...867...65C}
{Cendes}, Y., {Gaensler}, B.~M., {Ng}, C.~Y., {et~al.} 2018, ApJ, 867, 65

\bibitem[{{Chen} {et~al.}(2013){Chen}, {Zhou}, \& {Chu}}]{2013ApJ...769L..16C}
{Chen}, Y., {Zhou}, P., \& {Chu}, Y.-H. 2013, \apjl, 769, L16

\bibitem[{Churchwell {et~al.}(2009)Churchwell, Babler, Meade, Whitney, Benjamin, Indebetouw, Cyganowski, Robitaille, Povich, Watson, \& Bracker}]{Churchwell2009}
Churchwell, E., Babler, B.~L., Meade, M.~R., {et~al.} 2009, Publications of the Astronomical Society of the Pacific, 121, 213

\bibitem[{{Cotton} {et~al.}(2024){Cotton}, {Filipovi{\'c}}, {Camilo}, {Indebetouw}, {Alsaberi}, {Anih}, {Baker}, {Bastian}, {Boji{\v{c}}i{\'c}}, {Carli}, {Cavallaro}, {Crawford}, {Dai}, {Haberl}, {Levin}, {Luken}, {Pennock}, {Rajabpour}, {Stappers}, {van Loon}, {Zijlstra}, {Buchner}, {Geyer}, {Goedhart}, \& {Serylak}}]{2024MNRAS.529.2443C}
{Cotton}, W.~D., {Filipovi{\'c}}, M.~D., {Camilo}, F., {et~al.} 2024, MNRAS, 529, 2443

\bibitem[{{Das} {et~al.}(2022){Das}, {Brose}, {Meyer}, {Pohl}, {Sushch}, \& {Plotko}}]{2022A&A...661A.128D}
{Das}, S., {Brose}, R., {Meyer}, D. M.~A., {et~al.} 2022, \aap, 661, A128

\bibitem[{{Das} {et~al.}(2024){Das}, {Brose}, {Pohl}, {Meyer}, \& {Sushch}}]{2024A&A...689A...9D}
{Das}, S., {Brose}, R., {Pohl}, M., {Meyer}, D. M.~A., \& {Sushch}, I. 2024, \aap, 689, A9

\bibitem[{{Dokara} {et~al.}(2021){Dokara}, {Brunthaler}, {Menten}, {Dzib}, {Reich}, {Cotton}, {Anderson}, {Chen}, {Gong}, {Medina}, {Ortiz-Le{\'o}n}, {Rugel}, {Urquhart}, {Wyrowski}, {Yang}, {Beuther}, {Billington}, {Csengeri}, {Carrasco-Gonz{\'a}lez}, \& {Roy}}]{2021A&A...651A..86D}
{Dokara}, R., {Brunthaler}, A., {Menten}, K.~M., {et~al.} 2021, A\&A, 651, A86

\bibitem[{{Drimmel}(2000)}]{2000A&A...358L..13D}
{Drimmel}, R. 2000, A\&A, 358, L13

\bibitem[{Efremov(2011)}]{Efremov2011}
Efremov, Y.~N. 2011, Astronomy reports, 55, 108

\bibitem[{{Enokiya} {et~al.}(2023){Enokiya}, {Sano}, {Filipovi{\'c}}, {Alsaberi}, {Inoue}, \& {Oka}}]{2023PASJ...75..970E}
{Enokiya}, R., {Sano}, H., {Filipovi{\'c}}, M.~D., {et~al.} 2023, \pasj, 75, 970

\bibitem[{{Ferrand} \& {Safi-Harb}(2012)}]{Ferrand2012}
{Ferrand}, G., \& {Safi-Harb}, S. 2012, Advances in Space Research, 49, 1313

\bibitem[{Fesen {et~al.}(2023)Fesen, Schaefer, \& Patchick}]{Fesen_2023}
Fesen, R.~A., Schaefer, B.~E., \& Patchick, D. 2023, The Astrophysical Journal Letters, 945, L4

\bibitem[{{Filipovic} {et~al.}(2013){Filipovic}, {Horner}, {Crawford}, {Tothill}, \& {White}}]{2013SerAJ.187...43F}
{Filipovic}, M.~D., {Horner}, J., {Crawford}, E.~J., {Tothill}, N.~F.~H., \& {White}, G.~L. 2013, Serbian Astronomical Journal, 187, 43

\bibitem[{Filipovi{\'c} \& Tothill(2021)}]{book2}
Filipovi{\'c}, M.~D., \& Tothill, N. F.~H., eds. 2021, Multimessenger Astronomy in Practice, 2514-3433 (IOP Publishing), doi:10.1088/2514-3433/ac2256

\bibitem[{{Filipovi{\'c}} {et~al.}(2021{\natexlab{a}}){Filipovi{\'c}}, {Ili{\'c}}, {Jarrett}, {Payne}, {Uro{\v{s}}evi{\'c}}, {Tothill}, {Kavanagh}, {Longo}, {Crawford}, \& {Collier}}]{10542}
{Filipovi{\'c}}, M.~D., {Ili{\'c}}, M., {Jarrett}, T., {et~al.} 2021{\natexlab{a}}, European Journal of Science and Theology, 17, 11

\bibitem[{{Filipovi{\'c}} {et~al.}(2021{\natexlab{b}}){Filipovi{\'c}}, {Payne}, {Jarrett}, {Tothill}, {Uro{\v{s}}evi{\'c}}, {Kavanagh}, {Longo}, {Crawford}, {Collier}, \& {Ili{\'c}}}]{10541}
{Filipovi{\'c}}, M.~D., {Payne}, J.~L., {Jarrett}, T., {et~al.} 2021{\natexlab{b}}, European Journal of Science and Theology, 17, 147

\bibitem[{{Filipovi{\'c}} {et~al.}(2022{\natexlab{a}}){Filipovi{\'c}}, {Payne}, {Jarret}, {Tothill}, {Crawford}, {Uro{\v{s}}evi{\'c}}, {Longo}, {Collier}, {Kavanagh}, {Matthew}, \& {Ili{\'c}}}]{10543}
{Filipovi{\'c}}, M.~D., {Payne}, J.~L., {Jarret}, T., {et~al.} 2022{\natexlab{a}}, European Journal of Science and Theology, 18, 51

\bibitem[{{Filipovi{\'c}} {et~al.}(2022{\natexlab{b}}){Filipovi{\'c}}, {Payne}, {Alsaberi}, {Norris}, {Macgregor}, {Rudnick}, {Koribalski}, {Leahy}, {Ducci}, {Kothes}, {Andernach}, {Barnes}, {Boji{\v{c}}i{\'c}}, {Bozzetto}, {Brose}, {Collier}, {Crawford}, {Crocker}, {Dai}, {Galvin}, {Haberl}, {Heber}, {Hill}, {Hopkins}, {Hurley-Walker}, {Ingallinera}, {Jarrett}, {Kavanagh}, {Lenc}, {Luken}, {Mackey}, {Manojlovi{\'c}}, {Maggi}, {Maitra}, {Pennock}, {Points}, {Riggi}, {Rowell}, {Safi-Harb}, {Sano}, {Sasaki}, {Shabala}, {Stevens}, {van Loon}, {Tothill}, {Umana}, {Uro{\v{s}}evi{\'c}}, {Velovi{\'c}}, {Vernstrom}, {West}, \& {Wan}}]{2022MNRAS.512..265F}
{Filipovi{\'c}}, M.~D., {Payne}, J.~L., {Alsaberi}, R.~Z.~E., {et~al.} 2022{\natexlab{b}}, MNRAS, 512, 265

\bibitem[{{Filipovi{\'c}} {et~al.}(2023){Filipovi{\'c}}, {Dai}, {Arbutina}, {Hurley-Walker}, {Brose}, {Becker}, {Sano}, {Uro{\v{s}}evi{\'c}}, {Jarrett}, {Hopkins}, {Alsaberi}, {Alsulami}, {Bordiu}, {Ball}, {Bufano}, {Burger-Scheidlin}, {Crawford}, {English}, {Haberl}, {Ingallinera}, {Kapinska}, {Kavanagh}, {Koribalski}, {Kothes}, {Lazarevi{\'c}}, {Mackey}, {Rowell}, {Leahy}, {Loru}, {Macgregor}, {Nicastro}, {Norris}, {Riggi}, {Sasaki}, {Stupar}, {Trigilio}, {Umana}, {Vernstrom}, \& {Vukoti{\'c}}}]{2023AJ....166..149F}
{Filipovi{\'c}}, M.~D., {Dai}, S., {Arbutina}, B., {et~al.} 2023, AJ, 166, 149

\bibitem[{{Filipovi{\'c}} {et~al.}(2024){Filipovi{\'c}}, {Lazarevi{\'c}}, {Araya}, {Hurley-Walker}, {Kothes}, {Sano}, {Rowell}, {Martin}, {Fukui}, {Alsaberi}, {Arbutina}, {Ball}, {Bordiu}, {Brose}, {Bufano}, {Burger-Scheidlin}, {Anne Collins}, {Crawford}, {Dai}, {William Duchesne}, {Fuller}, {Hopkins}, {Ingallinera}, {Inoue}, {Jarrett}, {Silvia Koribalski}, {Leahy}, {Luken}, {Mackey}, {Macgregor}, {Norris}, {Payne}, {Riggi}, {Riseley}, {Sasaki}, {Smeaton}, {Sushch}, {Stupar}, {Umana}, {Uro{\v{s}}evi{\'c}}, {Velovi{\'c}}, {Vernstrom}, {Vukoti{\'c}}, \& {West}}]{2024PASA...41..112F}
{Filipovi{\'c}}, M.~D., {Lazarevi{\'c}}, S., {Araya}, M., {et~al.} 2024, \pasa, 41, e112

\bibitem[{{Finke} \& {Dermer}(2012)}]{FinkeDermer2012}
{Finke}, J.~D., \& {Dermer}, C.~D. 2012, ApJ, 751, 65

\bibitem[{{Foley} {et~al.}(2013){Foley}, {Challis}, {Chornock}, {Ganeshalingam}, {Li}, {Marion}, {Morrell}, {Pignata}, {Stritzinger}, {Silverman}, {Wang}, {Anderson}, {Filippenko}, {Freedman}, {Hamuy}, {Jha}, {Kirshner}, {McCully}, {Persson}, {Phillips}, {Reichart}, \& {Soderberg}}]{2013ApJ...767...57F}
{Foley}, R.~J., {Challis}, P.~J., {Chornock}, R., {et~al.} 2013, ApJ, 767, 57

\bibitem[{{Foster} {et~al.}(2013){Foster}, {Cooper}, {Reich}, {Kothes}, \& {West}}]{2013A&A...549A.107F}
{Foster}, T.~J., {Cooper}, B., {Reich}, W., {Kothes}, R., \& {West}, J. 2013, A\&A, 549, A107

\bibitem[{{Gaensler} {et~al.}(2010){Gaensler}, {Landecker}, {Taylor}, \& {POSSUM Collaboration}}]{2010AAS...21547013G}
{Gaensler}, B.~M., {Landecker}, T.~L., {Taylor}, A.~R., \& {POSSUM Collaboration}. 2010, in American Astronomical Society Meeting Abstracts, Vol. 215, American Astronomical Society Meeting Abstracts \#215, 470.13

\bibitem[{{Gaia Collaboration} {et~al.}(2016){Gaia Collaboration}, {Prusti}, {de Bruijne}, {Brown}, {Vallenari}, {Babusiaux}, {Bailer-Jones}, {Bastian}, {Biermann}, {Evans}, {Eyer}, {Jansen}, {Jordi}, {Klioner}, {Lammers}, {Lindegren}, {Luri}, {Mignard}, {Milligan}, {Panem}, {Poinsignon}, {Pourbaix}, {Randich}, {Sarri}, {Sartoretti}, {Siddiqui}, {Soubiran}, {Valette}, {van Leeuwen}, {Walton}, {Aerts}, {Arenou}, {Cropper}, {Drimmel}, {H{\o}g}, {Katz}, {Lattanzi}, {O'Mullane}, {Grebel}, {Holland}, {Huc}, {Passot}, {Bramante}, {Cacciari}, {Casta{\~n}eda}, {Chaoul}, {Cheek}, {De Angeli}, {Fabricius}, {Guerra}, {Hern{\'a}ndez}, {Jean-Antoine-Piccolo}, {Masana}, {Messineo}, {Mowlavi}, {Nienartowicz}, {Ord{\'o}{\~n}ez-Blanco}, {Panuzzo}, {Portell}, {Richards}, {Riello}, {Seabroke}, {Tanga}, {Th{\'e}venin}, {Torra}, {Els}, {Gracia-Abril}, {Comoretto}, {Garcia-Reinaldos}, {Lock}, {Mercier}, {Altmann}, {Andrae}, {Astraatmadja}, {Bellas-Velidis}, {Benson}, {Berthier}, {Blomme}, {Busso}, {Carry}, {Cellino}, {Clementini},
  {Cowell}, {Creevey}, {Cuypers}, {Davidson}, {De Ridder}, {de Torres}, {Delchambre}, {Dell'Oro}, {Ducourant}, {Fr{\'e}mat}, {Garc{\'\i}a-Torres}, {Gosset}, {Halbwachs}, {Hambly}, {Harrison}, {Hauser}, {Hestroffer}, {Hodgkin}, {Huckle}, {Hutton}, {Jasniewicz}, {Jordan}, {Kontizas}, {Korn}, {Lanzafame}, {Manteiga}, {Moitinho}, {Muinonen}, {Osinde}, {Pancino}, {Pauwels}, {Petit}, {Recio-Blanco}, {Robin}, {Sarro}, {Siopis}, {Smith}, {Smith}, {Sozzetti}, {Thuillot}, {van Reeven}, {Viala}, {Abbas}, {Abreu Aramburu}, {Accart}, {Aguado}, {Allan}, {Allasia}, {Altavilla}, {{\'A}lvarez}, {Alves}, {Anderson}, {Andrei}, {Anglada Varela}, {Antiche}, {Antoja}, {Ant{\'o}n}, {Arcay}, {Atzei}, {Ayache}, {Bach}, {Baker}, {Balaguer-N{\'u}{\~n}ez}, {Barache}, {Barata}, {Barbier}, {Barblan}, {Baroni}, {Barrado y Navascu{\'e}s}, {Barros}, {Barstow}, {Becciani}, {Bellazzini}, {Bellei}, {Bello Garc{\'\i}a}, {Belokurov}, {Bendjoya}, {Berihuete}, {Bianchi}, {Bienaym{\'e}}, {Billebaud}, {Blagorodnova}, {Blanco-Cuaresma}, {Boch},
  {Bombrun}, {Borrachero}, {Bouquillon}, {Bourda}, {Bouy}, {Bragaglia}, {Breddels}, {Brouillet}, {Br{\"u}semeister}, {Bucciarelli}, {Budnik}, {Burgess}, {Burgon}, {Burlacu}, {Busonero}, {Buzzi}, {Caffau}, {Cambras}, {Campbell}, {Cancelliere}, {Cantat-Gaudin}, {Carlucci}, {Carrasco}, {Castellani}, {Charlot}, {Charnas}, {Charvet}, {Chassat}, {Chiavassa}, {Clotet}, {Cocozza}, {Collins}, {Collins}, {Costigan}, {Crifo}, {Cross}, {Crosta}, {Crowley}, {Dafonte}, {Damerdji}, {Dapergolas}, {David}, {David}, {De Cat}, {de Felice}, {de Laverny}, {De Luise}, {De March}, {de Martino}, {de Souza}, {Debosscher}, {del Pozo}, {Delbo}, {Delgado}, {Delgado}, {di Marco}, {Di Matteo}, {Diakite}, {Distefano}, {Dolding}, {Dos Anjos}, {Drazinos}, {Dur{\'a}n}, {Dzigan}, {Ecale}, {Edvardsson}, {Enke}, {Erdmann}, {Escolar}, {Espina}, {Evans}, {Eynard Bontemps}, {Fabre}, {Fabrizio}, {Faigler}, {Falc{\~a}o}, {Farr{\`a}s Casas}, {Faye}, {Federici}, {Fedorets}, {Fern{\'a}ndez-Hern{\'a}ndez}, {Fernique}, {Fienga}, {Figueras}, {Filippi},
  {Findeisen}, {Fonti}, {Fouesneau}, {Fraile}, {Fraser}, {Fuchs}, {Furnell}, {Gai}, {Galleti}, {Galluccio}, {Garabato}, {Garc{\'\i}a-Sedano}, {Gar{\'e}}, {Garofalo}, {Garralda}, {Gavras}, {Gerssen}, {Geyer}, {Gilmore}, {Girona}, {Giuffrida}, {Gomes}, {Gonz{\'a}lez-Marcos}, {Gonz{\'a}lez-N{\'u}{\~n}ez}, {Gonz{\'a}lez-Vidal}, {Granvik}, {Guerrier}, {Guillout}, {Guiraud}, {G{\'u}rpide}, {Guti{\'e}rrez-S{\'a}nchez}, {Guy}, {Haigron}, {Hatzidimitriou}, {Haywood}, {Heiter}, {Helmi}, {Hobbs}, {Hofmann}, {Holl}, {Holland}, {Hunt}, {Hypki}, {Icardi}, {Irwin}, {Jevardat de Fombelle}, {Jofr{\'e}}, {Jonker}, {Jorissen}, {Julbe}, {Karampelas}, {Kochoska}, {Kohley}, {Kolenberg}, {Kontizas}, {Koposov}, {Kordopatis}, {Koubsky}, {Kowalczyk}, {Krone-Martins}, {Kudryashova}, {Kull}, {Bachchan}, {Lacoste-Seris}, {Lanza}, {Lavigne}, {Le Poncin-Lafitte}, {Lebreton}, {Lebzelter}, {Leccia}, {Leclerc}, {Lecoeur-Taibi}, {Lemaitre}, {Lenhardt}, {Leroux}, {Liao}, {Licata}, {Lindstr{\o}m}, {Lister}, {Livanou}, {Lobel}, {L{\"o}ffler},
  {L{\'o}pez}, {Lopez-Lozano}, {Lorenz}, {Loureiro}, {MacDonald}, {Magalh{\~a}es Fernandes}, {Managau}, {Mann}, {Mantelet}, {Marchal}, {Marchant}, {Marconi}, {Marie}, {Marinoni}, {Marrese}, {Marschalk{\'o}}, {Marshall}, {Mart{\'\i}n-Fleitas}, {Martino}, {Mary}, {Matijevi{\v{c}}}, {Mazeh}, {McMillan}, {Messina}, {Mestre}, {Michalik}, {Millar}, {Miranda}, {Molina}, {Molinaro}, {Molinaro}, {Moln{\'a}r}, {Moniez}, {Montegriffo}, {Monteiro}, {Mor}, {Mora}, {Morbidelli}, {Morel}, {Morgenthaler}, {Morley}, {Morris}, {Mulone}, {Muraveva}, {Musella}, {Narbonne}, {Nelemans}, {Nicastro}, {Noval}, {Ord{\'e}novic}, {Ordieres-Mer{\'e}}, {Osborne}, {Pagani}, {Pagano}, {Pailler}, {Palacin}, {Palaversa}, {Parsons}, {Paulsen}, {Pecoraro}, {Pedrosa}, {Pentik{\"a}inen}, {Pereira}, {Pichon}, {Piersimoni}, {Pineau}, {Plachy}, {Plum}, {Poujoulet}, {Pr{\v{s}}a}, {Pulone}, {Ragaini}, {Rago}, {Rambaux}, {Ramos-Lerate}, {Ranalli}, {Rauw}, {Read}, {Regibo}, {Renk}, {Reyl{\'e}}, {Ribeiro}, {Rimoldini}, {Ripepi}, {Riva}, {Rixon},
  {Roelens}, {Romero-G{\'o}mez}, {Rowell}, {Royer}, {Rudolph}, {Ruiz-Dern}, {Sadowski}, {Sagrist{\`a} Sell{\'e}s}, {Sahlmann}, {Salgado}, {Salguero}, {Sarasso}, {Savietto}, {Schnorhk}, {Schultheis}, {Sciacca}, {Segol}, {Segovia}, {Segransan}, {Serpell}, {Shih}, {Smareglia}, {Smart}, {Smith}, {Solano}, {Solitro}, {Sordo}, {Soria Nieto}, {Souchay}, {Spagna}, {Spoto}, {Stampa}, {Steele}, {Steidelm{\"u}ller}, {Stephenson}, {Stoev}, {Suess}, {S{\"u}veges}, {Surdej}, {Szabados}, {Szegedi-Elek}, {Tapiador}, {Taris}, {Tauran}, {Taylor}, {Teixeira}, {Terrett}, {Tingley}, {Trager}, {Turon}, {Ulla}, {Utrilla}, {Valentini}, {van Elteren}, {Van Hemelryck}, {van Leeuwen}, {Varadi}, {Vecchiato}, {Veljanoski}, {Via}, {Vicente}, {Vogt}, {Voss}, {Votruba}, {Voutsinas}, {Walmsley}, {Weiler}, {Weingrill}, {Werner}, {Wevers}, {Whitehead}, {Wyrzykowski}, {Yoldas}, {{\v{Z}}erjal}, {Zucker}, {Zurbach}, {Zwitter}, {Alecu}, {Allen}, {Allende Prieto}, {Amorim}, {Anglada-Escud{\'e}}, {Arsenijevic}, {Azaz}, {Balm}, {Beck}, {Bernstein},
  {Bigot}, {Bijaoui}, {Blasco}, {Bonfigli}, {Bono}, {Boudreault}, {Bressan}, {Brown}, {Brunet}, {Bunclark}, {Buonanno}, {Butkevich}, {Carret}, {Carrion}, {Chemin}, {Ch{\'e}reau}, {Corcione}, {Darmigny}, {de Boer}, {de Teodoro}, {de Zeeuw}, {Delle Luche}, {Domingues}, {Dubath}, {Fodor}, {Fr{\'e}zouls}, {Fries}, {Fustes}, {Fyfe}, {Gallardo}, {Gallegos}, {Gardiol}, {Gebran}, {Gomboc}, {G{\'o}mez}, {Grux}, {Gueguen}, {Heyrovsky}, {Hoar}, {Iannicola}, {Isasi Parache}, {Janotto}, {Joliet}, {Jonckheere}, {Keil}, {Kim}, {Klagyivik}, {Klar}, {Knude}, {Kochukhov}, {Kolka}, {Kos}, {Kutka}, {Lainey}, {LeBouquin}, {Liu}, {Loreggia}, {Makarov}, {Marseille}, {Martayan}, {Martinez-Rubi}, {Massart}, {Meynadier}, {Mignot}, {Munari}, {Nguyen}, {Nordlander}, {Ocvirk}, {O'Flaherty}, {Olias Sanz}, {Ortiz}, {Osorio}, {Oszkiewicz}, {Ouzounis}, {Palmer}, {Park}, {Pasquato}, {Peltzer}, {Peralta}, {P{\'e}turaud}, {Pieniluoma}, {Pigozzi}, {Poels}, {Prat}, {Prod'homme}, {Raison}, {Rebordao}, {Risquez}, {Rocca-Volmerange}, {Rosen},
  {Ruiz-Fuertes}, {Russo}, {Sembay}, {Serraller Vizcaino}, {Short}, {Siebert}, {Silva}, {Sinachopoulos}, {Slezak}, {Soffel}, {Sosnowska}, {Strai{\v{z}}ys}, {ter Linden}, {Terrell}, {Theil}, {Tiede}, {Troisi}, {Tsalmantza}, {Tur}, {Vaccari}, {Vachier}, {Valles}, {Van Hamme}, {Veltz}, {Virtanen}, {Wallut}, {Wichmann}, {Wilkinson}, {Ziaeepour}, \& {Zschocke}}]{GAIA2016}
{Gaia Collaboration}, {Prusti}, T., {de Bruijne}, J.~H.~J., {et~al.} 2016, A\&A, 595, A1

\bibitem[{{Gaia Collaboration} {et~al.}(2023){Gaia Collaboration}, {Vallenari}, {Brown}, {Prusti}, {de Bruijne}, {Arenou}, {Babusiaux}, {Biermann}, {Creevey}, {Ducourant}, {Evans}, {Eyer}, {Guerra}, {Hutton}, {Jordi}, {Klioner}, {Lammers}, {Lindegren}, {Luri}, {Mignard}, {Panem}, {Pourbaix}, {Randich}, {Sartoretti}, {Soubiran}, {Tanga}, {Walton}, {Bailer-Jones}, {Bastian}, {Drimmel}, {Jansen}, {Katz}, {Lattanzi}, {van Leeuwen}, {Bakker}, {Cacciari}, {Casta{\~n}eda}, {De Angeli}, {Fabricius}, {Fouesneau}, {Fr{\'e}mat}, {Galluccio}, {Guerrier}, {Heiter}, {Masana}, {Messineo}, {Mowlavi}, {Nicolas}, {Nienartowicz}, {Pailler}, {Panuzzo}, {Riclet}, {Roux}, {Seabroke}, {Sordo}, {Th{\'e}venin}, {Gracia-Abril}, {Portell}, {Teyssier}, {Altmann}, {Andrae}, {Audard}, {Bellas-Velidis}, {Benson}, {Berthier}, {Blomme}, {Burgess}, {Busonero}, {Busso}, {C{\'a}novas}, {Carry}, {Cellino}, {Cheek}, {Clementini}, {Damerdji}, {Davidson}, {de Teodoro}, {Nu{\~n}ez Campos}, {Delchambre}, {Dell'Oro}, {Esquej},
  {Fern{\'a}ndez-Hern{\'a}ndez}, {Fraile}, {Garabato}, {Garc{\'\i}a-Lario}, {Gosset}, {Haigron}, {Halbwachs}, {Hambly}, {Harrison}, {Hern{\'a}ndez}, {Hestroffer}, {Hodgkin}, {Holl}, {Jan{\ss}en}, {Jevardat de Fombelle}, {Jordan}, {Krone-Martins}, {Lanzafame}, {L{\"o}ffler}, {Marchal}, {Marrese}, {Moitinho}, {Muinonen}, {Osborne}, {Pancino}, {Pauwels}, {Recio-Blanco}, {Reyl{\'e}}, {Riello}, {Rimoldini}, {Roegiers}, {Rybizki}, {Sarro}, {Siopis}, {Smith}, {Sozzetti}, {Utrilla}, {van Leeuwen}, {Abbas}, {{\'A}brah{\'a}m}, {Abreu Aramburu}, {Aerts}, {Aguado}, {Ajaj}, {Aldea-Montero}, {Altavilla}, {{\'A}lvarez}, {Alves}, {Anders}, {Anderson}, {Anglada Varela}, {Antoja}, {Baines}, {Baker}, {Balaguer-N{\'u}{\~n}ez}, {Balbinot}, {Balog}, {Barache}, {Barbato}, {Barros}, {Barstow}, {Bartolom{\'e}}, {Bassilana}, {Bauchet}, {Becciani}, {Bellazzini}, {Berihuete}, {Bernet}, {Bertone}, {Bianchi}, {Binnenfeld}, {Blanco-Cuaresma}, {Blazere}, {Boch}, {Bombrun}, {Bossini}, {Bouquillon}, {Bragaglia}, {Bramante}, {Breedt},
  {Bressan}, {Brouillet}, {Brugaletta}, {Bucciarelli}, {Burlacu}, {Butkevich}, {Buzzi}, {Caffau}, {Cancelliere}, {Cantat-Gaudin}, {Carballo}, {Carlucci}, {Carnerero}, {Carrasco}, {Casamiquela}, {Castellani}, {Castro-Ginard}, {Chaoul}, {Charlot}, {Chemin}, {Chiaramida}, {Chiavassa}, {Chornay}, {Comoretto}, {Contursi}, {Cooper}, {Cornez}, {Cowell}, {Crifo}, {Cropper}, {Crosta}, {Crowley}, {Dafonte}, {Dapergolas}, {David}, {David}, {de Laverny}, {De Luise}, {De March}, {De Ridder}, {de Souza}, {de Torres}, {del Peloso}, {del Pozo}, {Delbo}, {Delgado}, {Delisle}, {Demouchy}, {Dharmawardena}, {Di Matteo}, {Diakite}, {Diener}, {Distefano}, {Dolding}, {Edvardsson}, {Enke}, {Fabre}, {Fabrizio}, {Faigler}, {Fedorets}, {Fernique}, {Fienga}, {Figueras}, {Fournier}, {Fouron}, {Fragkoudi}, {Gai}, {Garcia-Gutierrez}, {Garcia-Reinaldos}, {Garc{\'\i}a-Torres}, {Garofalo}, {Gavel}, {Gavras}, {Gerlach}, {Geyer}, {Giacobbe}, {Gilmore}, {Girona}, {Giuffrida}, {Gomel}, {Gomez}, {Gonz{\'a}lez-N{\'u}{\~n}ez},
  {Gonz{\'a}lez-Santamar{\'\i}a}, {Gonz{\'a}lez-Vidal}, {Granvik}, {Guillout}, {Guiraud}, {Guti{\'e}rrez-S{\'a}nchez}, {Guy}, {Hatzidimitriou}, {Hauser}, {Haywood}, {Helmer}, {Helmi}, {Sarmiento}, {Hidalgo}, {Hilger}, {H{\l}adczuk}, {Hobbs}, {Holland}, {Huckle}, {Jardine}, {Jasniewicz}, {Jean-Antoine Piccolo}, {Jim{\'e}nez-Arranz}, {Jorissen}, {Juaristi Campillo}, {Julbe}, {Karbevska}, {Kervella}, {Khanna}, {Kontizas}, {Kordopatis}, {Korn}, {K{\'o}sp{\'a}l}, {Kostrzewa-Rutkowska}, {Kruszy{\'n}ska}, {Kun}, {Laizeau}, {Lambert}, {Lanza}, {Lasne}, {Le Campion}, {Lebreton}, {Lebzelter}, {Leccia}, {Leclerc}, {Lecoeur-Taibi}, {Liao}, {Licata}, {Lindstr{\o}m}, {Lister}, {Livanou}, {Lobel}, {Lorca}, {Loup}, {Madrero Pardo}, {Magdaleno Romeo}, {Managau}, {Mann}, {Manteiga}, {Marchant}, {Marconi}, {Marcos}, {Marcos Santos}, {Mar{\'\i}n Pina}, {Marinoni}, {Marocco}, {Marshall}, {Martin Polo}, {Mart{\'\i}n-Fleitas}, {Marton}, {Mary}, {Masip}, {Massari}, {Mastrobuono-Battisti}, {Mazeh}, {McMillan}, {Messina}, {Michalik},
  {Millar}, {Mints}, {Molina}, {Molinaro}, {Moln{\'a}r}, {Monari}, {Mongui{\'o}}, {Montegriffo}, {Montero}, {Mor}, {Mora}, {Morbidelli}, {Morel}, {Morris}, {Muraveva}, {Murphy}, {Musella}, {Nagy}, {Noval}, {Oca{\~n}a}, {Ogden}, {Ordenovic}, {Osinde}, {Pagani}, {Pagano}, {Palaversa}, {Palicio}, {Pallas-Quintela}, {Panahi}, {Payne-Wardenaar}, {Pe{\~n}alosa Esteller}, {Penttil{\"a}}, {Pichon}, {Piersimoni}, {Pineau}, {Plachy}, {Plum}, {Poggio}, {Pr{\v{s}}a}, {Pulone}, {Racero}, {Ragaini}, {Rainer}, {Raiteri}, {Rambaux}, {Ramos}, {Ramos-Lerate}, {Re Fiorentin}, {Regibo}, {Richards}, {Rios Diaz}, {Ripepi}, {Riva}, {Rix}, {Rixon}, {Robichon}, {Robin}, {Robin}, {Roelens}, {Rogues}, {Rohrbasser}, {Romero-G{\'o}mez}, {Rowell}, {Royer}, {Ruz Mieres}, {Rybicki}, {Sadowski}, {S{\'a}ez N{\'u}{\~n}ez}, {Sagrist{\`a} Sell{\'e}s}, {Sahlmann}, {Salguero}, {Samaras}, {Sanchez Gimenez}, {Sanna}, {Santove{\~n}a}, {Sarasso}, {Schultheis}, {Sciacca}, {Segol}, {Segovia}, {S{\'e}gransan}, {Semeux}, {Shahaf}, {Siddiqui}, {Siebert},
  {Siltala}, {Silvelo}, {Slezak}, {Slezak}, {Smart}, {Snaith}, {Solano}, {Solitro}, {Souami}, {Souchay}, {Spagna}, {Spina}, {Spoto}, {Steele}, {Steidelm{\"u}ller}, {Stephenson}, {S{\"u}veges}, {Surdej}, {Szabados}, {Szegedi-Elek}, {Taris}, {Taylor}, {Teixeira}, {Tolomei}, {Tonello}, {Torra}, {Torra}, {Torralba Elipe}, {Trabucchi}, {Tsounis}, {Turon}, {Ulla}, {Unger}, {Vaillant}, {van Dillen}, {van Reeven}, {Vanel}, {Vecchiato}, {Viala}, {Vicente}, {Voutsinas}, {Weiler}, {Wevers}, {Wyrzykowski}, {Yoldas}, {Yvard}, {Zhao}, {Zorec}, {Zucker}, \& {Zwitter}}]{GAIA2023}
{Gaia Collaboration}, {Vallenari}, A., {Brown}, A.~G.~A., {et~al.} 2023, A\&A, 674, A1

\bibitem[{{Galvin} \& {Filipovic}(2014)}]{2014SerAJ.189...15G}
{Galvin}, T.~J., \& {Filipovic}, M.~D. 2014, Serbian Astronomical Journal, 189, 15

\bibitem[{{Ghavamian} {et~al.}(2003){Ghavamian}, {Rakowski}, {Hughes}, \& {Williams}}]{2003ApJ...590..833G}
{Ghavamian}, P., {Rakowski}, C.~E., {Hughes}, J.~P., \& {Williams}, T.~B. 2003, ApJ, 590, 833

\bibitem[{{Ghavamian} {et~al.}(2000){Ghavamian}, {Raymond}, {Hartigan}, \& {Blair}}]{2000ApJ...535..266G}
{Ghavamian}, P., {Raymond}, J., {Hartigan}, P., \& {Blair}, W.~P. 2000, ApJ, 535, 266

\bibitem[{{Green}(2022)}]{Green}
{Green}, D.~A. 2022, {A Catalogue of Galactic Supernova Remnants (2022 December version)}

\bibitem[{Green(2024)}]{Green2024_updatedcatalogue}
Green, D.~A. 2024, An updated catalogue of 310 Galactic supernova remnants and their statistical properties, arXiv:2411.03367

\bibitem[{Gupta {et~al.}(2022)Gupta, Huynh, Norris, Rosalind~Wang, Hopkins, Andernach, Koribalski, \& Galvin}]{Gupta2022}
Gupta, N., Huynh, M., Norris, R.~P., {et~al.} 2022, Publications of the Astronomical Society of Australia, 39, e051

\bibitem[{{Guzman} {et~al.}(2019){Guzman}, {Whiting}, {Voronkov}, {Mitchell}, {Ord}, {Collins}, {Marquarding}, {Lahur}, {Maher}, {Van Diepen}, {Bannister}, {Wu}, {Lenc}, {Khoo}, \& {Bastholm}}]{2019ascl.soft12003G}
{Guzman}, J., {Whiting}, M., {Voronkov}, M., {et~al.} 2019, {ASKAPsoft: ASKAP science data processor software}, Astrophysics Source Code Library, record ascl:1912.003

\bibitem[{{Hakobyan} {et~al.}(2017){Hakobyan}, {Barkhudaryan}, {Karapetyan}, {Mamon}, {Kunth}, {Adibekyan}, {Aramyan}, {Petrosian}, \& {Turatto}}]{2017MNRAS.471.1390H}
{Hakobyan}, A.~A., {Barkhudaryan}, L.~V., {Karapetyan}, A.~G., {et~al.} 2017, Mon. Not. R. Astron. Soc., 471, 1390

\bibitem[{{Hancock} {et~al.}(2018){Hancock}, {Trott}, \& {Hurley-Walker}}]{Hancock2018}
{Hancock}, P.~J., {Trott}, C.~M., \& {Hurley-Walker}, N. 2018, \pasa, 35, e011

\bibitem[{{H.E.S.S. Collaboration} {et~al.}(2018){H.E.S.S. Collaboration}, {Abdalla, H.}, {Abramowski, A.}, {Aharonian, F.}, {Ait Benkhali, F.}, {Angüner, E. O.}, {Arakawa, M.}, {Arrieta, M.}, {Aubert, P.}, {Backes, M.}, {Balzer, A.}, {Barnard, M.}, {Becherini, Y.}, {Becker Tjus, J.}, {Berge, D.}, {Bernhard, S.}, {Bernlöhr, K.}, {Blackwell, R.}, {Böttcher, M.}, {Boisson, C.}, {Bolmont, J.}, {Bonnefoy, S.}, {Bordas, P.}, {Bregeon, J.}, {Brun, F.}, {Brun, P.}, {Bryan, M.}, {Büchele, M.}, {Bulik, T.}, {Capasso, M.}, {Carrigan, S.}, {Caroff, S.}, {Carosi, A.}, {Casanova, S.}, {Cerruti, M.}, {Chakraborty, N.}, {Chaves, R. C. G.}, {Chen, A.}, {Chevalier, J.}, {Colafrancesco, S.}, {Condon, B.}, {Conrad, J.}, {Davids, I. D.}, {Decock, J.}, {Deil, C.}, {Devin, J.}, {deWilt, P.}, {Dirson, L.}, {Djannati-Ataï, A.}, {Domainko, W.}, {Donath, A.}, {Drury, L. O’C.}, {Dutson, K.}, {Dyks, J.}, {Edwards, T.}, {Egberts, K.}, {Eger, P.}, {Emery, G.}, {Ernenwein, J.-P.}, {Eschbach, S.}, {Farnier, C.}, {Fegan, S.},
  {Fernandes, M. V.}, {Fiasson, A.}, {Fontaine, G.}, {Förster, A.}, {Funk, S.}, {Füßling, M.}, {Gabici, S.}, {Gallant, Y. A.}, {Garrigoux, T.}, {Gast, H.}, {Gaté, F.}, {Giavitto, G.}, {Giebels, B.}, {Glawion, D.}, {Glicenstein, J. F.}, {Gottschall, D.}, {Grondin, M.-H.}, {Hahn, J.}, {Haupt, M.}, {Hawkes, J.}, {Heinzelmann, G.}, {Henri, G.}, {Hermann, G.}, {Hinton, J. A.}, {Hofmann, W.}, {Hoischen, C.}, {Holch, T. L.}, {Holler, M.}, {Horns, D.}, {Ivascenko, A.}, {Iwasaki, H.}, {Jacholkowska, A.}, {Jamrozy, M.}, {Jankowsky, D.}, {Jankowsky, F.}, {Jingo, M.}, {Jouvin, L.}, {Jung-Richardt, I.}, {Kastendieck, M. A.}, {Katarzyński, K.}, {Katsuragawa, M.}, {Katz, U.}, {Kerszberg, D.}, {Khangulyan, D.}, {Khélifi, B.}, {King, J.}, {Klepser, S.}, {Klochkov, D.}, {Kluźniak, W.}, {Komin, Nu.}, {Kosack, K.}, {Krakau, S.}, {Kraus, M.}, {Krüger, P. P.}, {Laffon, H.}, {Lamanna, G.}, {Lau, J.}, {Lees, J.-P.}, {Lefaucheur, J.}, {Lemière, A.}, {Lemoine-Goumard, M.}, {Lenain, J.-P.}, {Leser, E.}, {Lohse, T.}, {Lorentz,
  M.}, {Liu, R.}, {López-Coto, R.}, {Lypova, I.}, {Marandon, V.}, {Malyshev, D.}, {Marcowith, A.}, {Mariaud, C.}, {Marx, R.}, {Maurin, G.}, {Maxted, N.}, {Mayer, M.}, {Meintjes, P.J.}, {Meyer, M.}, {Mitchell, A. M. W.}, {Moderski, R.}, {Mohamed, M.}, {Mohrmann, L.}, {Morå, K.}, {Moulin, E.}, {Murach, T.}, {Nakashima, S.}, {de Naurois, M.}, {Ndiyavala, H.}, {Niederwanger, F.}, {Niemiec, J.}, {Oakes, L.}, {O’Brien, P.}, {Odaka, H.}, {Ohm, S.}, {Ostrowski, M.}, {Oya, I.}, {Padovani, M.}, {Panter, M.}, {Parsons, R. D.}, {Paz Arribas, M.}, {Pekeur, N. W.}, {Pelletier, G.}, {Perennes, C.}, {Petrucci, P.-O.}, {Peyaud, B.}, {Piel, Q.}, {Pita, S.}, {Poireau, V.}, {Poon, H.}, {Prokhorov, D.}, {Prokoph, H.}, {Pühlhofer, G.}, {Punch, M.}, {Quirrenbach, A.}, {Raab, S.}, {Rauth, R.}, {Reimer, A.}, {Reimer, O.}, {Renaud, M.}, {de los Reyes, R.}, {Rieger, F.}, {Rinchiuso, L.}, {Romoli, C.}, {Rowell, G.}, {Rudak, B.}, {Rulten, C. B.}, {Safi-Harb, S.}, {Sahakian, V.}, {Saito, S.}, {Sanchez, D. A.}, {Santangelo, A.},
  {Sasaki, M.}, {Schandri, M.}, {Schlickeiser, R.}, {Schüssler, F.}, {Schulz, A.}, {Schwanke, U.}, {Schwemmer, S.}, {Seglar-Arroyo, M.}, {Settimo, M.}, {Seyffert, A. S.}, {Shafi, N.}, {Shilon, I.}, {Shiningayamwe, K.}, {Simoni, R.}, {Sol, H.}, {Spanier, F.}, {Spir-Jacob, M.}, {Stawarz, Ł.}, {Steenkamp, R.}, {Stegmann, C.}, {Steppa, C.}, {Sushch, I.}, {Takahashi, T.}, {Tavernet, J.-P.}, {Tavernier, T.}, {Taylor, A. M.}, {Terrier, R.}, {Tibaldo, L.}, {Tiziani, D.}, {Tluczykont, M.}, {Trichard, C.}, {Tsirou, M.}, {Tsuji, N.}, {Tuffs, R.}, {Uchiyama, Y.}, {van der Walt, D. J.}, {van Eldik, C.}, {van Rensburg, C.}, {van Soelen, B.}, {Vasileiadis, G.}, {Veh, J.}, {Venter, C.}, {Viana, A.}, {Vincent, P.}, {Vink, J.}, {Voisin, F.}, {Völk, H. J.}, {Vuillaume, T.}, {Wadiasingh, Z.}, {Wagner, S. J.}, {Wagner, P.}, {Wagner, R. M.}, {White, R.}, {Wierzcholska, A.}, {Willmann, P.}, {Wörnlein, A.}, {Wouters, D.}, {Yang, R.}, {Zaborov, D.}, {Zacharias, M.}, {Zanin, R.}, {Zdziarski, A. A.}, {Zech, A.}, {Zefi, F.},
  {Ziegler, A.}, {Zorn, J.}, \& {Żywucka, N.}}]{HESS2018}
{H.E.S.S. Collaboration}, {Abdalla, H.}, {Abramowski, A.}, {et~al.} 2018, A\&A, 612, A1

\bibitem[{{H.E.S.S. Collaboration} {et~al.}(2020){H.E.S.S. Collaboration}, {Abdalla}, {Adam}, {Aharonian}, {Ait Benkhali}, {Ang{\"u}ner}, {Arakawa}, {Arcaro}, {Armand}, {Ashkar}, {Backes}, {Barbosa Martins}, {Barnard}, {Becherini}, {Berge}, {Bernl{\"o}hr}, {Blackwell}, {B{\"o}ttcher}, {Boisson}, {Bolmont}, {Bonnefoy}, {Bregeon}, {Breuhaus}, {Brun}, {Brun}, {Bryan}, {B{\"u}chele}, {Bulik}, {Bylund}, {Caroff}, {Carosi}, {Casanova}, {Cerruti}, {Chand}, {Chandra}, {Chaves}, {Chen}, {Colafrancesco}, {Cury{\l}o}, {Davids}, {Deil}, {Devin}, {deWilt}, {Dirson}, {Djannati-Ata{\"\i}}, {Dmytriiev}, {Donath}, {Doroshenko}, {Dyks}, {Egberts}, {Emery}, {Ernenwein}, {Eschbach}, {Feijen}, {Fegan}, {Fiasson}, {Fontaine}, {Funk}, {F{\"u}{\ss}ling}, {Gabici}, {Gallant}, {Gat{\'e}}, {Giavitto}, {Giunti}, {Glawion}, {Glicenstein}, {Gottschall}, {Grondin}, {Hahn}, {Haupt}, {Heinzelmann}, {Henri}, {Hermann}, {Hinton}, {Hofmann}, {Hoischen}, {Holch}, {Holler}, {Horns}, {Huber}, {Iwasaki}, {Jamrozy}, {Jankowsky}, {Jankowsky},
  {Jardin-Blicq}, {Jung-Richardt}, {Kastendieck}, {Katarzy{\'n}ski}, {Katsuragawa}, {Katz}, {Khangulyan}, {Kh{\'e}lifi}, {King}, {Klepser}, {Klu{\'z}niak}, {Komin}, {Kosack}, {Kostunin}, {Kreter}, {Lamanna}, {Lemi{\`e}re}, {Lemoine-Goumard}, {Lenain}, {Leser}, {Levy}, {Lohse}, {Lypova}, {Mackey}, {Majumdar}, {Malyshev}, {Malyshev}, {Marandon}, {Marcowith}, {Mares}, {Mariaud}, {Mart{\'\i}-Devesa}, {Marx}, {Maurin}, {Meintjes}, {Mitchell}, {Moderski}, {Mohamed}, {Mohrmann}, {Moore}, {Moulin}, {Muller}, {Murach}, {Nakashima}, {de Naurois}, {Ndiyavala}, {Niederwanger}, {Niemiec}, {Oakes}, {O'Brien}, {Odaka}, {Ohm}, {de Ona Wilhelmi}, {Ostrowski}, {Oya}, {Panter}, {Parsons}, {Perennes}, {Petrucci}, {Peyaud}, {Piel}, {Pita}, {Poireau}, {Priyana Noel}, {Prokhorov}, {Prokoph}, {P{\"u}hlhofer}, {Punch}, {Quirrenbach}, {Raab}, {Rauth}, {Reimer}, {Reimer}, {Remy}, {Renaud}, {Rieger}, {Rinchiuso}, {Romoli}, {Rowell}, {Rudak}, {Ruiz-Velasco}, {Sahakian}, {Sailer}, {Saito}, {Sanchez}, {Santangelo}, {Sasaki},
  {Schlickeiser}, {Sch{\"u}ssler}, {Schulz}, {Schutte}, {Schwanke}, {Schwemmer}, {Seglar-Arroyo}, {Senniappan}, {Seyffert}, {Shafi}, {Shiningayamwe}, {Simoni}, {Sinha}, {Sol}, {Specovius}, {Spir-Jacob}, {Stawarz}, {Steenkamp}, {Stegmann}, {Steppa}, {Takahashi}, {Tavernier}, {Taylor}, {Terrier}, {Tiziani}, {Tluczykont}, {Trichard}, {Tsirou}, {Tsuji}, {Tuffs}, {Uchiyama}, {van der Walt}, {van Eldik}, {van Rensburg}, {van Soelen}, {Vasileiadis}, {Veh}, {Venter}, {Vincent}, {Vink}, {V{\"o}lk}, {Vuillaume}, {Wadiasingh}, {Wagner}, {White}, {Wierzcholska}, {Yang}, {Yoneda}, {Zacharias}, {Zanin}, {Zdziarski}, {Zech}, {Zorn}, {{\.Z}ywucka}, \& {Bordas}}]{2020A&A...633A.102H}
{H.E.S.S. Collaboration}, {Abdalla}, H., {Adam}, R., {et~al.} 2020, \aap, 633, A102

\bibitem[{{H.E.S.S. Collaboration} {et~al.}(2024){H.E.S.S. Collaboration}, {Aharonian}, {Ait Benkhali}, {Aschersleben}, {Ashkar}, {Backes}, {Barbosa Martins}, {Batzofin}, {Becherini}, {Berge}, {Bernl{\"o}hr}, {B{\"o}ttcher}, {Boisson}, {Bolmont}, {de Bony de Lavergne}, {Borowska}, {Bouyahiaoui}, {Brose}, {Brown}, {Brun}, {Bruno}, {Bulik}, {Burger-Scheidlin}, {Caroff}, {Casanova}, {Celic}, {Cerruti}, {Chand}, {Chandra}, {Chen}, {Chibueze}, {Chibueze}, {Cotter}, {Damascene Mbarubucyeye}, {Devin}, {Djuvsland}, {Dmytriiev}, {Egberts}, {Einecke}, {Ernenwein}, {Fontaine}, {Funk}, {Gabici}, {Gallant}, {Glawion}, {Glicenstein}, {Goswami}, {Grolleron}, {Haerer}, {He{\ss}}, {Hofmann}, {Holch}, {Holler}, {Huang}, {Jamrozy}, {Jankowsky}, {Joshi}, {Jung-Richardt}, {Kasai}, {Katarzy{\'n}ski}, {Khangulyan}, {Khatoon}, {Kh{\'e}lifi}, {Klu{\'z}niak}, {Komin}, {Kosack}, {Kostunin}, {Kundu}, {Lang}, {Le Stum}, {Leitl}, {Lemi{\`e}re}, {Lemoine-Goumard}, {Lenain}, {Leuschner}, {Mackey}, {Malyshev}, {Mart{\'\i}-Devesa}, {Marx},
  {Mehta}, {Meintjes}, {Mitchell}, {Moderski}, {Mohrmann}, {Montanari}, {Moulin}, {Murach}, {de Naurois}, {Niemiec}, {Ohm}, {de Ona Wilhelmi}, {Ostrowski}, {Panny}, {Panter}, {Parsons}, {Pensec}, {Peron}, {Prokhorov}, {P{\"u}hlhofer}, {Punch}, {Quirrenbach}, {Regeard}, {Reimer}, {Reimer}, {Reis}, {Ren}, {Rieger}, {Rudak}, {Ruiz-Velasco}, {Sahakian}, {Salzmann}, {Santangelo}, {Sasaki}, {Sch{\"a}fer}, {Sch{\"u}ssler}, {Schutte}, {Shapopi}, {Spencer}, {Stawarz}, {Steenkamp}, {Steinmassl}, {Steppa}, {Streil}, {Sushch}, {Takahashi}, {Tanaka}, {Taylor}, {Terrier}, {Thorpe-Morgan}, {Tluczykont}, {Unbehaun}, {van Eldik}, {van Soelen}, {Vecchi}, {Venter}, {Vink}, {Wach}, {Wagner}, {Werner}, {Wierzcholska}, {Zacharias}, {Zdziarski}, {Zech}, \& {{\.Z}ywucka}}]{2024A&A...687A.219H}
{H.E.S.S. Collaboration}, {Aharonian}, F., {Ait Benkhali}, F., {et~al.} 2024, \aap, 687, A219

\bibitem[{{HI4PI Collaboration} {et~al.}(2016){HI4PI Collaboration}, {Ben Bekhti}, {Fl{\"o}er}, {Keller}, {Kerp}, {Lenz}, {Winkel}, {Bailin}, {Calabretta}, {Dedes}, {Ford}, {Gibson}, {Haud}, {Janowiecki}, {Kalberla}, {Lockman}, {McClure-Griffiths}, {Murphy}, {Nakanishi}, {Pisano}, \& {Staveley-Smith}}]{2016A&A...594A.116H}
{HI4PI Collaboration}, {Ben Bekhti}, N., {Fl{\"o}er}, L., {et~al.} 2016, A\&A, 594, A116

\bibitem[{Hou \& Han(2014)}]{Hou2014}
Hou, L.~G., \& Han, J.~L. 2014, A\&A, 569, A125

\bibitem[{{Hurley-Walker} {et~al.}(2017){Hurley-Walker}, {Callingham}, {Hancock}, {Franzen}, {Hindson}, {Kapi{\'n}ska}, {Morgan}, {Offringa}, {Wayth}, {Wu}, {Zheng}, {Murphy}, {Bell}, {Dwarakanath}, {For}, {Gaensler}, {Johnston-Hollitt}, {Lenc}, {Procopio}, {Staveley-Smith}, {Ekers}, {Bowman}, {Briggs}, {Cappallo}, {Deshpande}, {Greenhill}, {Hazelton}, {Kaplan}, {Lonsdale}, {McWhirter}, {Mitchell}, {Morales}, {Morgan}, {Oberoi}, {Ord}, {Prabu}, {Shankar}, {Srivani}, {Subrahmanyan}, {Tingay}, {Webster}, {Williams}, \& {Williams}}]{Hurley2017}
{Hurley-Walker}, N., {Callingham}, J.~R., {Hancock}, P.~J., {et~al.} 2017, \mnras, 464, 1146

\bibitem[{{Hurley-Walker} {et~al.}(2019{\natexlab{a}}){Hurley-Walker}, {Gaensler}, {Leahy}, {Filipovi{\'c}}, {Hancock}, {Franzen}, {Offringa}, {Callingham}, {Hindson}, {Wu}, {Bell}, {For}, {Johnston-Hollitt}, {Kapi{\'n}ska}, {Morgan}, {Murphy}, {McKinley}, {Procopio}, {Staveley-Smith}, {Wayth}, \& {Zheng}}]{2019PASA...36...48H}
{Hurley-Walker}, N., {Gaensler}, B.~M., {Leahy}, D.~A., {et~al.} 2019{\natexlab{a}}, PASA, 36, e048

\bibitem[{{Hurley-Walker} {et~al.}(2019{\natexlab{b}}){Hurley-Walker}, {Filipovi{\'c}}, {Gaensler}, {Leahy}, {Hancock}, {Franzen}, {Offringa}, {Callingham}, {Hindson}, {Wu}, {Bell}, {For}, {Johnston-Hollitt}, {Kapi{\'n}ska}, {Morgan}, {Murphy}, {McKinley}, {Procopio}, {Staveley-Smith}, {Wayth}, \& {Zheng}}]{2019PASA...36...45H}
{Hurley-Walker}, N., {Filipovi{\'c}}, M.~D., {Gaensler}, B.~M., {et~al.} 2019{\natexlab{b}}, PASA, 36, e045

\bibitem[{{Hurley-Walker} {et~al.}(2022){Hurley-Walker}, {Galvin}, {Duchesne}, {Zhang}, {Morgan}, {Hancock}, {An}, {Franzen}, {Heald}, {Ross}, {Vernstrom}, {Anderson}, {Gaensler}, {Johnston-Hollitt}, {Kaplan}, {Riseley}, {Tingay}, \& {Walker}}]{Hurley2022}
{Hurley-Walker}, N., {Galvin}, T.~J., {Duchesne}, S.~W., {et~al.} 2022, \pasa, 39, e035

\bibitem[{{Kavanagh} {et~al.}(2015){Kavanagh}, {Sasaki}, {Bozzetto}, {Filipovi{\'c}}, {Points}, {Maggi}, \& {Haberl}}]{2015A&A...573A..73K}
{Kavanagh}, P.~J., {Sasaki}, M., {Bozzetto}, L.~M., {et~al.} 2015, A\&A, 573, A73

\bibitem[{{Kavanagh} {et~al.}(2022){Kavanagh}, {Sasaki}, {Filipovi{\'c}}, {Points}, {Bozzetto}, {Haberl}, {Maggi}, \& {Maitra}}]{2022MNRAS.515.4099K}
{Kavanagh}, P.~J., {Sasaki}, M., {Filipovi{\'c}}, M.~D., {et~al.} 2022, \mnras, 515, 4099

\bibitem[{{Kavanagh} {et~al.}(2019){Kavanagh}, {Vink}, {Sasaki}, {Chu}, {Filipovi{\'c}}, {Ohm}, {Haberl}, {Manojlovic}, \& {Maggi}}]{2019A&A...621A.138K}
{Kavanagh}, P.~J., {Vink}, J., {Sasaki}, M., {et~al.} 2019, \aap, 621, A138

\bibitem[{{Kerr} \& {Lynden-Bell}(1986)}]{1986MNRAS.221.1023K}
{Kerr}, F.~J., \& {Lynden-Bell}, D. 1986, MNRAS, 221, 1023

\bibitem[{{Khabibullin} {et~al.}(2023){Khabibullin}, {Churazov}, {Bykov}, {Chugai}, \& {Sunyaev}}]{2023MNRAS.521.5536K}
{Khabibullin}, I.~I., {Churazov}, E.~M., {Bykov}, A.~M., {Chugai}, N.~N., \& {Sunyaev}, R.~A. 2023, MNRAS, 521, 5536

\bibitem[{Koribalski {et~al.}(2021)Koribalski, Norris, Andernach, Rudnick, Shabala, Filipović, \& Lenc}]{10.1093/mnrasl/slab041}
Koribalski, B.~S., Norris, R.~P., Andernach, H., {et~al.} 2021, MNRAS, 505, L11

\bibitem[{{Kosti{\'c}} {et~al.}(2024){Kosti{\'c}}, {Arbutina}, {Vukoti{\'c}}, \& {Uro{\v{s}}evi{\'c}}}]{Kosticetal2024}
{Kosti{\'c}}, P., {Arbutina}, B., {Vukoti{\'c}}, B., \& {Uro{\v{s}}evi{\'c}}, D. 2024, ApJ, 974, 236

\bibitem[{{Kothes} {et~al.}(2017){Kothes}, {Reich}, {Foster}, \& {Reich}}]{2017A&A...597A.116K}
{Kothes}, R., {Reich}, P., {Foster}, T.~J., \& {Reich}, W. 2017, A\&A, 597, A116

\bibitem[{{Kothes} \& {Reich}(2001)}]{Kothes2001}
{Kothes}, R., \& {Reich}, W. 2001, \aap, 372, 627

\bibitem[{{Lamer} {et~al.}(2021){Lamer}, {Schwope}, {Predehl}, {Traulsen}, {Wilms}, \& {Freyberg}}]{2021A&A...647A...7L}
{Lamer}, G., {Schwope}, A.~D., {Predehl}, P., {et~al.} 2021, A\&A, 647, A7

\bibitem[{{Lazarevi{\'c}} {et~al.}(2024){Lazarevi{\'c}}, {Filipovi{\'c}}, {Koribalski}, {Smeaton}, {Hopkins}, {Alsaberi}, {Velovi{\'c}}, {Ball}, {Kothes}, {Leahy}, \& {Ingallinera}}]{2024RNAAS...8..107L}
{Lazarevi{\'c}}, S., {Filipovi{\'c}}, M.~D., {Koribalski}, B.~S., {et~al.} 2024, Research Notes of the American Astronomical Society, 8, 107

\bibitem[{{Leahy} {et~al.}(2019){Leahy}, {Wang}, {Lawton}, {Ranasinghe}, \& {Filipovi{\'c}}}]{2019AJ....158..149L}
{Leahy}, D., {Wang}, Y., {Lawton}, B., {Ranasinghe}, S., \& {Filipovi{\'c}}, M. 2019, AJ, 158, 149

\bibitem[{Liu {et~al.}(2018)Liu, Chen, Chen, Zhou, Wang, \& Su}]{Liu_2018}
Liu, Q.-C., Chen, Y., Chen, B.-Q., {et~al.} 2018, The Astrophysical Journal, 859, 173

\bibitem[{{Lopez} {et~al.}(2011){Lopez}, {Ramirez-Ruiz}, {Huppenkothen}, {Badenes}, \& {Pooley}}]{2011ApJ...732..114L}
{Lopez}, L.~A., {Ramirez-Ruiz}, E., {Huppenkothen}, D., {Badenes}, C., \& {Pooley}, D.~A. 2011, \apj, 732, 114

\bibitem[{{L{\'o}pez-Sanjuan} {et~al.}(2019){L{\'o}pez-Sanjuan}, {Varela}, {Crist{\'o}bal-Hornillos}, {V{\'a}zquez Rami{\'o}}, {Carrasco}, {Tremblay}, {Whitten}, {Placco}, {Mar{\'\i}n-Franch}, {Cenarro}, {Ederoclite}, {Alfaro}, {Coelho}, {Civera}, {Hern{\'a}ndez-Fuertes}, {Jim{\'e}nez-Esteban}, {Jim{\'e}nez-Teja}, {Ma{\'\i}z Apell{\'a}niz}, {Sobral}, {V{\'\i}lchez}, {Alcaniz}, {Angulo}, {Dupke}, {Hern{\'a}ndez-Monteagudo}, {Mendes de Oliveira}, {Moles}, \& {Sodr{\'e}}}]{2019A&A...631A.119L}
{L{\'o}pez-Sanjuan}, C., {Varela}, J., {Crist{\'o}bal-Hornillos}, D., {et~al.} 2019, A\&A, 631, A119

\bibitem[{{Luken} {et~al.}(2020){Luken}, {Filipovi{\'c}}, {Maxted}, {Kothes}, {Norris}, {Allison}, {Blackwell}, {Braiding}, {Brose}, {Burton}, {De Horta}, {Galvin}, {Harvey-Smith}, {Hurley-Walker}, {Leahy}, {Ralph}, {Roper}, {Rowell}, {Sushch}, {Uro{\v{s}}evi{\'c}}, \& {Wong}}]{2020MNRAS.492.2606L}
{Luken}, K.~J., {Filipovi{\'c}}, M.~D., {Maxted}, N.~I., {et~al.} 2020, MNRAS, 492, 2606

\bibitem[{{Maggi} {et~al.}(2019){Maggi}, {Filipovi{\'c}}, {Vukoti{\'c}}, {Ballet}, {Haberl}, {Maitra}, {Kavanagh}, {Sasaki}, \& {Stupar}}]{2019A&A...631A.127M}
{Maggi}, P., {Filipovi{\'c}}, M.~D., {Vukoti{\'c}}, B., {et~al.} 2019, Astron. \& Astrophys., 631, A127

\bibitem[{{Mantovanini} {et~al.}(2025){Mantovanini}, {Hurley-Walker}, \& {Anderson}}]{2025PASA...42...21M}
{Mantovanini}, S., {Hurley-Walker}, N., \& {Anderson}, G. 2025, \pasa, 42, e021

\bibitem[{{Mattox} {et~al.}(1996){Mattox}, {Bertsch}, {Chiang}, {Dingus}, {Digel}, {Esposito}, {Fierro}, {Hartman}, {Hunter}, {Kanbach}, \& et~al.}]{1996ApJ...461..396M}
{Mattox}, J.~R., {Bertsch}, D.~L., {Chiang}, J., {et~al.} 1996, ApJ, 461, 396

\bibitem[{{McKee} \& {Truelove}(1995)}]{Mckee1995}
{McKee}, C.~F., \& {Truelove}, J.~K. 1995, \physrep, 256, 157

\bibitem[{{Meyer} {et~al.}(2015){Meyer}, {Langer}, {Mackey}, {Vel{\'a}zquez}, \& {Gusdorf}}]{2015MNRAS.450.3080M}
{Meyer}, D.~M.~A., {Langer}, N., {Mackey}, J., {Vel{\'a}zquez}, P.~F., \& {Gusdorf}, A. 2015, \mnras, 450, 3080

\bibitem[{Norris {et~al.}(2021)Norris, Crawford, \& Macgregor}]{galaxies9040083}
Norris, R.~P., Crawford, E., \& Macgregor, P. 2021, Galaxies, 9, doi:10.3390/galaxies9040083

\bibitem[{{Norris} {et~al.}(2011){Norris}, {Hopkins}, {Afonso}, {Brown}, {Condon}, {Dunne}, {Feain}, {Hollow}, {Jarvis}, {Johnston-Hollitt}, {Lenc}, {Middelberg}, {Padovani}, {Prandoni}, {Rudnick}, {Seymour}, {Umana}, {Andernach}, {Alexander}, {Appleton}, {Bacon}, {Banfield}, {Becker}, {Brown}, {Ciliegi}, {Jackson}, {Eales}, {Edge}, {Gaensler}, {Giovannini}, {Hales}, {Hancock}, {Huynh}, {Ibar}, {Ivison}, {Kennicutt}, {Kimball}, {Koekemoer}, {Koribalski}, {L{\'o}pez-S{\'a}nchez}, {Mao}, {Murphy}, {Messias}, {Pimbblet}, {Raccanelli}, {Randall}, {Reiprich}, {Roseboom}, {R{\"o}ttgering}, {Saikia}, {Sharp}, {Slee}, {Smail}, {Thompson}, {Urquhart}, {Wall}, \& {Zhao}}]{Norris2011}
{Norris}, R.~P., {Hopkins}, A.~M., {Afonso}, J., {et~al.} 2011, PASA, 28, 215

\bibitem[{{Norris} {et~al.}(2021){Norris}, {Marvil}, {Collier}, {Kapi{\'n}ska}, {O'Brien}, {Rudnick}, {Andernach}, {Asorey}, {Brown}, {Br{\"u}ggen}, {Crawford}, {English}, {Rahman}, {Filipovi{\'c}}, {Gordon}, {G{\"u}rkan}, {Hale}, {Hopkins}, {Huynh}, {HyeongHan}, {James Jee}, {Koribalski}, {Lenc}, {Luken}, {Parkinson}, {Prandoni}, {Raja}, {Reiprich}, {Riseley}, {Shabala}, {Sheil}, {Vernstrom}, {Whiting}, {Allison}, {Anderson}, {Ball}, {Bell}, {Bunton}, {Galvin}, {Gupta}, {Hotan}, {Jacka}, {Macgregor}, {Mahony}, {Maio}, {Moss}, {Pandey-Pommier}, \& {Voronkov}}]{Norris2021}
{Norris}, R.~P., {Marvil}, J., {Collier}, J.~D., {et~al.} 2021, PASA, 38, e046

\bibitem[{Norris {et~al.}(2021)Norris, Intema, Kapińska, Koribalski, Lenc, Rudnick, Alsaberi, Anderson, Anderson, Crawford, \& et~al.}]{Norris2021ORC}
Norris, R.~P., Intema, H.~T., Kapińska, A.~D., {et~al.} 2021, PASA, 38, e003

\bibitem[{{Oh} {et~al.}(2015){Oh}, {Kroupa}, \& {Pflamm-Altenburg}}]{OhKroPfl15}
{Oh}, S., {Kroupa}, P., \& {Pflamm-Altenburg}, J. 2015, \apj, 805, 92

\bibitem[{{Pavlovi{\'c}} {et~al.}(2018){Pavlovi{\'c}}, {Uro{\v{s}}evi{\'c}}, {Arbutina}, {Orlando}, {Maxted}, \& {Filipovi{\'c}}}]{2018ApJ...852...84P}
{Pavlovi{\'c}}, M.~Z., {Uro{\v{s}}evi{\'c}}, D., {Arbutina}, B., {et~al.} 2018, Astrophys. J., 852, 84

\bibitem[{{Poggio} {et~al.}(2018){Poggio}, {Drimmel}, {Lattanzi}, {Smart}, {Spagna}, {Andrae}, {Bailer-Jones}, {Fouesneau}, {Antoja}, {Babusiaux}, {Evans}, {Figueras}, {Katz}, {Reyl{\'e}}, {Robin}, {Romero-G{\'o}mez}, \& {Seabroke}}]{2018MNRAS.481L..21P}
{Poggio}, E., {Drimmel}, R., {Lattanzi}, M.~G., {et~al.} 2018, MNRAS, 481, L21

\bibitem[{Ranasinghe \& Leahy(2019)}]{Ranasinghe:2019quc}
Ranasinghe, S., \& Leahy, D. 2019, JHEP Grav. Cosmol., 5, 907

\bibitem[{{Ranasinghe} \& {Leahy}(2022)}]{2022ApJ...940...63R}
{Ranasinghe}, S., \& {Leahy}, D. 2022, ApJ, 940, 63

\bibitem[{{Ranasinghe} \& {Leahy}(2023)}]{2023ApJS..265...53R}
---. 2023, ApJs, 265, 53

\bibitem[{{Ranasinghe} {et~al.}(2021){Ranasinghe}, {Leahy}, \& {Stil}}]{2021Univ....7..338R}
{Ranasinghe}, S., {Leahy}, D., \& {Stil}, J. 2021, Universe, 7, 338

\bibitem[{{Reynolds} {et~al.}(2012){Reynolds}, {Gaensler}, \& {Bocchino}}]{2012SSRv..166..231R}
{Reynolds}, S.~P., {Gaensler}, B.~M., \& {Bocchino}, F. 2012, Space Science Reviews, 166, 231

\bibitem[{{Roper} {et~al.}(2018){Roper}, {Filipovic}, {Allen}, {Sano}, {Park}, {Pannuti}, {Sasaki}, {Haberl}, {Kavanagh}, {Yamane}, {Yoshiike}, {Fujii}, {Fukui}, \& {Seitenzahl}}]{2018MNRAS.479.1800R}
{Roper}, Q., {Filipovic}, M., {Allen}, G.~E., {et~al.} 2018, MNRAS, 479, 1800

\bibitem[{{Ross} {et~al.}(2024){Ross}, {Hurley-Walker}, {Galvin}, {Venville}, {Duchesne}, {Morgan}, {An}, {Gurkan}, {Hancock}, {Heald}, {Johnston-Hollitt}, \& {White}}]{Ross2024}
{Ross}, K., {Hurley-Walker}, N., {Galvin}, T.~J., {et~al.} 2024, arXiv e-prints, arXiv:2406.06921

\bibitem[{{Rudnick}(2002)}]{2002NewAR..46..101R}
{Rudnick}, L. 2002, \nar, 46, 101

\bibitem[{{Sano} {et~al.}(2017){Sano}, {Yamane}, {Voisin}, {Fujii}, {Yoshiike}, {Inaba}, {Tsuge}, {Babazaki}, {Mitsuishi}, {Yang}, {Aharonian}, {Rowell}, {Filipovi{\'c}}, {Mizuno}, {Tachihara}, {Kawamura}, {Onishi}, \& {Fukui}}]{2017ApJ...843...61S}
{Sano}, H., {Yamane}, Y., {Voisin}, F., {et~al.} 2017, ApJ, 843, 61

\bibitem[{Sano {et~al.}(2017)Sano, Reynoso, Mitsuishi, Nakamura, Furukawa, Mruganka, Fukuda, Yoshiike, Nishimura, Ohama, Torii, Kuwahara, Okuda, Yamamoto, Tachihara, \& Fukui}]{SANO20171}
Sano, H., Reynoso, E., Mitsuishi, I., {et~al.} 2017, Journal of High Energy Astrophysics, 15, 1

\bibitem[{Sano {et~al.}(2018)Sano, Yamane, Tokuda, Fujii, Tsuge, Nagaya, Yoshiike, Filipović, Alsaberi, Barnes, Onishi, Kawamura, Minamidani, Mizuno, Yamamoto, Tachihara, Maxted, Voisin, Rowell, Yamaguchi, \& Fukui}]{Sano_2018}
Sano, H., Yamane, Y., Tokuda, K., {et~al.} 2018, The Astrophysical Journal, 867, 7

\bibitem[{{Sasaki} {et~al.}(2025){Sasaki}, {Zangrandi}, {Filipovi{\'c}}, {Alsaberi}, {Collier}, {Haberl}, {Heywood}, {Kavanagh}, {Koribalski}, {Kothes}, {Lazarevi{\'c}}, {Maggi}, {Maitra}, {Points}, {Smeaton}, \& {Velovi{\'c}}}]{2025A&A...693L..15S}
{Sasaki}, M., {Zangrandi}, F., {Filipovi{\'c}}, M., {et~al.} 2025, \aap, 693, L15

\bibitem[{{Shabala} {et~al.}(2024){Shabala}, {Yates-Jones}, {Jerrim}, {Turner}, {Krause}, {Norris}, {Koribalski}, {Filipovi{\'c}}, {Rudnick}, {Power}, \& {Crocker}}]{Shabala2024}
{Shabala}, S.~S., {Yates-Jones}, P.~M., {Jerrim}, L.~A., {et~al.} 2024, PASA, 41, e024

\bibitem[{{Smeaton} {et~al.}(2024{\natexlab{a}}){Smeaton}, {Filipovi{\'c}}, {Koribalski}, {Lazarevi{\'c}}, {Alsaberi}, {Becker}, {Dage}, {Gordon}, {Hopkins}, {Kothes}, {Leahy}, \& {Mitras̆inovi{\'c}}}]{2024RNAAS...8..158S}
{Smeaton}, Z.~J., {Filipovi{\'c}}, M.~D., {Koribalski}, B.~S., {et~al.} 2024{\natexlab{a}}, Research Notes of the American Astronomical Society, 8, 158

\bibitem[{{Smeaton} {et~al.}(2024{\natexlab{b}}){Smeaton}, {Filipovi{\'c}}, {Lazarevi{\'c}}, {Alsaberi}, {Ahmad}, {Araya}, {Ball}, {Bordiu}, {Buemi}, {Bufano}, {Dai}, {Haberl}, {Hopkins}, {Ingallinera}, {Jarrett}, {Koribalski}, {Kothes}, {Kraan-Korteweg}, {Leahy}, {Lundqvist}, {Maitra}, {Martin}, {Payne}, {Rowell}, {Sano}, {Sasaki}, {Soria}, {Steyn}, {Umana}, {Uro{\v{s}}evi{\'c}}, {Velovi{\'c}}, {Vernstrom}, {Vukoti{\'c}}, \& {West}}]{2024MNRAS.534.2918S}
{Smeaton}, Z.~J., {Filipovi{\'c}}, M.~D., {Lazarevi{\'c}}, S., {et~al.} 2024{\natexlab{b}}, \mnras, 534, 2918

\bibitem[{{Smith} \& {Tombleson}(2015)}]{2015MNRAS.447..598S}
{Smith}, N., \& {Tombleson}, R. 2015, MNRAS, 447, 598

\bibitem[{{Soker}(2019)}]{2019NewAR..8701535S}
{Soker}, N. 2019, \nar, 87, 101535

\bibitem[{{Soker}(2024{\natexlab{a}})}]{2024RAA....24a5012S}
---. 2024{\natexlab{a}}, Research in Astronomy and Astrophysics, 24, 015012

\bibitem[{{Soker}(2024{\natexlab{b}})}]{2024OJAp....7E..31S}
---. 2024{\natexlab{b}}, The Open Journal of Astrophysics, 7, 31

\bibitem[{{Srivastav} {et~al.}(2022){Srivastav}, {Smartt}, {Huber}, {Chambers}, {Angus}, {Chen}, {Callan}, {Gillanders}, {McBrien}, {Sim}, {Fulton}, {Hjorth}, {Smith}, {Young}, {Auchettl}, {Anderson}, {Pignata}, {de Boer}, {Lin}, \& {Magnier}}]{2022MNRAS.511.2708S}
{Srivastav}, S., {Smartt}, S.~J., {Huber}, M.~E., {et~al.} 2022, MNRAS, 511, 2708

\bibitem[{Su {et~al.}(2009)Su, Chen, Yang, Koo, Zhou, Jeong, \& Zhang}]{Su_2009}
Su, Y., Chen, Y., Yang, J., {et~al.} 2009, The Astrophysical Journal, 694, 376

\bibitem[{{Sushch} {et~al.}(2022){Sushch}, {Brose}, {Pohl}, {Plotko}, \& {Das}}]{2022ApJ...926..140S}
{Sushch}, I., {Brose}, R., {Pohl}, M., {Plotko}, P., \& {Das}, S. 2022, \apj, 926, 140

\bibitem[{{Tingay} {et~al.}(2013){Tingay}, {Goeke}, {Bowman}, {Emrich}, {Ord}, {Mitchell}, {Morales}, {Booler}, {Crosse}, {Wayth}, {Lonsdale}, {Tremblay}, {Pallot}, {Colegate}, {Wicenec}, {Kudryavtseva}, {Arcus}, {Barnes}, {Bernardi}, {Briggs}, {Burns}, {Bunton}, {Cappallo}, {Corey}, {Deshpande}, {Desouza}, {Gaensler}, {Greenhill}, {Hall}, {Hazelton}, {Herne}, {Hewitt}, {Johnston-Hollitt}, {Kaplan}, {Kasper}, {Kincaid}, {Koenig}, {Kratzenberg}, {Lynch}, {Mckinley}, {Mcwhirter}, {Morgan}, {Oberoi}, {Pathikulangara}, {Prabu}, {Remillard}, {Rogers}, {Roshi}, {Salah}, {Sault}, {Udaya-Shankar}, {Schlagenhaufer}, {Srivani}, {Stevens}, {Subrahmanyan}, {Waterson}, {Webster}, {Whitney}, {Williams}, {Williams}, \& {Wyithe}}]{Tingay2013}
{Tingay}, S.~J., {Goeke}, R., {Bowman}, J.~D., {et~al.} 2013, \pasa, 30, e007

\bibitem[{{Uro{\v{s}}evi{\'c}}(2020)}]{2020NatAs...4..910U}
{Uro{\v{s}}evi{\'c}}, D. 2020, Nature Astronomy, 4, 910

\bibitem[{Vallée(2017)}]{Vallee2017}
Vallée, J.~P. 2017, Astronomical Review, 13, 113

\bibitem[{Verberne \& Vink(2021)}]{Verberne2021}
Verberne, S., \& Vink, J. 2021, Monthly Notices of the Royal Astronomical Society, 504, 1536

\bibitem[{{Vukoti{\'c}} {et~al.}(2019){Vukoti{\'c}}, {{\'C}iprijanovi{\'c}}, {Vu{\v{c}}eti{\'c}}, {Oni{\'c}}, \& {Uro{\v{s}}evi{\'c}}}]{2019SerAJ.199...23S}
{Vukoti{\'c}}, B., {{\'C}iprijanovi{\'c}}, A., {Vu{\v{c}}eti{\'c}}, M.~M., {Oni{\'c}}, D., \& {Uro{\v{s}}evi{\'c}}, D. 2019, Serbian Astronomical Journal, 199, 23

\bibitem[{{Vukoti{\'c}} {et~al.}(2021){Vukoti{\'c}}, {{\'C}irkovi{\'c}}, \& {Filipovi{\'c}}}]{2021map..book...11V}
{Vukoti{\'c}}, B., {{\'C}irkovi{\'c}}, M.~M., \& {Filipovi{\'c}}, M.~D. 2021, in Multimessenger Astronomy in Practice: Celestial Sources in Action, ed. M.~D. {Filipovi{\'c}} \& N.~F.~H. {Tothill} (IOP Publishing), 11--1

\bibitem[{{Wayth} {et~al.}(2015){Wayth}, {Lenc}, {Bell}, {Callingham}, {Dwarakanath}, {Franzen}, {For}, {Gaensler}, {Hancock}, {Hindson}, {Hurley-Walker}, {Jackson}, {Johnston-Hollitt}, {Kapi{\'n}ska}, {McKinley}, {Morgan}, {Offringa}, {Procopio}, {Staveley-Smith}, {Wu}, {Zheng}, {Trott}, {Bernardi}, {Bowman}, {Briggs}, {Cappallo}, {Corey}, {Deshpande}, {Emrich}, {Goeke}, {Greenhill}, {Hazelton}, {Kaplan}, {Kasper}, {Kratzenberg}, {Lonsdale}, {Lynch}, {McWhirter}, {Mitchell}, {Morales}, {Morgan}, {Oberoi}, {Ord}, {Prabu}, {Rogers}, {Roshi}, {Shankar}, {Srivani}, {Subrahmanyan}, {Tingay}, {Waterson}, {Webster}, {Whitney}, {Williams}, \& {Williams}}]{Wayth2015}
{Wayth}, R.~B., {Lenc}, E., {Bell}, M.~E., {et~al.} 2015, \pasa, 32, e025

\bibitem[{{Wayth} {et~al.}(2018){Wayth}, {Tingay}, {Trott}, {Emrich}, {Johnston-Hollitt}, {McKinley}, {Gaensler}, {Beardsley}, {Booler}, {Crosse}, {Franzen}, {Horsley}, {Kaplan}, {Kenney}, {Morales}, {Pallot}, {Sleap}, {Steele}, {Walker}, {Williams}, {Wu}, {Cairns}, {Filipovic}, {Johnston}, {Murphy}, {Quinn}, {Staveley-Smith}, {Webster}, \& {Wyithe}}]{Wayth2018}
{Wayth}, R.~B., {Tingay}, S.~J., {Trott}, C.~M., {et~al.} 2018, \pasa, 35, e033

\bibitem[{{Weaver} {et~al.}(1977){Weaver}, {McCray}, {Castor}, {Shapiro}, \& {Moore}}]{1977ApJ...218..377W}
{Weaver}, R., {McCray}, R., {Castor}, J., {Shapiro}, P., \& {Moore}, R. 1977, \apj, 218, 377

\bibitem[{{We{\ss}mayer} {et~al.}(2024){We{\ss}mayer}, {Urbaneja}, {Butler}, \& {Przybilla}}]{2024A&A...687L...7W}
{We{\ss}mayer}, D., {Urbaneja}, M.~A., {Butler}, K., \& {Przybilla}, N. 2024, A\&A, 687, L7

\bibitem[{West {et~al.}(2016)West, Safi-Harb, Jaffe, Kothes, Landecker, \& Foster}]{West2016}
West, J.~L., Safi-Harb, S., Jaffe, T., {et~al.} 2016, A\&A, 587, A148

\bibitem[{{Wright}(2020)}]{2020SerAJ.200....1W}
{Wright}, J.~T. 2020, Serbian Astronomical Journal, 200, 1

\bibitem[{{Yamane} {et~al.}(2021){Yamane}, {Sano}, {Filipovi{\'c}}, {Tokuda}, {Fujii}, {Babazaki}, {Mitsuishi}, {Inoue}, {Aharonian}, {Inaba}, {Inutsuka}, {Maxted}, {Mizuno}, {Onishi}, {Rowell}, {Tsuge}, {Voisin}, {Yoshiike}, {Fukuda}, {Kawamura}, {Bamba}, {Tachihara}, \& {Fukui}}]{2021ApJ...918...36Y}
{Yamane}, Y., {Sano}, H., {Filipovi{\'c}}, M.~D., {et~al.} 2021, ApJ, 918, 36

\bibitem[{{Yao} {et~al.}(2017){Yao}, {Manchester}, \& {Wang}}]{2017ApJ...835...29Y}
{Yao}, J.~M., {Manchester}, R.~N., \& {Wang}, N. 2017, ApJ, 835, 29

\bibitem[{{Zekovi{\'c}} {et~al.}(2024){Zekovi{\'c}}, {Spitkovsky}, \& {Hemler}}]{Zekovicetal2024}
{Zekovi{\'c}}, V., {Spitkovsky}, A., \& {Hemler}, Z. 2024, arXiv e-prints, arXiv:2408.02084

\bibitem[{{Zhang} {et~al.}(2018){Zhang}, {Tian}, \& {Wu}}]{2018ApJ...867...61Z}
{Zhang}, M.~F., {Tian}, W.~W., \& {Wu}, D. 2018, ApJ, 867, 61

\end{thebibliography}

\end{document}